\documentclass[12pt,a4paper,oneside]{report}
\usepackage[utf8]{inputenc}
\usepackage{amsmath}
\usepackage{amsfonts}
\usepackage{amssymb}
\usepackage{makeidx}
\usepackage[toc,page]{appendix} 
\usepackage{pdflscape} 
\usepackage{hyperref} 
\usepackage[left=4cm,right=2cm,top=2cm,bottom=2cm]{geometry} 

\usepackage{verbatim}
\usepackage{listing}
\usepackage{parskip} 
\usepackage{setspace} 

\usepackage{graphicx}
\usepackage{subfig}

\usepackage{placeins} 
\usepackage{floatflt}
\usepackage{wrapfig}
\usepackage{rotating}
\usepackage{dcolumn}

\usepackage{cite}	
\usepackage{chicago}
\begin{document}
\sloppy 

\begin{titlepage}
\begin{center}
\noindent
\vspace{50 mm}

{\LARGE \textbf{ Entropy Analysis of Financial Time Series}}

\bigskip
\bigskip
\bigskip

{\large A thesis submitted to The University of Manchester for the degree of Doctor in Business Administration.}

\vfill

{\Large \textbf{2015}}

\vfill

{\large \textbf{Stephan Schwill}}

\bigskip
\bigskip
\bigskip

{\large \textbf{Manchester Business School}}
\vspace{30 mm}

\end{center}

\end{titlepage}
\addtocounter{page}{1}

\clearpage
\tableofcontents
\setcounter{tocdepth}{2}
\vfill

\listoftables
\listoffigures

\clearpage
\begin{center}
{\large Entropy Analysis of Financial Time Series}

\bigskip

{The University of Manchester}

\medskip

{Stephan Schwill}

\medskip

{Doctor of Business Administration}

\medskip

\the\year
\end{center}

\section*{\centering \large Abstract}

This thesis applies entropy as a model independent measure to address research questions concerning the dynamics of various financial time series. The thesis consists of three main studies as presented in chapters \ref{chap_paper1}, \ref{chap_paper2} and \ref{chap_paper3}.  Chapters \ref{chap_paper1} and \ref{chap_paper2} apply an entropy measure to conduct a bivariate analysis of drawdowns and drawups in foreign exchange rates. Chapter \ref{chap_paper3} investigates the dynamics of investment strategies of hedge funds using entropy of realised volatility in a conditioning model. 

In all three studies, methods from information theory are applied in novel ways to financial time series. As Information Theory and its central concept of entropy are not widely used in the economic sciences, a methodology chapter was therefore included in chapter \ref{chap_entropy} that gives an overview on the theoretical background and statistical features of the entropy measures used in the three main studies. In the first two studies the focus is on mutual information and transfer entropy. Both measures are used to identify dependencies between two exchange rates. The chosen measures generalise, in a well defined manner, correlation and Granger causality. A different entropy measure, the approximate entropy, is used in the third study to analyse the serial structure of S\&P realised volatility.

The study of drawdowns and drawups has so far been concentrated on their univariate characteristics. Encoding the drawdown information of a time series into a time series of discrete values, Chapter \ref{chap_paper1} uses entropy measures to analyse the correlation and cross correlations of drawdowns and drawups.  The method to encode the drawdown information is explained and applied to daily and hourly EUR/USD and GBP/USD exchange rates from 2001 to 2012. For the daily series, we find evidence of dependence among the largest draws (i.e. 5\% and 95\% quantiles), but it is not as strong as the correlation between the daily returns of the same pair of FX rates. There is also dependence between lead/lagged values of these draws. Similar and stronger findings were found among the hourly data. We further use transfer entropy to examine the spill over and lead-lag information flow between drawup/drawdown of the two exchange rates. Such information flow is indeed detectable in both daily and hourly data. The amount of information transferred is considerably higher for the hourly than the daily data. Both daily and hourly series show clear evidence of information flowing from EUR/USD to GBP/USD and, slightly stronger, in the reverse direction. Robustness tests, using effective transfer entropy, show that the information measured is not due to noise. 

Chapter \ref{chap_paper2} uses state space models of volatility to investigate volatility spill overs between exchange rates. Our use of entropy related measures in the investigation of dependencies of two state space series is novel. A set of five daily exchange rates from emerging and developed economies against the dollar over the period 1999 to 2012 is used. We find that among the currency pairs, the co-movement of EUR/USD and CHF/USD volatility states show the strongest observed relationship. With the use of transfer entropy, we find evidence for information flows between the volatility state series of AUD, CAD and BRL. 

Chapter \ref{chap_paper3} uses the entropy of S\&P realised volatility in detecting changes of volatility regime in order to re-examine the theme of market volatility timing of hedge funds.  A one-factor model is used, conditioned on information about the entropy of market volatility, to measure the dynamic of hedge funds equity exposure. On a cross section of around 2500 hedge funds with a focus on the US equity markets we find that, over the period from 2000 to 2014, hedge funds adjust their exposure dynamically in response to changes in volatility regime. This adds to the literature on the volatility timing behaviour of hedge fund manager, but using entropy as a model independent measure of volatility regime.
Finally, chapter \ref{chap_conclusion} summarises and concludes with some suggestions for future research.

\clearpage

\section*{\centering \large Declaration}
No portion of the work referred to in the thesis has been submitted in support of an application for another degree or 
qualification of this or any other university or other institute of learning.

\clearpage

\section*{\centering Copyright}
The author of this thesis (including any appendices and/or schedules to this thesis) owns certain copyright or related rights in it (the ``Copyright'') and s/he has given The University of Manchester certain rights to use such Copyright, including for administrative purposes.

Copies of this thesis, either in full or in extracts and whether in hard or electronic copy, may be made only in accordance with the Copyright, Designs and Patents Act 1988 (as amended) and regulations issued under it or, where appropriate, in
accordance with licensing agreements which the University has from time to time. This page must form part of any such copies made.

The ownership of certain Copyright, patents, designs, trade marks and other intellectual property (the ``Intellectual Property'') and any reproductions of copyright works in the thesis, for example graphs and tables (``Reproductions''), which may be described in this thesis, may not be owned by the author and may be owned by third parties. Such Intellectual Property and Reproductions cannot and must not be made available for use without the prior written permission of the owner(s) of the relevant Intellectual Property and/or Reproductions.

Further information on the conditions under which disclosure, publication and commercialisation of this thesis, the Copyright and any Intellectual Property and/or Reproductions described in it may take place is available in the University IP Policy (see \url{http://documents.manchester.ac.uk/DocuInfo.aspx?DocID=487}), in any relevant Thesis restriction declarations deposited in the University Library, The University Library's regulations (see \url{http://www.manchester.ac.uk/library/aboutus/regulations}) and in The University's policy on Presentation of Theses.

\clearpage

\section*{\centering Acknowledge}
My utmost thanks and sincere gratitude go to my supervisor, Prof. Dr. Ser-Huang Poon, who guided and supported me throughout this journey.

I would like to dedicate this work to my love, Doreen, whose unwavering support, and encouragement during the ups and downs of a long study have made this thesis possible. And to my parents, who instilled me with a sense of wonder and curiosity about the world around me.

\clearpage

\doublespacing

\chapter{Introduction}\label{chap_intro}
The work of \citeN{Shannon:1948wk} on the theory of communication, and \citeN{Jaynes:1963ua} ideas on the maximum entropy principle, to name just two authors in this area, has led to the development of concepts that have been applied to many research areas. Questions about information, uncertainty, entropy and ignorance play a role in the theory at a fundamental level; \textit{information} is a decrease in ambiguity, \textit{uncertainty} is a state of knowledge in which logical reasoning is impossible, \textit{entropy} is expected information and \textit{ignorance} is not knowing that there is uncertainty. At a more concrete level, these concepts have been developed and used to generalize important results in statistics, such as the Cram\'er-Rao inequality \citeN{Kullback:1954ua}\footnote{For an excellent and comprehensive review of Information Theory and Entropy Econometrics with its history, please consult \citeN{Golan:uk}.} and led to a better understanding of non-linear time series, dynamical systems and complexity.

This thesis consists of three studies in two research topics of finance, which have, as a common thread, the application of tools from information theory. Extreme market events, spillover and interdependence between foreign exchange (FX) markets is the first research topic. In chapter 3, crashes in currency pairs are analysed via drawdowns and drawups, for which a concept for the bivariate analysis is developed that uses mutual information and transfer entropy as information theoretic tools. FX markets are also studied in chapter 4, focusing on volatility spillovers between currency pairs, which are measured using the same entropy tools. Chapter 5 is devoted to analysing a hedge fund manager's ability to time changes in market volatility regimes. A conditional factor model is built, using approximate entropy to identify volatility regime changes in equity markets. As information theory and its central concept of entropy are not widely used in the economic sciences, we provide in chapter 2 an overview of the theoretical background and statistical features of the entropy measures used in the three main studies. The following subsections explain the research motivation and findings of chapters 3, 4 and 5.

\section{Bivariate Analysis of Drawdowns}
In financial time series, drawdowns (drawups) are defined as the cumulative return from a local maximum (minimum) to the next local minimum (maximum). Draws, including both drawdowns and drawups, is a series that is irregularly spaced in time. It provides a statistical measure for dependence in different market regimes. The concept of draws is important, not only from a theoretical point of view but in its use in various contexts in the finance industry, e.g. in asset management, where drawdowns tend to drive redemptions. For buy and hold portfolios that are subject to mark to market limits, extended drawdowns can force unwanted liquidations and readjustments of portfolios. Furthermore, large drawdowns could trigger feedback mechanisms through portfolio insurance and fire sales at funds, which have the tendency to drive prices further down. Many empirical properties of drawdowns in financial time series of various asset classes have been documented in a series of papers, such as \citeN{sornette-2003-378},  \citeN{springerlink:10.1007/s100510070147}, and \citeN{Johansen_largestock}.
The characteristic size of draws is stable within an asset class but varies across asset classes. For virtually all markets, the authors found large draws are outliers to the parametric model (\citeN{Johansen_largestock}) of the draw distribution.  \citeN{Johansen_largestock} lay out a crash theory that explains drawdown/drawup that appeas to fit almost all time series. They claim that drawdown/drawup can be explained by the onset of a well-defined micro-structural phase transition, which is characterized by the emergence of the sudden persistence of daily drops, coupled with an increase in the correlated amplitude of the drops. \citeN{Rebonato:2006xs} confirm these findings also for the US Treasury market. 
 
\subsection*{Research Motivation and Findings}
Many cases are known where a large market event triggered a reaction in another market. Volatility spillover and contagion are well documented phenomena in financial markets (see \citeN{Engle:1990go}). The question of cross-correlation and interdependence is also a natural question in the case of draws. To our knowledge, there has been no study done on the relationship of large drawdowns (drawups) with financial time series. 
\citeN{Rebonato:2006xs} qualitatively examined on a case-by-case basis if large draw-downs/drawups from different maturity blocks of US interest rates coincide with each other. But no quantitative, systematic measure has been proposed so far on how to analyse them. One obstacle arising when moving from fixed interval returns to draws is that time is compressed and is no longer equally spaced. Instead of having time series of synchronous measurements, the draws resulting from two time series will in general be clocked very differently. A further obstacle is that it is often argued that large drawdowns/drawups are generated from nonlinear mechanisms (see \citeN{sornette-2003-378}), thus making conventional parametric models and statistical inference no longer applicable. Converting draws into symbolic time series and using entropy quantities as non-linear, model independent and flexible tools, we were able to answer the question: can a large draw over possibly multiple periods that is unfolding in one market help to predict a large draw in another market around the same period?

Encoding the drawdown information of a time series into a time series of discrete values, we use entropy to analyse the correlation and cross correlations of draw-downs and drawups between two key exchange rates. The method we used to encode the drawdown information is novel and enables us to study, for the first time, draws in a bivariate setting. We investigate daily and hourly EUR/USD and GBP/USD exchange rates from 2001 to 2012. For the hourly and the daily data, we found evidence of dependence among the largest draws. We were also able to identify information flows between drawup/drawdown of the two exchange rates. Both daily and hourly series show clear evidence of information flowing from EUR/USD to GBP/USD and, slightly more pronounced, in the reverse direction. Robustness tests, using effective transfer entropy, show that the information measured is not due to noise. By using the 'relative explanation added' for the information gain of the entropy of the process, we conclude that there is a measurable information transfer between the two exchange rates, which can be potentially useful for forecasting and risk management.

\section{Information Flows between FX Volatility Regimes}
History provides many examples of financial crisis and how turbulence in one financial market spread across other markets within a short time period. Evidence for volatility transmissions has been found in many markets. One of the first studies on volatility spillover focussing on foreign exchange markets is \citeN{Engle:1990go}. 
Using the Japanese yen and US dollar exchange rate, the authors found evidence of intraday volatility spillover across different  markets. \citeN{Hong:2001ci} investigated the German mark and Japanese yen denominated in US dollars and found evidence of unidirectional volatility spillover from the German mark to the Japanese yen.

The methods to identify spillover, co-movement or interdependence vary considerably. \citeN{Engle:1990go} employed a GARCH model in a vector autoregression model to test if conditional variances are affected by the squared innovations (i.e. news, information) in other markets. \citeN{Cheung:1996fg} used the cross-correlation function between squared residuals, that are standardized by their individual conditional variance estimators, as test for volatility spillover. \citeN{Hamilton:1989jg}  introduced a model in which changes in regimes are governed by a hidden state process, which is Markovian. It gives more flexibility to the different adjustment speeds of volatility, and the persistence of the estimated volatility appears to be significantly different from that of the standard single-regime GARCH estimated volatility. Further models have been proposed, e.g. the SWARCH model in \citeN{Hamilton:1994bq} that has a Markov-modulated GARCH process, and the bivariate Markov switching process in \citeN{Biaikowski:2005kr}, to name a few.

\subsection*{Research Motivation and Findings}
In this study, we use a model similar to \citeN{Hamilton:1989jg} to identify volatility regimes in a set of currencies all against the US dollar. We define a two-state hidden Markov model with high and normal volatility states. The estimated state process contains discrete values, which lends itself to an analysis with information theoretic tools. Following the procedures in the previous chapter, we use the entropy framework with mutual information and transfer entropy to examine the interdependence and volatility spillover among the exchange rates in our sample in a novel way.

We found evidence for various volatility regime relationships between the currency pairs in the sample. Among the European currencies (viz. EUR/USD, GBP/USD, CHF/USD), volatility regime co-movements are identified. CAD/USD and AUD/USD were found to exhibit volatility co-movement and, to a lesser extent, interdependency with a time lag. We were able to find evidence for information flows between the volatility states of currency pairs. Most notable is an information flow from EUR/USD to GBP/USD, which indicates a causal volatility spillover relationship, confirming the findings in \citeN{Inagaki:2006kj} using a different model and set up. The work showcases the usefulness of the concepts of mutual information and transfer entropy when studying volatility spillover between markets.

\section{Volatility Timing of Hedge Funds and Entropy}
Hedge fund managers are in general free to change trading strategies, to allocate capital to asset classes and to choose leverage. Hedge fund managers often reallocate their capital dynamically in response to a change in the market environment. To pursue their investment goals, hedge fund managers scan the markets for signals and information that can give support to their analyses regarding the current and the future states of the market. From the information gathered, the fund manager forms an expectation of the future states of the market and develops a trading strategy to make profitable use of this knowledge. Absolute and relative price levels, market volatility and trend are information that fund manager uses in determining the investment opportunity set.

\citeN{Treynor:1966ui} were among the first to study whether investment managers are able to profitably time changes in market conditions. The authors extended the CAPM by a quadratic term to capture the nonlinear exposure to the market depending on an expected high or low market return. They found, from a sample of 57 mutual funds, only one fund for which the hypothesis of market timing ability was acceptable. \citeN{Henriksson:1981gu} formulated a market timing model modelled as an option payoff on the market return, with a strike price equal to the risk free rate. \citeN{Ferson:1996je} defined a model of market timing where the risk exposure evolves linearly with the market observables. In their analysis, various variables that are public information are used as market condition. Using a monthly data set of mutual funds, the authors found evidence for changes in risk exposure in response to public information on the economy for several state variables.

Market volatility timing has also been actively researched. \citeN{Busse:1999hi} uses daily returns of mutual funds to investigate the fund's ability to time market volatility. His findings suggest that funds decrease market exposure when market volatility is high. The value of this timing ability is in higher Sharpe ratios for successful volatility timing managers. The author further showed that surviving funds show sensitivity to market volatility, where non-surviving funds do not show this sensitivity. The economic value of volatility timing for investors has been investigated by \citeN{Fleming:2001jt}. Using conditional analysis on short horizon asset allocation strategies, the authors found volatility timing strategies outperformed unconditional static strategies.

\subsection*{Research Motivation and Findings}
Market environments where the prices and volatility develop with a high degree of irregularity inhibit managers' ability to form a hypothesis on the further development of the market. Fund managers are therefore more likely to choose a lower level of exposure when there is no clear view on market expected level of volatility.

We extend the market timing model of \citeN{Ferson:1996je}, which assumes risk exposures evolve as a linear function of market observables. Various modifications of this baseline model are tested using the approximate entropy (ApEn) of market volatility series as the chosen state variable for the market condition. ApEn, which quantifies serial correlation patterns, is sensitive to volatility regime changes. Depending on the state of the market, the manager then adjusts the fund's equity exposure via its market beta. Volatility is an important determinant for the level of market beta that a fund is choosing. Funds wishing to keep their volatility stable over time often target a specific Sharpe ratio and are more likely to reduce the fund's beta when a higher volatility regime is expected.

The current analysis adds to the market volatility timing research on hedge funds. Through simulations, we demonstrate  the use of ApEn in measuring changes in the serial structure of volatility. Popular econometric models of volatility are used to show the measure's usefulness in distinguishing different volatility patterns. In a sample of 1,903 hedge funds, we found that managers adjust a fund's market beta when volatility is settling into a new regime. Dead funds show no such market beta adjustment in response to a change in volatility regularity, differing from live funds where such an adjustment can be identified. The results are consistent with the findings from \citeN{Busse:1999hi} on mutual funds. The results also show the usefulness of entropy as a model independent measure of volatility regime and patterns.

\chapter{Entropy Measure and Information Theory}\label{chap_entropy}
\section*{\centering \large Abstract}
As information theory and the concept of entropy are not widely used in the economic sciences, we provide in this chapter an overview of the theoretical background and statistical features of the entropy measures used in the three main chapters. A brief introduction to the concept of entropy, as \citeN{Shannon:1948wk} developed it, is given. This is followed by an overview of the basic features of \textit{Mutual Information} and \textit{Transfer Entropy}, which are used in chapters 3 and 4. Both measures are employed to identify dependencies between two exchange rates. A different entropy measure, the \textit{Approximate Entropy}, is outlined in the last section. It is used in chapter 5 to analyse the serial structure of S\&P 500 volatility.
 
\clearpage
\section{Introduction}
The seminal work of \citeN{Shannon:1948wk} on a theory of communication and \citeN{Jaynes:1963ua} on the maximum entropy principle form one line of research concerning information and uncertainty that has led, in recent years, to rich research in statistical sciences. Applications have touched a wide array of sciences including Econometrics, where it is sometimes termed \textit{Information and Entropy Econometrics} \citeN{Golan:uk}.

\citeN{Shannon:1948wk} developed his theory of communication in thermodynamics, where entropy is a measure of disorder in a physical system. In information theory, the interest is in finding an 'optimal' coding given the source data. Shannon solved this by introducing a quantity, later known as the \textit{Shannon Entropy}, that corresponds to the average number of bits needed to optimally encode the source data. This concept and the ideas of Jaynes were used to enrich and complement the classical toolbox in econometrics. In our research, we are particularly interested in two developments in this research area, viz. \textit{Mutual Information} and \textit{Approximate entropy}, which we will introduce later in this chapter.   Mutual information and its generalisation, transfer entropy, captures any dependency relations between two time series. Approximate entropy captures changes in a systems complexity.

The structure of this chapter is as follows: In the first section, we introduce the concept of Shannon Entropy, which forms the basis for all other tools we use. The second section introduces \textit{Mutual Information} and \textit{Transfer Entropy}. Both concepts are used as generalized cross-correlation measures, which we apply to time series of exchange rates. In the last section, the complexity measure, \textit{Approximate Entropy}, is explained. We use it to measure patterns in time series and classify the time series by regularity or irregularity.

\section{Entropy}
\label{subsec_paper2shannon}

\citeN{Shannon:1948wk} developed his theory of communication based on concepts in thermodynamics, where entropy is a measure of disorder in a physical system. Entropy, in a physical system, is the number of allowed microscopic states compatible with a given observed macro state. Shannon used this in his work to relate to the average number of bits to optimally encode a source.

Entropy is a measure of uncertainty in random variables. For a random variable $X$ over a probability space $\Omega$ and probability distribution $p(x)$, entropy is defined, according to \citeN{Shannon:1948wk}, as:
\begin{align}
H(X)_{cont} &= -\int_{\Omega} p(x) \cdot log_2(p(x)) dx \label{eq_shannoncont} \\
H(X)_{discr}&= - \sum_{x \in \Omega} p(x) \cdot log_2(p(x)) \label{eq_shannon}
\end{align}
where $H(X)_{cont}$ and $H(X)_{discr}$ are for, respectively, continuous random variables and discrete probability distribution. Entropy, as defined in the above equations, and all the concepts described in this chapter can be applied to random variables on any probability space. The formulation on discrete probability spaces is a special case, but the concepts which we introduce in this section for discrete values can be generalised for continuous randam variables in a straightforward way. We will therefore use the discrete version of the concepts only. 
A neat and brief version of equations (\ref{eq_shannoncont}) and (\ref{eq_shannon}), is shown below: 
\begin{equation}
H(X) = \text{E}_p \text{log} \frac{1}{p(X)} \label{eq_shannonlog}
\end{equation}
$H(X)$ in (\ref{eq_shannonlog}) expresses entropy as the expected value of the random variable $\text{log} (\frac{1}{p(X)})$. There are two important conventions implicit in the formula. First, historically, the logarithm has been taken with base two, but any base would fulfil the requirements of an information measure, as described by \citeN{Shannon:1948wk}. Second, when computing entropy, the state with $p(x) = 0$ is set to 0, which can be justified by $\lim_{p \rightarrow 0} p \cdot log_2(p) = 0$. For entropy, the units of measurement are bits.

\begin{figure}[ht]
\centering
\includegraphics[scale=0.4]{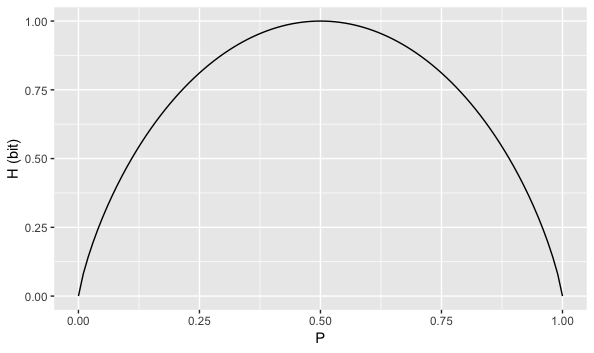}
\caption{Coin Toss Entropy}
 \begin{quote}
Entropy of coin-tossing experiment. At $p=0.5$, the coin has the highest entropy. Biased coins have lower entropies. \label{fig_entropycoin}
 \end{quote}
\label{fig:apentar1}
\end{figure}

As an example, a coin that has an equal probability for head and tail has $H(X) = - 2 \cdot \frac{1}{2} log_2 (\frac{1}{2}) = 1$. The maximum entropy of 1 for a discrete variable with two states is reached in this case. On the other hand, if the coin was biased and always produced head (or tail), then $H(X) = -1 \cdot log_2(1) - 0 \cdot log_2(0) = 0$; the uncertainty is zero. Figure \ref{fig_entropycoin} shows the level of entropy dependent on the probability of head (or tail) coming up when a coin is tossed.

The uncertainty, as measured by entropy, can be understood as the average number of questions, each with 'yes/no' answer, needed to determine the value of the random variable. In the coin example, the questions are 'head?' (or 'tail'?). For 50\% of the cases with a fair coin, the answer will be 'yes' and the questioning stops. In the other half of the cases, we are required to ask again. Entropy measures, in this sense, represents the average number of binary (yes/no) questions, expressed in units of bits. The higher the entropy the higher the uncertainty and the more questions we need to ask to determine the correct answer.

Entropy is bounded and the following inequality holds for discrete distributions $0 \leq H(X) \leq log_2( \vert $X$ \vert) $, where $\vert $X$ \vert$ denotes the number of discrete values of $X$, and the equality holds if $X$ has a uniform distribution. 

The joined entropy and conditional entropy of two variables $X$ and $Y$ with probability space $\mathcal{X},\mathcal{Y}$ are defined analogously:
\begin{align}
H(X,Y) &= - \sum_{x \in \mathcal{X}, y \in \mathcal{Y}} p(x,y) \cdot log_2(p(x,y)) \label{eq_joinedentropy} \\
H(X \vert Y) &= - \sum_{x \in \mathcal{X}, y \in \mathcal{Y}} p(x \vert y) \cdot log_2(p(x \vert y)) \label{eq_conditionalentropy} 
\end{align}
The general case of joined entropy for multiple variables $X_1, \ldots X_n$, $H(X_1, \ldots , X_n)$ is of importance when introducing transfer entropy below. 

If $X$ and $Y$ are independent random variables, the conditional entropy is the same as the entropy of the variable:
\begin{align}
H(X \vert Y) &= - \sum_{x \in \mathcal{X}, y \in \mathcal{Y}} p(x \vert y) \cdot log_2(p(x \vert y)) \nonumber \\
&= - \sum_{x \in \mathcal{X}, y \in \mathcal{Y}} p(x) \cdot log_2(p(x)) = H(X) \label{independententropy}
\end{align}

Conditional and joined entropy are related as follows:
\begin{equation}
H(X,Y) = H(X) + H(Y \vert X) = H(Y) + H(X \vert Y)
\end{equation}
In the case of independent variables $X$ and $Y$, equation (\ref{independententropy}) is a special case of this general chain rule for entropy (\citeN{Cover:awdrr_-9}). 

\section{Mutual Information and Transfer Entropy}
\label{subsec_paper2info}

\subsection{Definition}
The joint and conditional entropy enable us to define simple measures of dependency between $X$ and $Y$. In particular, Mutual Information  $I(X;Y )$ of two random variables, $X$ and $Y$, with joint distribution $p(x,y)$ is defined below. We also show, in the deduction below, how mutual information is related to the entropies of $X$ and $Y$.

\begin{align}
\label{eq_mutualinfo}
I(X;Y) &\equiv - \sum_{x \in X , y \in Y} p(x,y) \cdot log_2 \frac{p(x,y)}{p(x)p(y)}\\
 &= - \sum_{x \in X , y \in Y} p(x,y) \cdot log_2 \frac{p(x \vert y)}{p(y)}\\
 &= - \sum_{x \in X , y \in Y} p(x,y) \cdot log_2 p(x) + \sum_{x \in X , y \in Y} p(x,y) \cdot log_2 p(x \vert y) \\
 &= - \sum_{x \in X} p(x) \cdot log_2 p(x) - \Bigl( - \sum_{x \in X , y \in Y} p(x,y) \cdot log_2 p(x \vert y) \Bigr) \\
 &= H(X) - H(X \vert Y)
\end{align}

Mutual information measures the reduction of uncertainty about $X$ from observing $Y$. In the extreme case, where $X$ and $Y$ are independent such that $p(x,y) = p(x) \cdot p(y)$ then $I(X,Y) = 0$, there is no mutual information.
So the information gained is due to some dependence between $X$ and $Y$. In fact for Gaussian variables $X$ and $Y$ $\backsimeq \mathcal{N} (0, \sigma)$ correlated at $\rho$,  mutual information $I(X,Y)$ has a simple analytic solution  $- \frac{1}{2}log(1- \rho^2)$ (see \citeN{Cover:awdrr_-9}, p.252). 
With the chain rule, one can easily show that mutual information is symmetric $I(X;Y) = I(Y;X)$ and, hence, mutual information cannot be used to determine the direction of an information flow. 
Mutual information is positive $I(X;Y) \ge 0$  and equal to $0$ if and only if $X$ and $Y$ are independent. This important feature is used in a closely-related concept. The Kullback-Leibler distance, or the relative entropy, between two probability distributions is zero if and only if the two distributions are equal.

\begin{figure}[ht]
\begin{minipage}[b]{0.4\linewidth}
\centering
\includegraphics[width=2.4in]{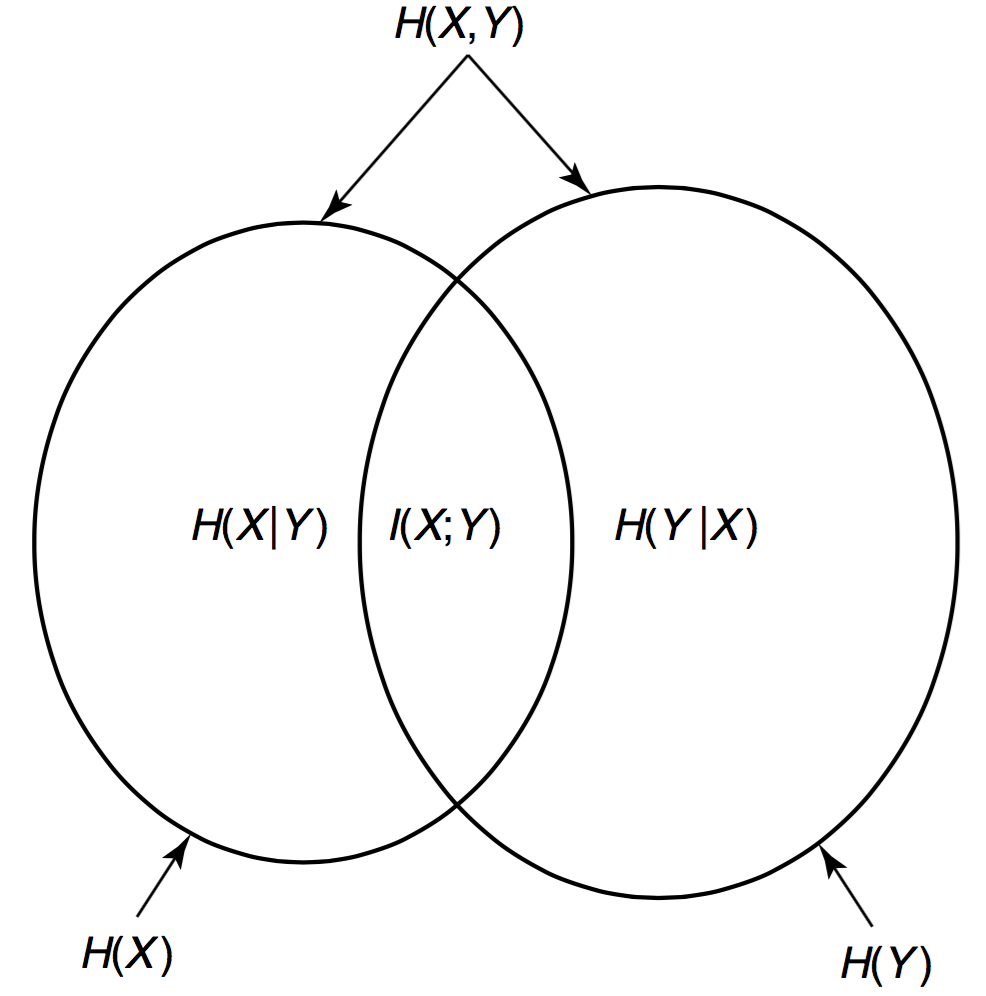}
\end{minipage}
\hspace{0.1cm}
\begin{minipage}[b]{0.5\linewidth}
\centering
   \includegraphics[width=3in]{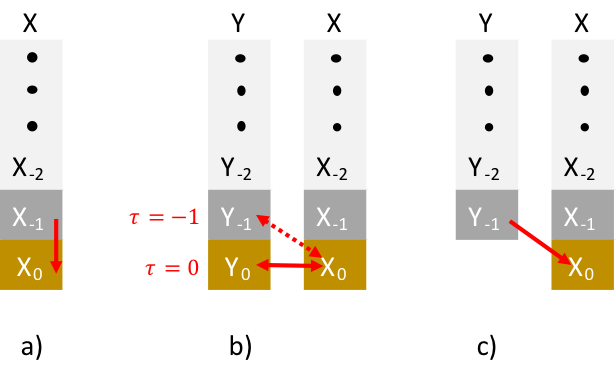}
\end{minipage}
\caption{Entropy Measures}
\begin{quote}
The left diagram shows the relationship between the various entropies and mutual information. The right diagrams show a) $h_{\infty}$ and $H(X_0 \vert X_{-1})$, b) $I(X_{t},Y_{t}), I(X_{t},Y_{t-1})$, and c) $T_{Y  \rightarrow X } (1,1)$
\end{quote}
\label{fig_entropymethods}
\end{figure}

Mutual information can be modified to include lead-lag relationships:
\begin{equation}
\label{eq_laggedmutualinfo}
I(X;Y)_{\tau} = - \sum_{x_{n- \tau} \in X , y_n \in Y} p(x_{n- \tau},y_n) \cdot log_2 \frac{p(x_{n- \tau},y_n)}{p(x_n)q(y_n)}
\end{equation}
Mutual information with and without lag is depicted in Figure \ref{fig_entropymethods} (b).

It is important to note that mutual information does not imply causality. As pointed out by \citeN{Schreiber:2000jx}, introducing a time delay in one of the observations does not distinguish information that is actually generated from a common response to input signal or a common history driven by an external factor. Transfer entropy was introduced by the author as an information measure that corrects for this shortcoming of mutual information.

Let $p(x_1, \ldots x_n)$ denote the probability of observing the subsequence $(x_1, \ldots x_n)$, \citeN{Schreiber:2000jx} defines transfer entropy as 
\begin{align}
\label{eq_transfer1}
T_{Y \rightarrow X} (m,l)  &=  \sum p( x_{t_{1}} , \ldots x_{t_{m}} ,y_{t_{m-l+1}} , \ldots , y_{t_{m}}) \notag \\
 & \cdot log_2 \frac{p(x_{t_{m+1}} \vert x_{t_{1}} , \ldots x_{t_{m}} ,y_{t_{m-l+1}} , \ldots , y_{t_{m}})}{p(x_{t_{m+1}} \vert x_{t_{1}} , \ldots x_{t_{m}})}
\end{align}
where $x_t$ and $y_t$ represent the discrete states of $X$ and $Y$ at time $t$. The parameters $m$ and $l$ indicate the number of  past observations included in $X$ and $Y$ respectively (see Figure \ref{fig_entropymethods} (c)).  In the absence of an information flow from $Y$ to $X$, i.e. $X$ and $Y$ are independent, then 
\begin{equation*}
p(x_{t_{m+1}} \vert x_{t_{1}} , \ldots x_{t_{m}} ,y_{t_{m-l+1}} , \ldots , y_{t_{m}}) = p(x_{t_{m+1}} \vert x_{t_{1}} , \ldots x_{t_{m}} ) 
\end{equation*}
and $T_{Y \rightarrow X} (m,l) = 0$.
Transfer entropy in (\ref{eq_transfer1}) can also be expressed as conditional entropies of the different blocks of history.
\begin{align}
\label{eq_transferconditional}
 T_{Y \rightarrow X} (m,l)  = 
  &  H(x_{t_{m+1}} | x_{t_{m}} , \ldots x_{t_{1}}  ) - H(x_{t_{m+1}} | x_{t_{m}} , \ldots x_{t_{1}} , y_{t_{m}} , \ldots y_{t_{m-l+1}})   \notag \\
  = & \left( H(x_{t_{m+1}} ,x_{t_{m}} , \ldots x_{t_{1}}  ) - H(x_{t_{m}} , \ldots x_{t_{1}}  ) \right)  \notag \\
  & -  \left( H(x_{t_{m+1}}  \ldots x_{t_{1}} , y_{t_{m}} , \ldots y_{t_{m-l+1}})  - H(x_{t_{m}} , \ldots x_{t_{1}} , y_{t_{m}} , \ldots y_{t_{m-l+1}}) \right)
\end{align}
In this way, transfer entropy can be understood as difference in the gain of information about $x_{t_{m+1}} $ conditional on both its own history and the history of $y$, and that conditional on its own history only. If the history of $X$ does not inform us of $x_{t_{m+1}}$, but the history of $Y$ completely determines $x_{t_{m+1}}$, then $  T_{Y \rightarrow X} (m,l) = H(x_{t_{m+1}} \vert x_{t_m}, \ldots x_{t_1}) = H(x_{t_{m+1}})$. In the absence of information flow from $Y$ to $X$, and the difference between the third and fourth terms in (\ref{eq_transferconditional}) is zero.
Hence we have $0 \leq  T_{Y \rightarrow X} (m,l)  \leq H(X)$.
While mutual information quantifies the deviation from independent $X$ and $Y$, transfer entropy quantifies the deviation from $X$ being determined by its own history only (via conditional probabilities).
Unlike mutual information, transfer entropy $T_{Y \rightarrow X} (m,l)$ is not symmetric and takes into account only the statistical dependencies originating from the variable $Y$ and not those from a common signal. 

Another way to look at transfer entropy it is to understand it as the resolution of uncertainty. This is similar to the interpretation of Granger causality in terms of prediction. The transfer entropy from $Y$ to $X$ is the degree to which $Y$ makes less uncertain the future of $X$ beyond the degree that $X$ is already giving information on resolving the future of $X$.
One can actually show that Granger causality for Gaussian variables is equivalent to transfer entropy (\citeN{Barnett:2009uu}). 

 \FloatBarrier
\subsection{Entropy Estimation, Statistical Features}
\label{subsec_discrete}
There is a multitude of methods for estimating entropy measures. One technique uses the kernel density method. In particular, transfer entropy and mutual information have both been estimated in the literature based on the kernel density method (see \citeN{Schreiber:2000jx}; and \citeN{Blumentritt:2011wf}). The maximum likelihood method (\citeN{Paninski:2003vz}) has also been applied successfully. In the following, we will focus on estimation methods which assume a discrete co-domain of the random variables, or the domains have been 'discretized'.

The discretization of the values in a given time series (e.g. return series, drawdown series) is a mapping of the continuous values of a random variable $X$ to a discrete set by partitioning the support of $X$. These discretized values are often referred to as 'letters' or symbols. Entropy is then approximated by the finite sum:

\begin{equation*}
H(X) \approx \hat{H}_{binned}(X) = - \sum_{i} \hat{p}(X \in \Delta_i) \cdot \text{log}_2(\hat{p}(X \in \Delta_i))
\end{equation*}
where $\hat{p}(X \in \Delta_i)$ denotes the estimated probability of the random variable having values in $\Delta_i$. As the partitions get finer with smaller bin sizes, $\hat{H}_{binned}(X)$ converges to $H(X)$ for well behaved distributions (see \citeN{2003physics...7138G} for a discussion).

In the literature\footnote{See for example \citeN{HlavackovaSchindler:2007tw}.} different ways of partitioning discrete data have been introduced. In some instances, the partitions are formed by dividing the distribution into equal parts, producing an equal marginal probability for every symbol defined. \citeN{Marschinski:2002dq} chose such a partition scheme to avoid 'undesirable' effects due to very inhomogenous histograms. As an example of an equal marginal probability discretization, consider a return series $\{r_1, \ldots r_n \}$ with quantiles $q(r_t,c)$ for $c \in (0,1)$. The following maps all returns into three symbols depending on whether the returns are below, between or above the $33\%$ and $66\%$ quantiles, forming an equiprobable binning scheme:
\begin{equation}
\label{eq_discreteddmag}
d_r(r_t) =
  \begin{cases}
  0 & \text{if } r_t < q(r_t,0.33) \\
  1 & \text{if } q(r_t,0.33) \leq r_t \leq q(r_t,0.66) \\ 
   2   & \text{if } r_t  > q(r_t,0.66)
  \end{cases}
\end{equation}

Other authors\footnote{See for instance \citeN{Jizba:2011dq} and \citeN{Peter:2010fr}.} prefer unequal marginal probabilities partition scheme. In our studies, equal marginal probability is not guaranteed and will depend on the distribution of the draw size.

Let $X \rightarrow \mathcal{A}$ be the  discretized random variable where all values are binned into $M = \vert \mathcal{A} \vert$ boxes.  The probability $p_i$ of $X$ falling into bin $i$, $1\le i \le M$, would be estimated by counting the number of times $X$ is in bin $i$, $n_i = \vert \{X \in \Delta_i \} \vert$, divided by the sample size $N$, so $p_i = \frac{n_i}{N}$. 
The estimator, which we will call the na\"{i}ve entropy estimator, can be calculated as follows.
\begin{align}
\label{eq_naiveestimator1}
\hat{H}_{naive}  &= - \sum_{i =1}^{M} p_i \cdot log_2(p_i) 
= - \sum_{i =1}^{M} \frac{ n_i}{N} \cdot log_2(\frac{ n_i}{N})  \notag \\
&= log_2 (N) - \frac{1}{N} \sum_i n_i \cdot log_2(n_i) 
\end{align}

\begin{figure}[ht]
\begin{minipage}[b]{0.5\linewidth}
\centering
\includegraphics[width=3in]{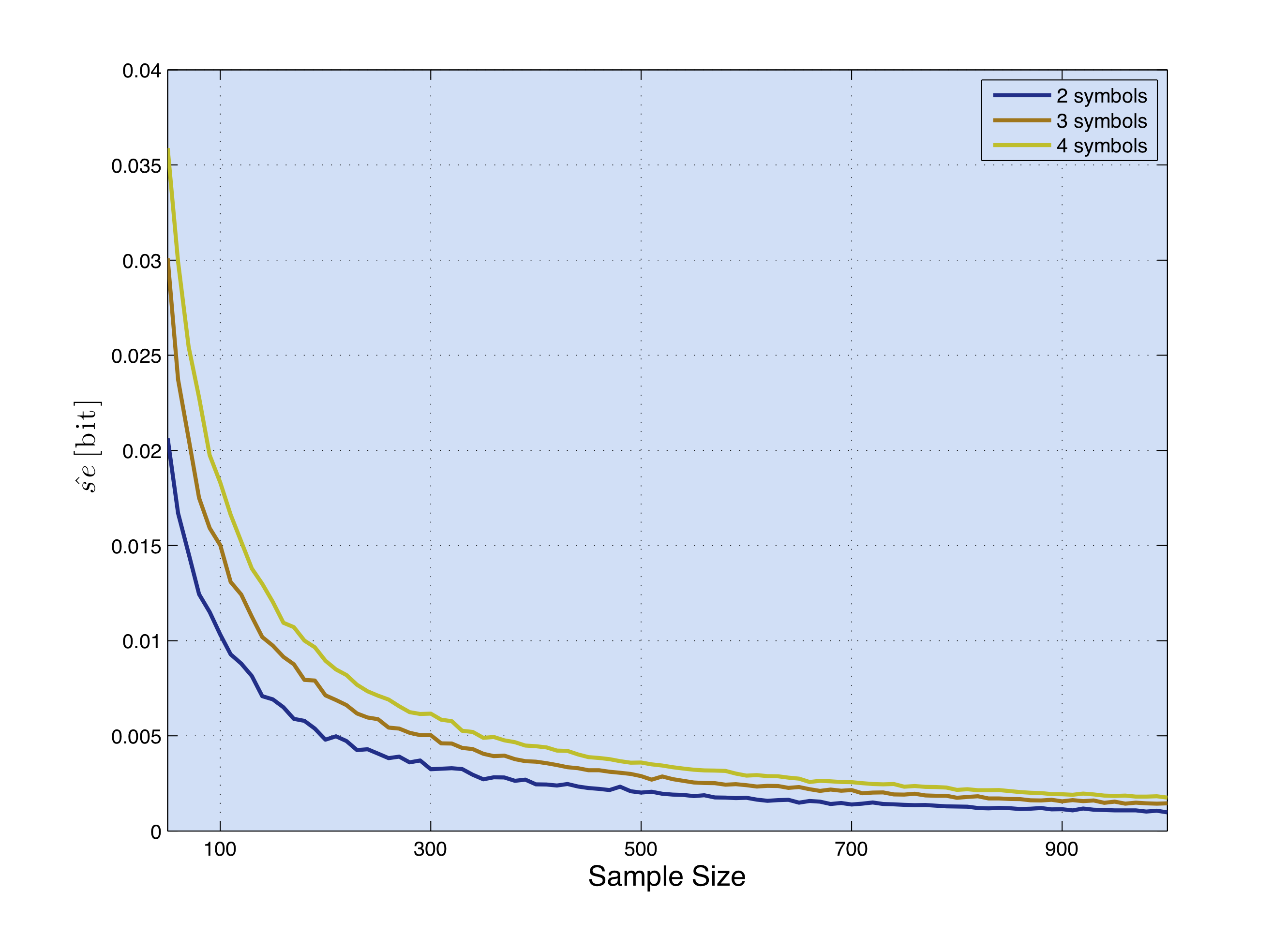}
\end{minipage}
\hspace{0.2cm}
\begin{minipage}[b]{0.5\linewidth}
\centering
   \includegraphics[width=3in]{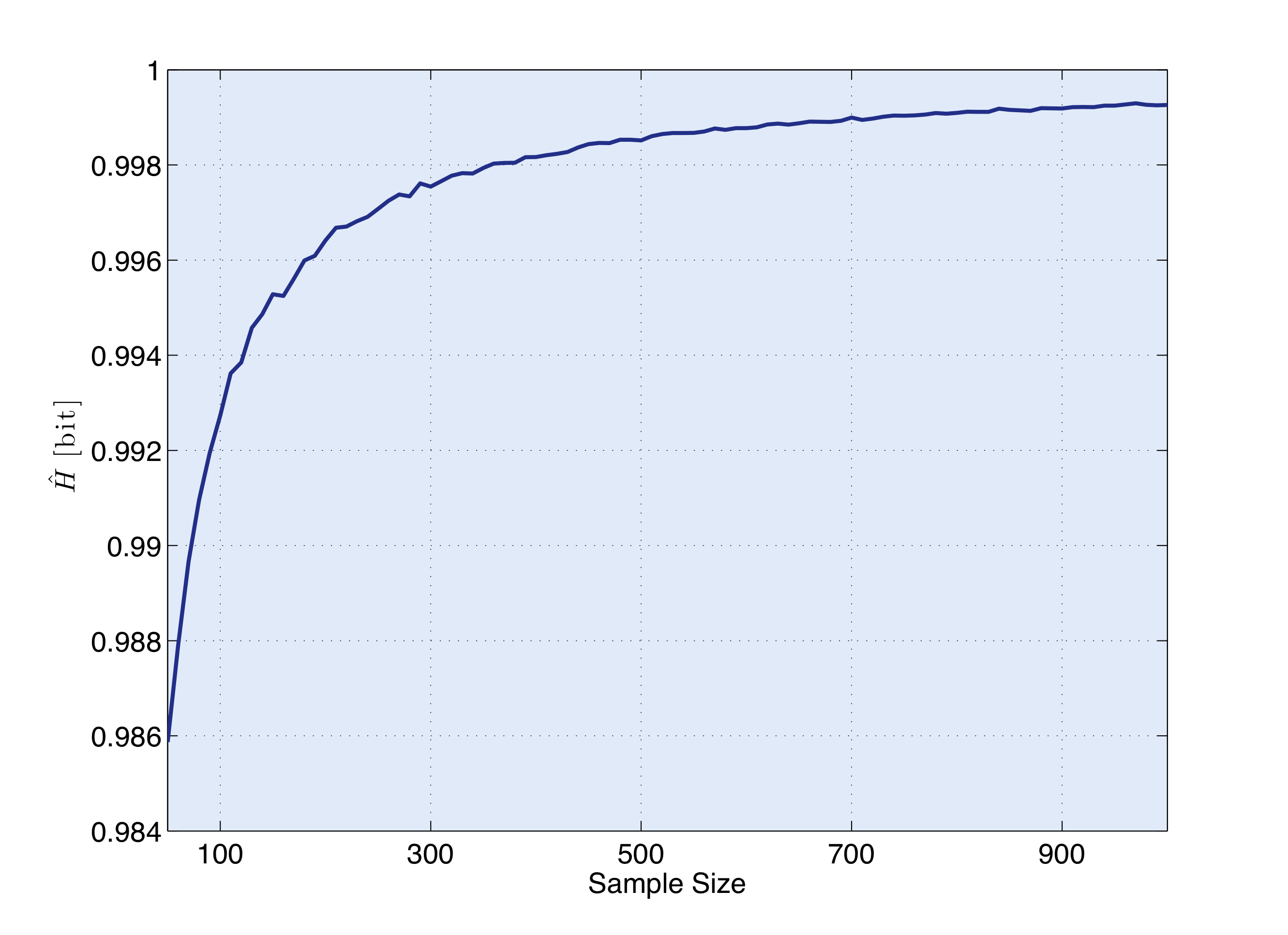}
\end{minipage}
\caption{Entropy Estimation}
\begin{quote}
The graphs show the entropy estimate and standard error in relation to sample size between 50-1000 data points. The graph on the left shows standard error $\hat{se}$ for uniform distribution with $2,3,4$ symbols. On the right, the entropy estimate $\hat{H}_{naive}$ for a fair coin (uniform distribution with two symbols) is shown.
\end{quote}
\label{fig_entropyestimationerror}
\end{figure}

For small samples, the binned distribution is likely to be less uniform and, hence, has a lower and downward-biased entropy $E[\hat{H}_{\text{naive}}] - H < 0$. In figure \ref{fig_entropyestimationerror}, the average na\"{i}ve entropy estimate for a fair coin ($H = 1$ bit) is shown for different samples sizes. In a Monte Carlo simulation, a fair coin is tossed over sequences with lengths ranging from 50 to 1000, each sampled 5000 times. For each of the simulated samples of a specific size, the entropy is estimated with the na\"{i}ve estimator in equation (\ref{eq_naiveestimator1}), and the mean of the estimate $E[\hat{H}_{\text{naive}}]$ is calculated.
$\hat{H}_{\text{naive}}$ turns out to be below 1 for all sample sizes, which illustrates the downward bias of the na\"{i}ve estimator. The effect is more pronounced for small sample sizes (left diagram of figure  \ref{fig_entropyestimationerror}). \citeN{2003physics...7138G}  offers a correction to this 'small sample' bias. Assuming that all $p_i \ll 1$, the proposed new estimator $H_{\psi}$ is as follows:
\begin{align}
\label{eq_grassberger}
\hat{H}_{\psi} &= ln N - \frac{1}{N}\sum_{i =1}^{M} n_i \cdot \psi (n_i) \\ 
\psi(x) &=\frac{d(ln \Gamma (x))}{dx}, \qquad \Gamma(0,x) = \int_{1}^{\infty} \frac{e^{-xt}}{t} dt
\end{align}

We will use this estimator $\hat{H}_{\psi}$, in particular when estimating transfer entropy for larger blocks of history. Transfer entropy, as we have seen in equation (\ref{eq_transferconditional}), is calculated with block entropies. The block time series is a series of symbol set representing the combinations of letters in each cell of the block. Since the blocks overlapped in the original time series, the block time series has the same size as the original sequence minus the block length. For alphabets of two to three letters that we use here, the assumption $p_i \ll 1$ will hold when estimating block entropies, and hence the Grassberger estimator can be used.

Statistical fluctuations in samples, and in particular small samples, will artificially induce deviations to the entropy estimates. The magnitude of the 'noise' depends on the sample size and the size of the symbol set $M$, or in the case of more than one discrete random variable, the size of the co-domain. For some distributions, there exists analytical expressions for difference between the expected value of $\hat{H}$ and the actual value $H$ (\citeN{2004JPhA...37L.295S}). At this point, we would like to highlight the variance of the na\"{i}ve estimator with the coin tossing example we used above. For the Monte Carlo simulations of a specific length, the standard error $\hat{se}(\hat{H}_{{\text naive}})$ of the entropy estimates is calculated. The standard errors are plotted in figure  \ref{fig_entropyestimationerror}. Also plotted in figure \ref{fig_entropyestimationerror} are the standard errors of entropy estimates of a uniform distributed random variable with three and four symbols (analogous to a coin with three or four sides). The figure shows that the standard error is slightly higher for a given sample size when the symbol set is larger. The difference is greater the smaller the sample size.

For a given size of alphabet, it seems logical to choose the largest posible block length in order to find invariant values and to detect the patterns in history that offer the most information in forecasting. But for block entropies, the length of the block is limited by the sample size (and the size of the symbol set). The larger the block length, the larger the noise. As block length approaches the sample size, the estimated block entropy drops to zero as the symbol combinations concentrate in fewer and fewer symbols.

As a way to control the noise in the transfer entropy estimates, an approach pioneered by \citeN{Marschinski:2002dq} is to estimate the noise itself by a bootstrapping method. For two processes $X$ and $Y$, the information flow from $Y$ to $X$ is measured by Transfer Entropy $T_{Y \rightarrow X}$. For the information source $Y$, we form a second set by shuffling the original time series. If $\{Y_1, \ldots , Y_n \}$ is the original sample and $\pi$ is a permutation of the tuple $(1, \ldots ,n)$, then a shuffled series is $\{Y_{\pi(1)}, \ldots , Y_{\pi(n)} \}$. Through shuffling the information source, all potential correlations between the two time series are destroyed and hence the observed transfer entropy should be zero. Repeating the shuffling process many times, the mean $\mu({\hat{T}_{sh}})$ and the standard error $\sigma({\hat{T}_{sh}})$ of the transfer entropy is estimated. Since there is no structure in the reshuffled data, any non-zero estimate must be an artefact of the finite sample size.

To account for the 'noise' in the Transfer Entropy calculation, the \textit{Effective Transfer Entropy} (ET) is calculated by deducting the mean $\mu({\hat{T}_{sh}})$ from the Transfer Entropy estimate as follows: 
\begin{equation} \label{eq_ete}
ET_{Y \rightarrow X} (m,l) \equiv T_{Y \rightarrow X} (m,l) - \mu({\hat{T}_{sh}}) 
\end{equation}

For $ET_{Y \rightarrow X} (m,l)$, $m$ amount of past history of $X$ and $l$ amount of past history of $Y$ are used in predicting $X$. 

\FloatBarrier
\subsection{Process Entropy}
\label{subsec_entropyblock}
For a sequence of identical distributed and independent random variables, the entropy of one variable conditional on the others has a particular simple form. We have seen for two independent variables $X$ and $Y$ that the conditional entropy is $H(X \vert Y) = H(X)$. Generalizing this to the case of $n$ iid variables gives $H(X_{0} \vert X_{-1}, \ldots X_{-n})=H(X_0)$. The block entropy in this case is $H(X_{0}, X_{-1}, \ldots X_{-n}) = n \cdot H(X_i)$ with $0 \leq i \leq n$. 

For non-independent random variables, which is often the case in empirical studies, we would like to have some understanding of how both conditional entropy and block entropy behave over different block lengths.
Given a process $\{X_i\}$, the limit of the conditional entropy 
\begin{equation}
h_{\infty} = \lim_{n \rightarrow \infty} H(X_{0}\vert X_{-1} , \ldots X_{-(n+1)}  ) 
\end{equation}
is, in the literature, referred to as \textit{entropy rate} or \textit{entropy of the source} (\citeN{Cover:awdrr_-9}).  It quantifies the average information needed to predict a future observation $X_{0}$ given an $m$ period of history $(X_{-1} , \ldots X_{-(m+1)})$.  In the univariate case, it quantifies the level of predictability of the series conditional on a specific pattern in the history. Similarly, the transfer entropy $T_{Y \rightarrow X}$ can be expressed as the entropy of $X$ conditional on its own history and on the history of $Y$. See figure \ref{fig_entropymethods} for a depiction of process entropy and its relation to other entropy measures. We will discuss here some special cases that are relevant in the subsequent chapters.

If a process is periodic,\footnote{A process is periodic if there exists a $k \in \mathbb{N} $, the periodicity for which $X_t = X_{t+k}$ for all $t$.} then $h_{\infty} = 0$ for block length longer than the periodicity, and the conditional entropy goes to $0$.  
For any \textit{iid} process $\{X_i\}$, $h_{\infty} = H(X)$, with $H(X)$ being the entropy of the symbol distribution. This means that there is no information gain no matter how much history is included in the calculation. The entropy rate for any given process lies between these two extremes: $0 \leq h_{\infty} \leq H(X)$. 

For Markov processes, there is an analytic expression for $h_{\infty}$. Let $X$ be a Markov process with transition matrix $P = (P_{ij})$. The stationary distribution $\pi$ of $X$, is the limiting distribution (if it exists) to which the Markov process $X$ converges. It is the solution of the eigenvalue problem $\pi = P \cdot \pi$ and has the form $\pi_j = \sum_{i} \pi_i \cdot P_{ij}$ ($\pi_i$ is the probability of $X$ being in state $i$). The stationary distribution is so called because if the initial state of a Markov chain is drawn according to a stationary distribution, the Markov chain forms a stationary process. One can show that if the Markov chain is irreducible and aperiodic, the stationary distribution is unique, and from any starting distribution, the distribution of $X_n$ tends to the stationary distribution as $n \rightarrow \infty$.

For a stationary distribution of a Markov process, the entropy rate can be simplified as
\begin{align}
h_{\infty} &= \lim_{n \rightarrow \infty} H(X_{n}\vert X_{n-1} , \ldots X_{1}  ) \notag \\
 &= \lim_{n \rightarrow \infty} H(X_{n}\vert X_{n-1} ) = H(X_2 \vert X_1)
\end{align}
That is, only the immediate history $X_{n-1}$ is needed for predicting $X_{n}$. Combining this with the expression for the stationary distribution, the entropy rate of a stationary markov process with stationary distribution $\pi_i$ and transition matrix $P_{ij}$ can be simplified as:  
\begin{equation}
h_{\infty} = \sum_{i} \pi_i \cdot \left( \sum_j - P_{ij} \cdot log_2 P_{ij} \right)
\end{equation}

\subsection{Entropy Magnitude}

To measure the amount of information gain due to the transfer entropy, we follow \citeN{Marschinski:2002dq} by relating the information gain to the entropy of the series based on the \textit{REA} (\textit{relative explanation added}) measure defined as follows:
\begin{equation}
REA (m,l) \equiv \frac{T_{Y \rightarrow X} (m,l)}{H(X_{0}\vert X_{-1} , \ldots X_{-(m)}  ) }
\end{equation}
Here the information flow from $Y$ to $X$, measured by the transfer entropy, is related to the total flow of information based the conditional block entropy of $X$. Put differently, it measures the additional information gain by observing the history of $X$ and $Y$ when already observing $X$. 
Using the definition of transfer entropy, we arrive at the following form
\begin{equation}
REA (m,l) = \frac{H(X_{0}\vert X_{-1} , \ldots X_{-m} , Y_{-1} , \ldots Y_{-l}) }{H(X_{0}\vert X_{-1} , \ldots X_{-m}  ) } - 1 
\end{equation}
If the past history of $Y$ is not adding information to predicting $X$, then $H(X_{0}\vert X_{-1} , \ldots X_{-m} , Y_{-1} , \ldots Y_{-l}) = H(X_{0}\vert X_{-1} , \ldots X_{-m})$ and the \textit{REA} is zero.

To give a sense of the levels expected for \textit{REA}, we cite directly the results of \citeN{Marschinski:2002dq}. The \textit{REA} for using DAX to predict the Dow Jones Industrial, using one minute tick data, is estimated to be 0.4\% for a three symbol discretization. 

\section{Approximate Entropy (ApEn)}
\label{sec_apent}
Various information measures have been introduced to complement our understanding of stochastic time series. 
In a series of articles,\footnote{See for instance \citeN{1991PNAS...88.2297P}, \citeN{Pincus:1995bb}, \citeN{Pincus:1992fn}.} \citeN{1991PNAS...88.2297P} developed a new mathematical approach to measuring regularity in a time series based on the \textit{Approximate Entropy}.

\subsection{ApEn Definition}
\label{subsec_apent}
Let $u$ be a sequence of numbers, $u(1), \ldots u(N)$, of length $N$.
Given a non-negative number $m$ with $m \leq N$, we form $m$-blocks of subsequences $x(i) \equiv (u(i), u(i+1) \ldots , u(i+m-1))$. The distance between two blocks is measured by $d(x(i),x(j)) \equiv \text{max}_{k=1,2, \ldots m} (\vert u(i+k-1) - u(j+k-1) \vert)$, so it is the maximum of the point-wise difference between the blocks. Given a positive real number $r$, we count for a given block $x(i)$ the fraction of blocks $x(j)$ that have a distance of less than $r$ and name it $C^m_i (r) $. The formal definition is shown in equation (\ref{eq_cm}), which is based on the Heaviside function $\Theta$ counting the instances where the distance $d$ is below the threshold $r$: 
\begin{align}
C^m_i (r) &= \frac{1}{N-m+1} \sum_{j=1}^{N-m+1} \Theta (r - d(x(i),x(j)) ) \label{eq_cm} \\ 
\Phi^m(r) &= \frac{1}{N-m+1} \sum_{i=1}^{N-m+1} \text{log} C_i^m (r) \label{eq_phi}
\end{align}
With $\Phi_m(r)$ defined in equation (\ref{eq_phi}), the approximate entropy, $\text{ApEn}(m,r,N)$, is defined as:
\begin{align}
\text{ApEn}(m,r,N)(u) &= \Phi^m(r) - \Phi^{m+1}(r) \qquad , \thickspace m\geq 0 \label{eq_apen1} \\
\text{ApEn}(0,r,N)(u) &= - \Phi^1(r) 
\end{align}
ApEn($m,r,N$) is always well defined. In equation (\ref{eq_cm}),  when a fraction of block is $<r$,  the distance of the block with itself, $d(x_i,x_i)$, is always counted. So $C_i^m(r) \geq 1$, and hence the logarithmic of $C_i^m(r)$ is never undefined.
The following expansion shows that ApEn measures how close specific patterns are.
\begin{align}
- \text{ApEn}(m,r,N)(u) &= \Phi^{m+1}(r) - \Phi^m(r) \nonumber \\
  &=  \frac{1}{N-m} \sum_{i=1}^{N-m} \text{log} C_i^{m+1} (r) - \frac{1}{N-m+1} \sum_{i=1}^{N-m+1} \text{log} C_i^m (r) \nonumber \\
   &\approx \frac{1}{N-m}  \sum_{i=1}^{N-m}  \left( \text{log} C_i^{m+1} (r) - \text{log} C_i^{m} (r) \right) \nonumber \\
 &= \frac{1}{N-m}  \sum_{i=1}^{N-m} \text{log} \left( \frac{C_i^{m+1} (r)}{C_i^{m} (r)} \right) \label{eq_exploglikelihood}
\end{align}
The last line is the average, over all $m$-blocks, of the log of the conditional probability of $\vert u(j+m) - u(i+m) \vert < r $, given that  $\vert u(j+k) - u(i+k) \vert < r $ for all $k = 0,1, \ldots m-1$. So, approximate entropy measures the logarithmic likelihood that sequences of patterns that are close for $m$ observations remain close on next comparisons. ApEn measures persistence, correlation and regularity. If $C_i^{m+1} (r)$ equal $C_i^{m} (r)$, then ApEn=0 and there is strong serial dependence. Hence, lower ApEn values correspond to higher persistence and autocorrelation. Higher ApEn values imply less persistence and more independent observations.
\begin{figure}[ht]
\centering
\includegraphics[scale=0.5]{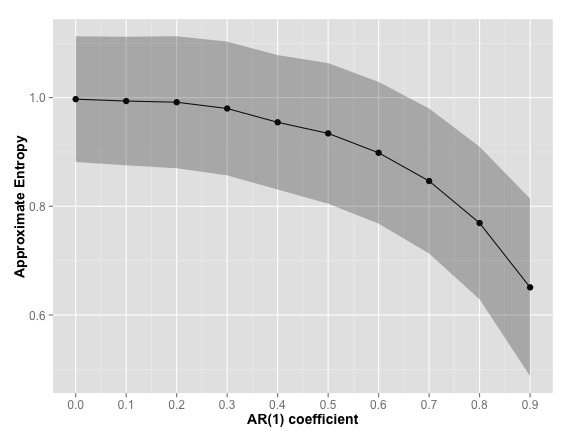}
\caption{ApEn for AR(1) process.}
 \begin{quote}
 Approximate Entropy $\text{ApEn}(N=200,m=1)$ for AR(1) process with various autoregressive coefficients $\rho=0, \ldots , 0.9$.
 \end{quote}
\label{fig:apentar1}
\end{figure}

We can derive another useful representation of approximate entropy. We have shown, in equation (\ref{eq_exploglikelihood}), that the approximate entropy for $m=1$, $\text{ApEn}(m=1,r,N)$, can be reformulated as expected log-likelihood of two blocks staying close when extended by one element.  Assuming the stationarity of the stochastic process, we can reduce the number of calculations in $\text{ApEn}(m=1,r,N)$, as follows: 
\begin{align}
ApEn(1,r,N)(u) &= \Phi^1 - \Phi^2 \\
&\substack{(\ref{eq_exploglikelihood}) \\ =} - \frac{1}{N-1}  \sum_{i=1}^{N-1} \text{log} \left( \frac{C_i^{2} (r)}{C_i^{1} (r)} \right) \\
&\substack{N \rightarrow \infty \\ =}  \quad -\text{E} \left( \text{log} \left( \frac{C_1^2(r)}{C_1^1(r)}  \right) \right) 
\end{align}

Furthermore, \citeN{1991PNAS...88.2297P} found analytical expression for ApEn for a number of processes. For \textit{iid} distributed random variables, ApEn(m,r) is given by the following theorem.

\underline{Theorem} For an \textit{iid} process with density function \textit{f(x)}, for any \textit{m},
\begin{equation}
\label{eq_apeniid}
\text{ApEn}(m,r) = - \int f(y) \cdot \text{log} \left( \int_{z=y-r}^{z=y+r} f(z) dz \right) dy 
\end{equation}

For example, consider the autoregressive process of order one, AR(1), below:
\begin{equation}
X_k = \rho \cdot X_{k-1} + \epsilon_k \qquad k=0,1, 2 \ldots 
\end{equation}
For different values of $\rho = 0, 0.1, \ldots 0.9$, a Monte Carlo simulation of 1000 paths of length 200 are drawn. The resulting approximate entropy estimation with standard error is depicted in figure \ref{fig:apentar1}. For higher correlation values, $\rho$, the approximate entropy is lower.  The highest ApEn value is for the random walk process with $\rho = 0$. For $\rho = 0.9$, ApEn declined to $0.65$ [bit]. This shows that persistence is a feature that the approximate entropy measure is sensitive to.

\subsection{ApEn Estimation}\label{subsec_apenestimate}

\begin{figure}[ht]
\begin{minipage}[b]{0.5\linewidth}
\centering
\includegraphics[width=3.3in]{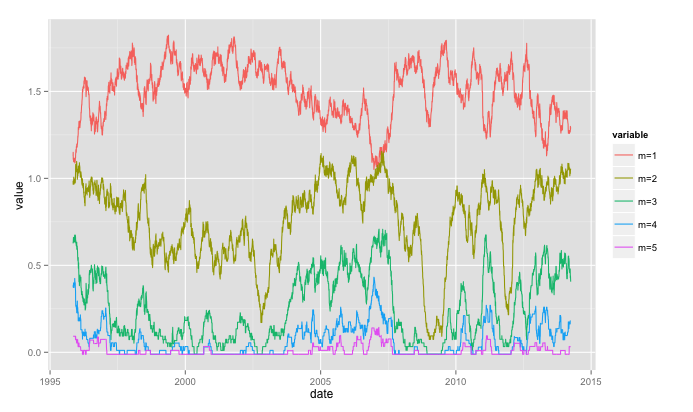}
\end{minipage}
\hspace{0.2cm}
\begin{minipage}[b]{0.5\linewidth}
\centering
   \includegraphics[width=2.6in]{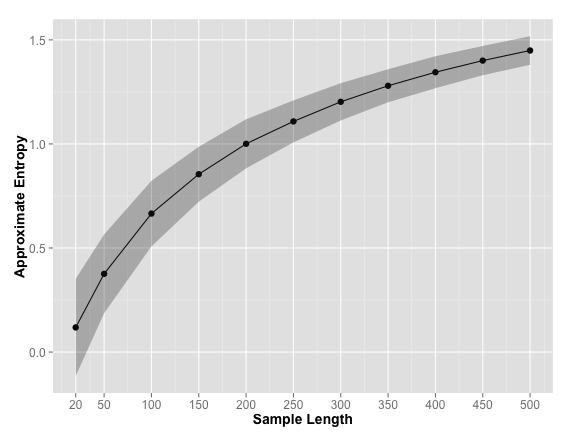}
\end{minipage}
\caption{ApEn with $m=1,\ldots, 5$ and $N=20, \ldots 500$. }
\begin{quote}
Left: rolling 90-day window $\widehat{\text{ApEn}}(m=1,\ldots 5, r=0.2 \hat{\sigma})(\text{SP500}) $ time series. Right: graph of estimated ApEn of a random walk with different sample sizes and standard error.
\end{quote}
\label{fig:apentestimation}
\end{figure}

\citeN{Rukhin:2000tj} shows that ApEn converges asymptotically to a chi-squared distribution for $N \rightarrow \infty$. Specifically for an $iid$ binary-valued sequence of size $N$, a block size $m$ and a constant $c= \frac{-3}{ln2}$, ApEn($m$,$N$) converges to a $\chi^2$ distribution with $2^{m-2}$ degrees of freedom as $N \rightarrow \infty$. For finite sample, ApEn is a biased statistic. 
The small sample bias can be illustrated with a normal distributed random variable, $N(0,\sigma)$, for sample length ranging from $20$ to $500$, each with 1000 simulated paths. In the right of figure \ref{fig:apentestimation}, the mean and standard error of the ApEn estimates for this random walk are presented. The threshold $r$ is set equal to 20\% of the sample standard deviation. The mean value of ApEn increases from $\widehat{\text{ApEn}}(m=2, r=0.2 \hat{\sigma}, N=20) = 0.11$ [bit] to $\widehat{\text{ApEn}}(m=2, r=0.2 \hat{\sigma}, N=500) = 1.44$ [bit] as the sample length, $N$, increases. 

For small samples, the underestimation of ApEn is visible. We can see from figure \ref{fig:apentestimation} that as sample length approaches $1$, ApEn approaches 0. In view of this small sample bias, one may increase the threshold $r$ as a counter measure. But there is a tradeoff, with a larger $r$ information is lost from the distribution. 
The small sample bias also implies that comparisons between ApEn make sense only if the sample length $N$ is the same.Furthermore, \citeN{Pincus:2008tb} studied $\text{ApEn}(m,r,N)$ for various equity indices and concluded that the parameters $m=1,2$ and the filter threshold $0.1 \hat{\sigma} \leq r \leq 0.25 \hat{\sigma}$, with $\hat{\sigma}$ being the standard deviation of the sample, produce good results. 

To construct statistical bounds, the bootstrapping method is employed in chapter 5. The bootstrapping method normally involves reshuffling the data several times, and the approximate entropy estimate averaged over the bootstrapped simulations is used to estimate the bias and calculate the effective ApEn. However, reshuffling the data destroys the dependencies within the univariate time series as well. The more approapriate method is a standard block bootstrap, where block resampling with a fixed block length is used. This is the approach adopted here. In estimating ApEn for a random walk in figure \ref{fig:apentestimation}, the bootstrapped standard error is shown as grey band around the ApEn estimate in the figure. The smaller sample lengths are associated with bigger standard errors.

\chapter{Bivariate Analysis of Drawdowns}\label{chap_paper1}
\section*{\centering \large Abstract}
In this chapter, we use mutual information and transfer entropy to analyse the dependence relation between large drawdowns/drawups in EUR/USD and GBP/USD. Despite various crisis episodes documented in our sample period, we did not find evidence of extreme dependence in this currency pair. Large drawdowns/drawups in EUR/USD and GBP/USD are weakly dependent. Transfer entropy shows information flows between EUR/USD and GBP/USD however.
\clearpage
\section{Introduction}
\label{sec_intro}
Financial time series drawdowns (drawups) are defined as the cumulative return from a local maximum (minimum) to the next local minimum (maximum). Draws, which include both drawdowns and drawups, are a time series feature that has no regular time stamp. It provides aspect of statistical dependence in different market regimes. The concept of draws is important, not only from a theoretical point of view but also for its use in various contexts in the finance industry, e.g. in asset management where drawdowns tend to drive redemptions. For buy and hold portfolios that are subject to mark to market limits, extended drawdowns can force unwanted liquidations and the readjustments of portfolios. Furthermore, large drawdowns could trigger feedback mechanisms through portfolio insurance and fire sales at funds, which have the tendency to drive prices further down. The study of drawdowns has drawn considerable attention, and many empirical properties of drawdowns in financial time series have been documented. 

\citeN{Johansen_largestock}
lay out a crash theory that explains drawdowns/drawups that appears to fit almost all time series. They claim that drawdowns/drawups can be explained by the onset of a well-defined, micro-structural transition phase, which is characterized by the emergence of the sudden persistence of daily drops, coupled with an increase in the correlated amplitude of the drops. This explains why price changes appear to be independent for most of the time, but in exceptional circumstances, serial co-dependence and amplification may set in, creating drawdown/drawup outliers. To our knowledge, there has been no study conducted on the relationship of drawdowns (drawups) in financial time series.
\citeN{Rebonato:2006xs} examined qualitatively, on a case-by-case basis, if large drawdowns/drawups from different maturity blocks of US interest rates coincide with each other. But no quantitative measure has been proposed so far to analyse these drawdowns/drawups.  

Detecting and measuring interactions between financial time series is a key area of research in finance and economics. With regard to linear dependence, there are, for instance, correlation and cross-correlation measures and their more general formulations in vector autoregressive models.\footnote{For more details see \citeN{Tsay:2010tu}.} More recently, copula methods have been used, with some degree of success, to model the more general form of dependence between time series (\citeN{McNeil:2005wm}).
But these tools are not suited for modelling the dependence of drawdowns/drawups. One obstacle, arising when moving from returns to draws, is that time is compressed and is no longer equally spaced. Instead of having time series of synchronous measurements, the draws resulting from two time series will in general be clocked very differently. A further obstacle is that it is often argued that large drawdowns/drawups are generated from nonlinear mechanisms. The herding effects of market participants under stressed market conditions, positive feedback from portfolio insurance and liquidity squeeze in times of market stress can all lead to non-linear price dynamics (\citeN{sornette-2003-378}).
 
As an alternative to the standard models, information theoretic measures have been applied successfully to finance and other areas to cope with the aforementioned problems.\footnote{See, for example, \citeN{Marschinski:2002dq}, \citeN{Kwon:2008be}, \citeN{Jizba:2011dq}.} Transfer entropy, in particular, measures information flows between series and detects the subtle dependencies between them. It is a model-free measure, which is based on the Kullback-Leibler distance (\citeN{Schreiber:2000jx}) of transition probabilities.  Mutual information, on the other hand, quantifies correlation and cross-correlation type dependencies of time series. It has also been used in finance extensively (\citeN{Golan:uk}, \citeN{Dionisio:2004ua}). 
In this chapter, we use information theoretic tools to analyse how large drawdowns and large drawups in EUR/USD and GBP/USD are related to each other. Specifically, the mutual information measure is used to analyse the contemporaneous dependence (i.e. same hour or same day) or for a specific time lag. Transfer entropy is used to measure how large drawdowns and drawups in one time series influence large drawdowns and drawups in another time series. Our results show that the relationship between large draws of EUR/USD and GBP/USD is weaker than the dependence between the returns of the same series. The strength of dependence drops markedly when time lags between the series are considered. On the other hand, transfer entropy suggests an information flow between the large draws in the two series. Such information spillover is slightly stronger from GBP/USD to EUR/USD than in the reverse direction.

This chapter is structured as follows. Section \ref{sec_relwork} reviews the relevant literature pertaining to the analysis. It covers the major research on the drawdown/drawup and some recent finance applications of entropy measure and information theoretic tools. 
Section \ref{sec_data_draw} describes the data used and discusses its statistical properties. 
Various information theoretic tools used in this study, which have been introduced in Chapter \ref{chap_entropy}, are briefly reviewed in section \ref{sec_depanalysis}. The method used to translate draws into sequences of symbols is also explained in this section.
The results from the dependence analysis are presented and discussed in Section \ref{sec_results}. We summarise and conclude in Section \ref{sec_conclusion} and suggest possible directions for future research.

\section{Related Literature}
\label{sec_relwork}
 
\subsection{Drawdowns Literature}
\label{subsec_draw}
In the microstructure model proposed by \citeN{Johansen00crashesas}, agents respond differently during normal and crisis market conditions, producing different patterns of imitation, herding and positive feedback during market rallies and crashes, leaving behind different signatures of 'auto-correlation' and dependencies. The authors claim that outliers in drawdowns commence with the emergence of sudden and persistent daily price drops, coupled with an increase in the correlated amplitude of the drops. So price changes appear to be independent for most of the time and under 'normal' market conditions, but serial co-dependence and amplification may set in in exceptional circumstances, creating the drawdown outliers.

Given a time series, a drawdown is defined as a consecutive series of negative returns. A drawup is accordingly defined as an uninterrupted sequence of positive returns. Specifically, a drawdown (drawup) is the cumulative return from one local maximum (minimum) to the next local minimum (maximum). For the remainder of this chapter, unless otherwise stated, we use draws to refer to both drawdowns and drawups in general.

Draws play a significant role in Sornette's theory in capturing market dependence under different market regimes, as laid out by \citeN{Johansen00crashesas} and \citeN{Johansen_largestock}. Following \citeN{Johansen_largestock}, and starting with drawdowns of a specific length, the probability of observing a drawdown of magnitude $D$ having lasted exactly $n$ units of time is proportional to
\begin{equation}{\label{eq:magndistcond}}
p_n (D) \propto \int_{-\infty}^{0} dx_1p(x_1) \dotsb \int_{-\infty}^{0} dx_n  p(x_n) \delta \biggl( D - \sum_{i=1}^{n} x_i  \biggr)
\end{equation}
where $\delta$ is the delta function,\footnote[1]{$\delta$ is a function with $\int_{-\infty}^{+ \infty} \delta (x-a) f(x) dx = f(a)$} $\delta \left( D - \sum_{i=1}^{n} x_i  \right) $ on the right hand side of equation (\ref{eq:magndistcond}) ensures that the sum is over all possible run durations $n \in \mathbb{N} - \{ 0 \}$.

As drawdowns can, in principle, last any number of periods, we need to sum over all possible durations to derive the probability of a drawdown of magnitude $D$ unconditional on $n$ as follows:
\begin{equation}{\label{eq:magndist}}
p(D) =   \frac{p_u}{p_d}  \biggl[ \sum_{n=1}^{\infty} \int_{-\infty}^{0} dx_1p(x_1) \dotsb \int_{-\infty}^{0} dx_n  p(x_n) \delta \biggl( D - \sum_{i=1}^{n} x_i  \biggr) \biggr]
\end{equation}
where $p_d = \int_{-\infty}^{0} p(x)dx$ is the probability of observing a negative price change, $p_u = 1- p_d$ is the positive counterpart and the term $\frac{p_u}{p_d}$ ensures the normalisation of the distribution.

Through asymptotic analysis and assuming \textit{iid} returns, the analytical expression for the drawdown probability density function $p(D)$ has an exponential distribution (\citeN{Johansen_largestock}):
\begin{equation}
\label{eq_simpleexp}
p(D) = \frac{1}{D_0} \exp ( - \frac{ |D|}{D_0})   ; \quad   D_0 = - \frac{1}{p_u p_d} \int_{-\infty}^{0} x p(x) dx 
\end{equation}
$D_0$ has a natural interpretation as the characteristic size of a drawdown. Sornette et al. extend the exponential distribution to the stretched exponential distribution in equation (\ref{E:drawdown4}) below to account for the slower decaying tails using extra terms in the Taylor series expansion in the asymptotic analysis:
\begin{equation} \label{E:drawdown4}
P_{se}(D) = D_0 e^{- \left( \frac{\vert D \vert}{\chi} \right)^z}
\end{equation}
Equation (\ref{E:drawdown4}) can fit distributions with smaller ($z >1$) or larger ($z < 1$) kurtosis than the simple exponential distribution. 

The parameters $z$ and $ \chi$ characterize the distribution concisely: the larger the value of $\chi$, the larger the 'typical' drawdowns; the smaller the value of $z$, the fatter the tails and the greater the relative likelihood of large drawdowns.
\citeN{Rebonato:2006xs} show, for identical and independent Gaussian returns, the expected drawdown magnitude is $E[D]_{\text{Gauss}} = \frac{4 \sigma}{\sqrt{2 \pi}}$, and the expected daily drop $d$ within a draw, $E[d]_{\text{Gauss}} = \frac{2 \sigma}{\sqrt{2 \pi}}$. 

A concept closely related to drawdowns and drawups is that of a {\it run}, defined as a sequence of positive (or negative) price changes. Results on the length of a run (which is independent of the duration of a draw) have been obtained by \citeN{Mood:1940tu}  and others. 
We denote the length of a run or drawdown by $l_d$. The probability that an arbitrarily chosen drawdown consists of exactly $l_d = n $ price moves can be calculated as the probability, given an initial down move, of further $n-1$ moves and a concluding up move as follows:
\begin{equation}
\label{eq_lengthdd}
P(l_d = n) = p_u \cdot p_d^{n-1}
\end{equation}
where $p_u$ the probability of an up move and $p_d$ is the probability of a down move, with $p_u + p_d = 1$, in a return series $\{x_0, \ldots x_n \}$. 

Applying the more general results from \citeN{Mood:1940tu}, one can calculate the expected number of runs  $E[N_{run}]$ in a time series of $n$ price changes. This is given as
\begin{equation}
E[N_{run}] = 2 n p_u (1-p_u) + p_u^2 + (1-p_u)^2
\end{equation}
For the case of a fair coin, $p_u = p_d = \frac{1}{2}$, the expected number of runs is therefore given as
$E[N_{run}]  = 2 n \frac{1}{4} + \frac{1}{4} + \frac{1}{4} = \frac{n+1}{2}$. For large $n$ this is approximately $ \frac{n}{2}$.

For the independent tossing of a fair coin with $p_u=p_d = \frac{1}{2}$, the expected length and variance of a draw are
\begin{align}
E[l_d]_{p_u = p_d = 1/2} & =   2 \\
var[l_d]_{p_u = p_d = 1/2} & = 2
\end{align}

Having reviewed the fundamental theoretical results of drawdowns, we review the empirical findings on financial market drawdowns as documented in a series of papers by Johansen, Ledoit and Sornette (JLS)\footnote{See, for example, \citeN{sornette-2003-378}, \citeN{Johansen_largestock} and \citeN{springerlink:10.1007/s100510070147}.} on the daily changes of Dow Jones Industrial Average, FTSE 100, some single-name stocks in different markets, foreign exchange US\$/DM, US\$/Yen, US\$/CHF and Gold price.
JLS found the modified exponential distribution introduced in equation (\ref{E:drawdown4}) is appropriate for up to  $98\%$ of drawdowns and drawups in equity and most financial returns.
The remaining $1\%$ to $2\%$ of the largest drawdowns are not explained by the simple exponential or by the stretched exponential. Drawdowns up to three times larger than expected are prevalent in equity indices, stock prices, FX rates and commodity prices, with the CAC40 being among the very few exceptions. JLS suggest that these draws are outliers which belong to a different class. Only half of the time series studied exhibit outliers in the drawups. Furthermore, in contrast to drawdowns drawups have a $\hat{z}$ (see Table \ref{t_zparam}) closer to the simple exponential case. 
The characteristic size of draws, $\chi$, is stable within an asset class but varies across asset classes. 
FX, as a group, has the smallest $\chi$, and has $z$ values $ \approx 0.84$ to $ 0.91$ (see Table  \ref{t_zparam}), which corresponds to the fat tail distribution. For equity indices, the typical price movement has $\chi \approx 1.05\%$ to $2.1\%$ for drawdowns and drawups, which is larger than that for FX. The $z$ coefficient for drawdowns is in the same range as for currencies i.e. 0.80 to 0.90, but for drawups, it is close to $1$, i.e. the simple exponential distribution. For individual stocks, the characteristic size is much larger, with $\chi \approx 1.91 \%$ to $7.61\%$. The exponents $z$ are close to $1$ for both drawups and drawdowns.
The $z < 1$ exponent of the equity indices, gold and FX rates, is compatible with a daily return distribution with fat tails. Diverging from this is the distribution of drawdowns of large US stocks, with $z \approx 1$, compatible with a daily return distribution with normal tail size.

 \begin{table}[!h]
\centering
{\scriptsize
\begin{tabular}{ l c  c  c  c  c  c  c  }
\hline
  & Drawdowns & Drawups &   &   & Drawdowns & Drawups   \\
 \hline
&&&&&&\\
\textit{FX and US Equity}  &  &  &   & \textit{US Interest Rate}  &  &    \\
FX Japanese Yen/US Dollar &  $0.90 \pm 0.02$  & $0.89 \pm 0.02$   &&  3m & $1.006 \pm 0.034$ &  $0.999 \pm 0.032$ \\
SP 500  & $0.90 \pm 0.01$  & $1.03 \pm 0.02$   && 6m & $1.058 \pm 0.032$ & $1.053 \pm 0.031$ \\
NASDAQ Composite &  $0.80 \pm 0.02$  & $0.90 \pm 0.02$   && 1y & $1.188 \pm 0.035$ & $1.211 \pm 0.030$ \\
Dow Jones & $0.84 \pm 0.01$  & $0.99 \pm 0.01$     && 2y & $1.053 \pm 0.023$ & $0.987 \pm  0.022$ \\
Microsoft &  $1.04 \pm 0.03$  & $1.01 \pm 0.03$   && 5y & $1.127 \pm 0.025$ &  $1.050 \pm 0.024$ \\
Cisco &  $1.16 \pm 0.04$  & $1.21 \pm 0.04$   && 10y & $ 1.154 \pm 0.025$ & $1.052 \pm 0.023$ \\
General Electric &  $1.02  \pm 0.02$  & $1.02 \pm 0.02$   && 20y & $ 1.188 \pm 0.025$ &  $1.060 \pm 0.024$ \\
Intel  & $1.06 \pm 0.03$  & $1.21 \pm 0.03$   && 30y & $1.144 \pm 0.026$ & $1.063 \pm 0.023$ \\
  \hline
\end{tabular}}

\caption{Draws in FX, Equity and Interest Rate markets}
\begin{quote}
Reported fit of the exponential distribution $\hat{z}$ ($\pm$ standard error) for drawdowns and drawups on FX, selected US equities and interest rates daily returns from \citeN{Johansen_largestock}.
\end{quote}
\label{t_zparam}
\end{table}

JLS argue that the anomalously large drawdowns can only be explained by rare and successive daily drops with correlated magnitude. These occur when the market enters a critical stage, where a phase transition leads to a different behaviour among rational, informed traders and noisy, imitative traders. Herding behaviour and market participants imitating the majority as an optimal strategy are cited in \citeN{sornette-2003-378} as examples of different traits of the market participants. It is the change of mode of interaction between market participants that triggers a regime change in price moves, producing different serial co-dependence and magnitudes. Other industry practices, such as portfolio insurance and margin requirements, further aggravate the herding behaviour during periods of market stress.

\citeN{Rebonato:2006xs} conducted an extensive study of draws in US LIBOR and swap rates and found a significantly larger number of outliers among the interest rates with shorter maturities, for both drawdowns and drawups. By and large, when the $z$ coefficient  for drawups is close to $1$ and for drawdowns higher than $1$, thin tails are suggested (see Table \ref{t_zparam}). The runs of length 1 (immediate price reversals) are more often than expected, when compared to the null hypothesis of independent price moves. Runs of lengths $2$ and $3$ are, on the other hand, less often than expected. 


\citeN{Rebonato:2006xs} study in the USD interest rate market is the first and perhaps also the only bivariate analysis of draws. It is well known that the first eigenvector from a PCA analysis accounts for 80\% - 90\% of the movements of the interest rate term structure. The study provides some evidence of coinciding drawdowns within the blocks of the short rates (3m, 6m and 1y), the mid-term rates (2y, 5y and 10y) and the long-term interest rates (20y and 30y), but not between the blocks. In particular, none of the nine large drawdowns among the short rates coincides with any of the large drawdowns in the other two blocks.

 \FloatBarrier

\subsection{Entropy Econometric Literature}
\label{subsec_infotheory}

An important time series tool based on the information theory of \citeN{Shannon:1948wk} has started to gain attention among the finance research community. The key in this line of research is to find optimal encodings of data with respect to the noise. It is a non-parametric way to analyse information or relationships under uncertainty.

\citeN{Darbellay:2000vw} use mutual information to estimate the dependencies in the 30-minute returns on the DEM/USD exchange rate, and the daily returns on Dow Jones industrial average stock index. The authors show that returns and volatility, measured over a rolling fixed window, are not independent.

\citeN{Schreiber:2000jx} extends mutual information into the concept of transfer entropy, which takes into account transition probabilities. Transfer entropy is particularly useful in detecting information flows between systems and has, since then, found applications in the multivariate analysis of time series in different epistemic domains, e.g. neurology and ecology.

Transfer entropy has been applied to financial time series by \citeN{Marschinski:2002dq}, who find evidence of information flow from the Dow Jones Industrial Average index to the DAX stock index, sampled at the one-minute interval between May 2000 and June 2001. Any information flow from the reverse direction is considerably less pronounced. 

\citeN{Kwon:2008be} use transfer entropy and find empirical evidence for information flows between the daily returns of 25 composite stock indices. The value of outgoing transfer entropy from the American and European markets is the highest and appears to flow mainly to the Asian and Pacific regions. Intra-continent information flows are less pronounced than inter-continent flows. Similarly, \citeN{Kwon:2008vl} found empirical evidence for information flows between daily returns of 25 stock indices and individual single-name stock returns in the respective markets.
 
\citeN{Dimpfl:2011tp} used transfer entropy and found evidence of bi-directional information flows between CDS and bond spread, with the information transfer from the CDS market to the bond markets being slightly higher. The data could not be modelled using a Vector Error Correction Model due to the lack of a cointegration relation. Furthermore, a unidirectional information flow from the VIX to the iTraxx is also established.
 
\citeN{Jizba:2011dq} extend the transfer entropy into the R\'{e}nyi transfer entropy, using
\textit{R\'{e}nyi Entropy}, $S_q^{(R)}$.  $S_q^{(R)}$, defined below, is a family of entropy measures for a distribution $P$ on a finite set $X$ and indexed by an order $q > 0$:
\begin{equation}
\label{eq_renyi}
S_q^{(R)}(P) = \frac{1}{1-q} log_2 \sum_{x \in X} p^q(x)
\end{equation}
Unlike the Shannon entropy, with the control of the parameter $q$, the R\'{e}nyi entropy can place emphasis on different parts of the distribution, not necessarily taking into account equally the entire underlying empirical price distribution. A $q \in (0,1)$ focuses on the tail of the distribution, and with $q>1$, the focus is on the main part of the distribution. 
The R\'{e}nyi  transfer entropy was estimated on daily returns of 11 stock indices from Europe, the US and Asia over the period from 1990 to 2009. The authors found that US and European markets are more influenced by price turbulence in the Asia - Pacific region than the other way around.

\section{Data}
\label{sec_data_draw}

Log returns of daily closing and hourly EUR/USD and GBP/USD exchange rates were calculated for the period from 3rd January 2001 to 27th July 2012. There are 3,017 daily and 71,039 hourly observations.\footnote{Daily closing rates were from Bloomberg. The hourly data were downloaded from www.fxhistoricaldata.com.}
 
\begin{figure}[ht]
\centering
\includegraphics[width=4.5in]{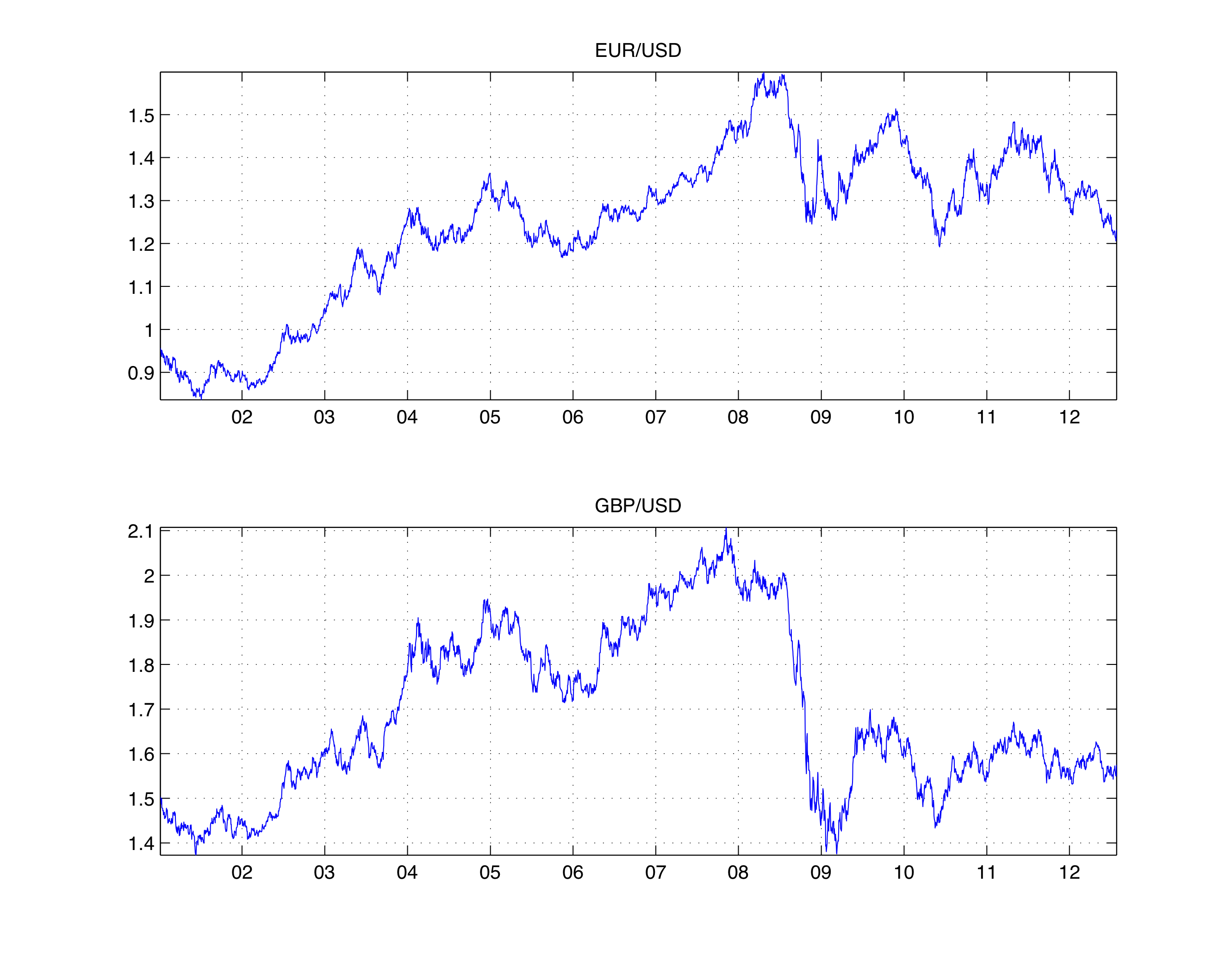}
\caption{EUR/USD, GBP/USD time series}
\begin{quote}
Daily EUR/USD and GBP/USD exchange rates for the period from 3rd January 2001 to 27th July 2012.
\end{quote}
\label{fig:rtnallcurrencies}
\end{figure}
 
The sample period includes different market regimes and a number of significant economic events, such as the financial crisis of 2007. Table  \ref{T:basicstats} presents statistics for exchange rate returns and drawdowns/drawups. 
The first four lines of Table \ref{T:basicstats} report the four first central moments ($\mu_1$, $\mu_2$, $\mu_3$, $\mu_4$) of the respective return series. The mean for the daily series are comparable to the mean of the respective hourly series for both currency pairs, sampled 24 hours within a day. Within the chosen sample period, both EUR and GBP appreciated against the USD (more so for the EUR than the GBP) resulting in positive means for both currency pairs.
The Volatility ($\mu_2$) 
of the EUR/USD hourly series is comparable to the daily series (applying a square root of time scaling). For the GBP/USD series, the hourly series is more volatile than the daily series. EUR/USD is positive skewed, reflecting the general appreciation of the EUR during the sample period despite the European 
sovereign crisis.
GBP/USD is negatively skewed, mainly due to the large movements in 2008. All the series have a kurtosis higher than four, which is more pronounced for the hourly than the daily returns. The unconditional correlation $\hat{\rho}$ between the two exchange rates is $0.673$ for both the daily and hourly data. 
 
\begin{table}[t]
\centering
{\scriptsize
\begin{tabular}{ l  c  c  c  c  }
\hline
  & EURUSD (h) & GBPUSD (h) & EURUSD (d) & GBPUSD (d) \\
 \hline
 \textit{Panel A: Sample moments} &&&& \\
$\mu_1$ [$10^{-6}$] & 3.624 & 0.644 & 86.413 & 15.937 \\
$\hat{\mu}_2$ & 0.002 & 0.002 & 0.007 & 0.006 \\
$\hat{\mu}_3$ & 0.07 & -0.263 & 0.006 & -0.306 \\
$\hat{\mu}_4$ & 12.814 & 15.481 & 4.201 & 5.33 \\
$\hat{\rho}$ &0.6730  & & 0.6736 & \\
 \hline
 \textit{Panel B: Gaussian Model} &&&& \\
$E[D]_{\text{Gauss}} = 4\sigma / \sqrt{2 \pi}$ $[10^{-3}]$ & 2.15 & 2.04 & 10.44 & 9.41 \\
$E[d]_{\text{Gauss}} = 2\sigma / \sqrt{2 \pi}$ $[10^{-3}]$ & 1.08 & 1.02 & 5.22 & 4.71 \\
\hline
 \textit{Panel C: Statistics on Drawdowns, Drawups} &&&& \\
$E[D]/E[d] = E[l_{d}]$ & 1.98 & 1.96 & 1.85 & 1.98 \\
$E[U]/E[u] = E[l_{u}]$ & 2.02 & 2.01 & 1.96 & 2.04 \\
$E[D]$ $[10^{-3}]$ & 1.7566 & 1.6288 & 9.2321 & 8.8101 \\
$E[U]$ $[10^{-3}]$ & 1.7711 & 1.6314 & 9.5508 & 8.8623 \\
$\sigma(D)$ $[10^{-3}]$ & 2.0492 & 1.9979 & 9.2219 & 9.4067 \\
$\sigma(U)$ $[10^{-3}]$ & 2.0533 & 1.9449 & 8.9 & 8.41 \\
$\sigma(l_{d})$ & 1.33 & 1.31 & 1.26 & 1.36 \\
\hline
 \textit{Panel D: Largest Drawdowns, Drawups} &&&& \\
maximum d.d. $[\%]$ & 2.68 & 3.93 & 7.94 & 10.68 \\
date of max. d.d. & 05-Jan-09 08:00 & 24-Oct-08 07:00 & 29-Sep-08 & 17-Oct-08 \\
maximum d.u. $[\%]$ & 3.11 & 3.3 & 10.93 & 5.62 \\
date of max. d.u. & 13-Nov-08 19:00 & 28-Oct-08 18:00& 10-Dec-08 & 28-Oct-08 \\
 \hline
\end{tabular}}
\caption{EUR/USD and GBP/USD return and draw statistics.}
\begin{quote}
The first four sample moments $\hat{\mu_1}, \ldots \hat{\mu_4}$ of the hourly (h) and daily (d) time series are estimated. $\hat{\rho}$ is correlation between the currency pairs in the hourly and daily series. 
In panel B we report the expected draw magnitude $E[D]_{\text{Gauss}}$ and daily drop $E[d]_{\text{Gauss}}$ assuming iid Gaussian returns. Panel C lists for drawdowns (drawups) expected length $E[l_{d}$ ($E[l_{u}$) magnitude $E[D]$ ($E[U]$), standard deviation $\sigma(D)$ ($\sigma(U)$) and standard deviation of length $\sigma(l_{d})$. The largest drawdowns and drawups in the time series with magnitude and start time are reported in panel D.
\end{quote}
\label{T:basicstats}
\end{table}
The statistics concerning the drawdowns and drawups in the four time series are reported in the lower panel of Table \ref{T:basicstats}. 
For all series, the average draw magnitude is lower than if the series were a normally distributed \textit{iid} process, i.e. $E[D] \le E[D]_{Gauss}$,  $E[U] \le E[D]_{Gauss}$. The average length of the drawdown is lower than the Gaussian case with $E[l_d] \le 2$, but higher for drawups, with the exception of the daily EUR/USD series. 
The bulk of the drawdowns have a length $\vert l_d \vert \le  E[l_d] + \sigma(l_d) \approx 3.3$ periods, and similar numbers hold for drawups. The average drawdown magnitude is marginally lower than that of the average drawup $E[D] < E[U]$. 
For both currency pairs, the largest draws in the daily series occurred in autumn 2008, at the height of the financial crisis, which started in 2007 (\textit{Panel D} in Table \ref{T:basicstats}). The largest drawdown of the daily EUR/USD series started on September 29, 2008, lasting six days, and was triggered by renewed concerns over the European  financial system. Two months later, the largest drawup, lasting six days, started on December 10, 2008, in the week when the Federal Reserve Bank cut its short term interest rates to historical lows.  On October 17, 2008, the largest drawdown of the daily GBP/USD was observed. It lasted seven days and was set in a period of sell off of the pound due to the macroeconomic environment and concerns over the British banking sector. Immediately following it, there was a market correction on October 28 lasting three days, which formed the largest drawup in the GBP/USD daily return series. The expected drawdown length for all series is lower than the Gaussian case with $E[l_{d}] < 2$, which shows a tendency for price reversal. With the exception of the daily EUR/USD, all drawups had a marginally higher expected length than in the Gaussian case with $E[l_{d}] > 2$.

\begin{table}[t]
\centering
{\scriptsize
\begin{tabular}{ l  c  c  c  c  }
\hline
 & EUR/USD (h) & GBP/USD (h) & EUR/USD (d) & GBP/USD (d) \\
\hline
\textit{Panel A: Drawdowns} &&&&\\
$\hat{z}(D)$ & 0.993 & 0.98 & 1.145 & 1.121 \\
$\sigma_{\hat{z}}(D)$ & 0.005 & 0.005 & 0.029 & 0.025 \\
$\hat{\chi}(D)$ [$10^{-3}$] & 1.765 & 1.616 & 10.03 & 9.255 \\
$\sigma_{\hat{\chi}}(D)$ [$10^{-3}$] & 0.012 & 0.011 & 0.254 & 0.249 \\
\hline
\textit{Panel B: Drawups} &&&&\\
$\hat{z}(U)$ & 0.986 & 0.959 & 1.026 & 1.033 \\
$\sigma_{\hat{z}}(U)$ & 0.005 & 0.005 & 0.022 & 0.025 \\
$\hat{\chi}(U)$ [$10^{-3}$] & 1.745 & 1.595 & 9.332 & 8.932 \\
$\sigma_{\hat{\chi}}(U)$ [$10^{-3}$] & 0.011 & 0.011 & 0.268 & 0.256 \\
\hline
\end{tabular}}
\caption{Stretched exponential distribution fit}
\begin{quote}
For drawdowns (drawups), the stretched exponential $P_{se}(D) = D_0 e^{- \left( \frac{\vert D \vert}{\chi} \right)^z}$ is estimated using the maximum likelihood method. The shape parameter,  $\hat{z}$, and size parameter, $\hat{\chi}$,  of the hourly and daily series are reported for drawdowns in \textit{Panel A} and drawups in \textit{Panel B}. Following \citeN{Rebonato:2006xs}, we calculate the standard error of the estimates using a Jackknife procedure. 
\end{quote}
\label{T:exponentialfit}
\end{table}

Table \ref{T:exponentialfit} shows the shape parameter, $\hat{z}$, and the size parameter, $\hat{\chi}$, of the modified exponential distribution fitted the drawdowns and drawups, and based on the maximum likelihood method. 
This method is known to be asymptotically unbiased but requires large samples. The estimation error $\hat{\sigma}$ is calculated by the bootstrapping technique (see \citeN{Rebonato:2006xs}). New data sets are generated repeatedly by replacing a subset of the data by data drawn from an exponential distribution specified by the estimated parameters. The daily returns have a draw shape parameter $\hat{z} >1$ by more than two standard deviations for drawdowns. This suggests thinner tails than the simple exponential distribution. On the other hand, the drawups of the daily return series exhibit a shape $\hat{z}$ which is within two standard deviations of the simple exponential distribution. Similarly, the estimated shape $\hat{z}(D)$ of the hourly drawdown is within two standard deviations of the simple exponential. For the drawups of the hourly series, $\hat{z}(U) < 1$ by more than two standard deviations, which indicates a thicker tail than the simple exponential.

For drawdowns in both daily and hourly series, the characteristic scale $\hat{\chi}(D)$, determining the expected draw $E[D]$, is within two standard deviations to the mean, i.e. $\vert E[D] - \hat{\chi}(D) \vert < 2 \cdot \sigma_{\hat{\chi}}(D)$. The same is true for the daily drawups, but not for the hourly drawups, which is lower than the expected drawup $E[U]$. 

\FloatBarrier
\section{Information Theoretic Method}
\label{sec_depanalysis}

\subsection{Entropy Measures}
\label{subsec_paper1info}

In Chapter \ref{chap_entropy}, we introduced the entropy measures, \textit{Mutual Information} and \textit{Transfer Entropy}. These allows us to detect correlation and cross correlation effects between time series. 
Mutual Information  $I(X;Y )$ of two random variables $X$ and $Y$ with joint distribution $p(x,y)$ is defined as
\begin{equation}
\label{eq_mutualinfo2}
I(X;Y) = - \sum_{x \in X , y \in Y} p(x,y) \cdot log_2 \frac{p(x,y)}{p(x)p(y)} 
\end{equation}
Mutual information measures the reduction of uncertainty about $X$ from observing $Y$. 
We further saw in Chapter \ref{chap_entropy} that introducing a time delay in one of the observations does not distinguish information that is actually exchanged from a common response to input signal or a common history driven by an external factor. To detect information spillover, we use transfer entropy instead. 
Let $p(x_1, \ldots x_n)$ denote the probability of observing the subsequence $(x_1, \ldots x_n)$, then transfer entropy is defined as
\begin{align}
\label{eq_transferconditional2}
T_{Y \rightarrow X} (m,l)  &=  \sum p( x_{t_{1}} , \ldots x_{t_{m}} ,y_{t_{m-l+1}} , \ldots , y_{t_{m}}) \notag \\
 & \cdot log_2 \frac{p(x_{t_{m+1}} \vert x_{t_{1}} , \ldots x_{t_{m}} ,y_{t_{m-l+1}} , \ldots , y_{t_{m}})}{p(x_{t_{m+1}} \vert x_{t_{1}} , \ldots x_{t_{m}})}
\end{align}
where $x_t$ and $y_t$ represent the discrete states of $X$ and $Y$ at time $t$. The parameters $m$ and $l$ indicate the number of  past observations included in $X$ and $Y$ respectively.  
While mutual information quantifies the deviation from $X$  (or $Y$) being independent, transfer entropy quantifies the deviation from $X$ being determined by its own history only (via conditional probabilities) or by the history of  $Y$.
Unlike mutual information, transfer entropy $T_{Y \rightarrow X} (m,l)$ is not symmetric, and takes into account only the statistical dependencies originating in the variable $Y$ and not those from a common signal. 

In the literature, both transfer entropy and mutual information have been estimated for continuous variables based on the kernel density methods \citeN{Schreiber:2000jx}, \citeN{Blumentritt:2011wf} and the maximum likelihood methods \citeN{Paninski:2003vz}. 
In this chapter, we are dealing with market-observed discrete returns, and the number of discrete values is too high for any practical purposes and needs to be considerably reduced before calculating the entropy measures. 
In this chapter, the partition used is motivated by economic considerations focussing on specific market situations, such as large drawdowns and drawups, which may or may not lead to a equal marginal probability for each partition.
\begin{center}
\begin{figure}
   \includegraphics[width=6in]{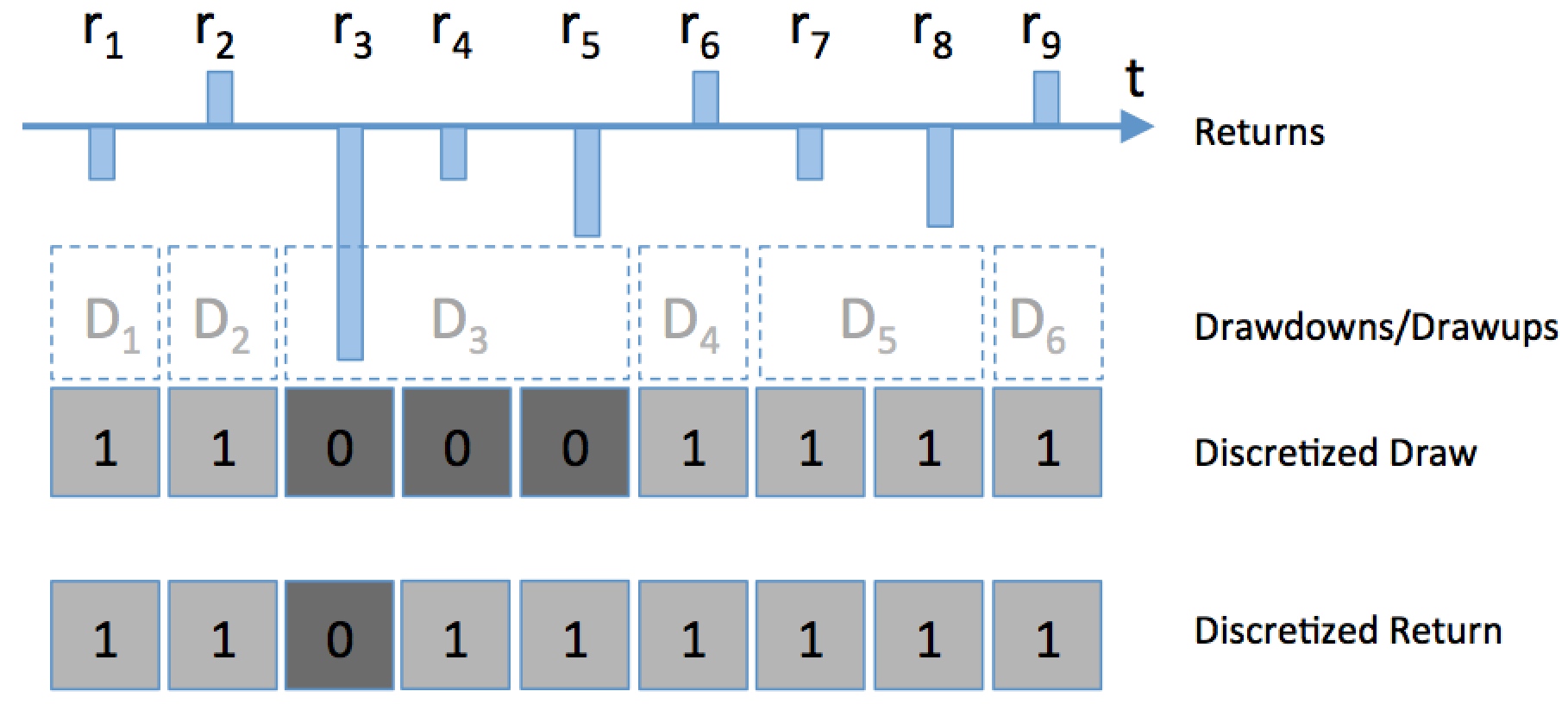}
      \caption{Discretization example}
      \begin{quote}
Example nine returns with draw states and discretized outcomes.  \label{fig_discretization}
\end{quote}
   \end{figure}
\end{center}
Take the example of a return series, displayed in Figure \ref{fig_discretization} with nine time steps. The returns $\{r_1, \ldots , r_9\}$ form drawdowns, drawups $\{ D_1, \ldots , D_{6} \}$, which are depicted in the second line of the figure. For instance, the consecutive negative returns $r_3 , r_4 , r_5$ form one drawdown $D_3$.
These nine returns are part of a larger time series that we wish to put into partitions. Here, there are at least two different approaches to partitioning the data. The first approach is to follow the literature (such as discussed in section \ref{subsec_infotheory}) by mapping individual returns to a partition (or letter) according to the size of the returns. Let $q(a,b)$ denote the quantile of $a$ at $b$ confidence level. 
The first approach is to map the returns as follows:
\begin{equation}
\label{eq_discreteddmag}
d_r(r_t) =
  \begin{cases}
  0 & \text{if } r_t < q(r_t,0.33) \\
  1 & \text{if } q(r_t,0.33) \leq r_t \leq q(r_t,0.66) \\ 
   2   & \text{if } r_t  > q(r_t,0.66)
  \end{cases}
\end{equation}
Since $r_3$ is the only return that is negative and large in absolute term, it is likely to be the only return in the group of nine returns to be mapped to the letter '0'.
The other eight returns are small in absolute term, and are likely to be mapped to letter '1'. The outcome of this mapping is shown in Figure \ref{fig_discretization} as 'Discretized Returns'.
Alternatively, since we are interested in large drawdowns and drawups, we could map returns according to their membership of a particular type of draw instead of the magnitude of the individual returns:
\begin{equation}
\label{eq_discreteddmag}
d_D(r_t) =
  \begin{cases}
  0 & \text{if } D(r_t) < q(D,0.05) \\
  1 & \text{if } q(D,0.05) \leq D(r_t) \leq q(D,0.95) \\ 
   2   & \text{if } D(r_t)  > q(D,0.95)
  \end{cases}
\end{equation}
where $D(r_t)$ denotes the size of the draw, $q(D,0.05)$ is the 5\% quantile of the draw distribution and similarly $q(D,0.95)$ is in the 95\% percentile. This means that the letter '0' is associated with a large drawdown, the letter '2' is associated with a large drawup and the letter '1' has all the draws with small draw size. 
So, it is likely that in our example, all the returns in $D_3$ (i.e. $r_3$, $r_4$ and $r_5$) will be mapped to the letter '0' and the other six returns ($r_1$, $r_2$, $r_6$, $r_7$, $r_8$ and $r_9$) will be mapped to the letter '1'. There is no large drawup among the nine returns. 
The outcome of this mapping is shown in Figure \ref{fig_discretization} as 'Discretized Draw'.
In the previous case, the marginal probability for each partition (or letter) will be the same. For the second approach, which is the approach we have adopted here, equal marginal probability is not guaranteed and will depend on the distribution of the draw size. 

Let $X \rightarrow \mathcal{A}$ be the  discretized random variable where all values are binned into $M = \vert \mathcal{A} \vert$ boxes. 
The entropy of a discrete probability distribution can be calculated by counting the relative frequencies of $X$ falling into each bin. The estimator, which we will call the na\"{i}ve entropy estimator, can be calculated as follows.
\begin{align}
\label{eq_naiveestimator2}
\hat{H}_{naive}  &= - \sum_{i =1}^{M} p_i \cdot log_2(p_i) 
= - \sum_{i =1}^{M} \frac{ n_i}{N} \cdot log_2(\frac{ n_i}{N})  \notag \\
&= log_2 (N) - \frac{1}{N} \sum_i n_i \cdot log_2(n_i) 
\end{align}

If the sample size, $N$, is small, statistical fluctuations will artificially induce deviations to the entropy estimates producing a downward bias.  \citeN{2003physics...7138G}  offers a correction to this 'small sample' bias. Assuming that all $p_i \ll 1$, the proposed new estimator $H_{\psi}$ is as follows:
\begin{align}
\label{eq_grassberger}
\hat{H}_{\psi} &= ln N - \frac{1}{N}\sum_{i =1}^{M} n_i \cdot \psi (n_i) \\ 
\psi(x) &=\frac{d(ln \Gamma (x))}{dx}, \qquad \Gamma(0,x) = \int_{1}^{\infty} \frac{e^{-xt}}{t} dt
\end{align}
We will use this estimator in the Section \ref{subsec_infoflow}.

\FloatBarrier
\label{subsec_paper1_entropyblock}
Depending on the particular choice of partition, the entropies of the individual processes will change. Herein, the choice of quantile $q_1$ in the discretization scheme (\ref{eq_discreteddmag}) determines which symbols the draw states in the process will get mapped.
With a low $q_1$ the individual processes are expected to have low entropies as the symbols are more unequally distributed. Higher $q_1$ quantiles lead to correspondingly higher entropies. The maximal entropy in this set up of symbol sequences with three letters is given by the entropy of the uniform three symbol distribution: $- 3 \cdot \frac{1}{3} \cdot log_2 (\frac{1}{3}) = 1.585$. The entropy estimates for the two exchange rates are shown in Table \ref{T:symbolentropy}. 
\begin{table}[t]
\centering
{\scriptsize
\begin{tabular}{ l  c  c  c  c  c  c  c  c  c  c  }
\hline
 & \multicolumn{10}{c}{\textit{Quantile} $q_1$:}\\ 
 &&&&& &&&&& \\
 & 0.025 & 0.050 & 0.075 & 0.100 & 0.125 & 0.150 & 0.175 & 0.200 & 0.225 & 0.250 \\
\hline
\textit{Panel A: Daily Series} &&&&& &&&&& \\
&&&&& &&&&& \\
$\hat{H}_{\text{EUR/USD}}$(d) & 0.674 & 0.968 & 1.164 & 1.308 & 1.411 & 1.483 & 1.531 & 1.561 & 1.58 & 1.585 \\
$\hat{H}_{\text{GBP/USD}}$ (d) & 0.66 & 0.968 & 1.163 & 1.315 & 1.412 & 1.478 & 1.532 & 1.562 & 1.58 & 1.585 \\
\hline
\textit{Panel B: Hourly Series} &&&&& &&&&& \\
&&&&& &&&&& \\
$\hat{H}_{\text{EUR/USD}}$(h) & 0.597 & 0.89 & 1.095 & 1.247 & 1.362 & 1.446 & 1.507 & 1.549 & 1.574 & 1.585 \\
$\hat{H}_{\text{GBP/USD}}$ (h) & 0.593 & 0.888 & 1.088 & 1.242 & 1.357 & 1.442 & 1.504 & 1.547 & 1.572 & 1.584 \\
\hline
\end{tabular}}
\caption{Entropy $\hat{H}$ of daily and hourly EUR/USD and GBP/USD}
\begin{quote} 
The partitions are defined as in equation (\ref{eq_discreteddmag}), with a range of values for quantile $q_1$:

$d_D(r_t) =
  \begin{cases}
  0 & \text{if } D(r_t) < q(D,q_1) \\
  1 & \text{if } q(D,q_1) \leq D(r_t) \leq q(D,1-q_1) \\ 
   2   & \text{if } D(r_t)  > q(D,q_1)
  \end{cases}$
\end{quote} 
\label{T:symbolentropy}
\end{table}
From Table \ref{T:symbolentropy}, the maximal entropy is reached with partitions of quantiles $q_1 \approx 0.225$ and $q_2 = 1-q_1 \approx 0.775$ which are lower than $q_1 = \frac{1}{3}$. This is the case as the quantiles of the draw distribution were used in the discretization and not the quantile of the draw state distribution. As larger draws tend to be longer than smaller draws, moving from a partition defined by draw magnitude to that based on the discretized draw states will lead to a partition thats is 'skewed' towards the larger draws.

\begin{figure}[ht]
\begin{minipage}[b]{0.5\linewidth}
\centering
\includegraphics[width=2.95in]{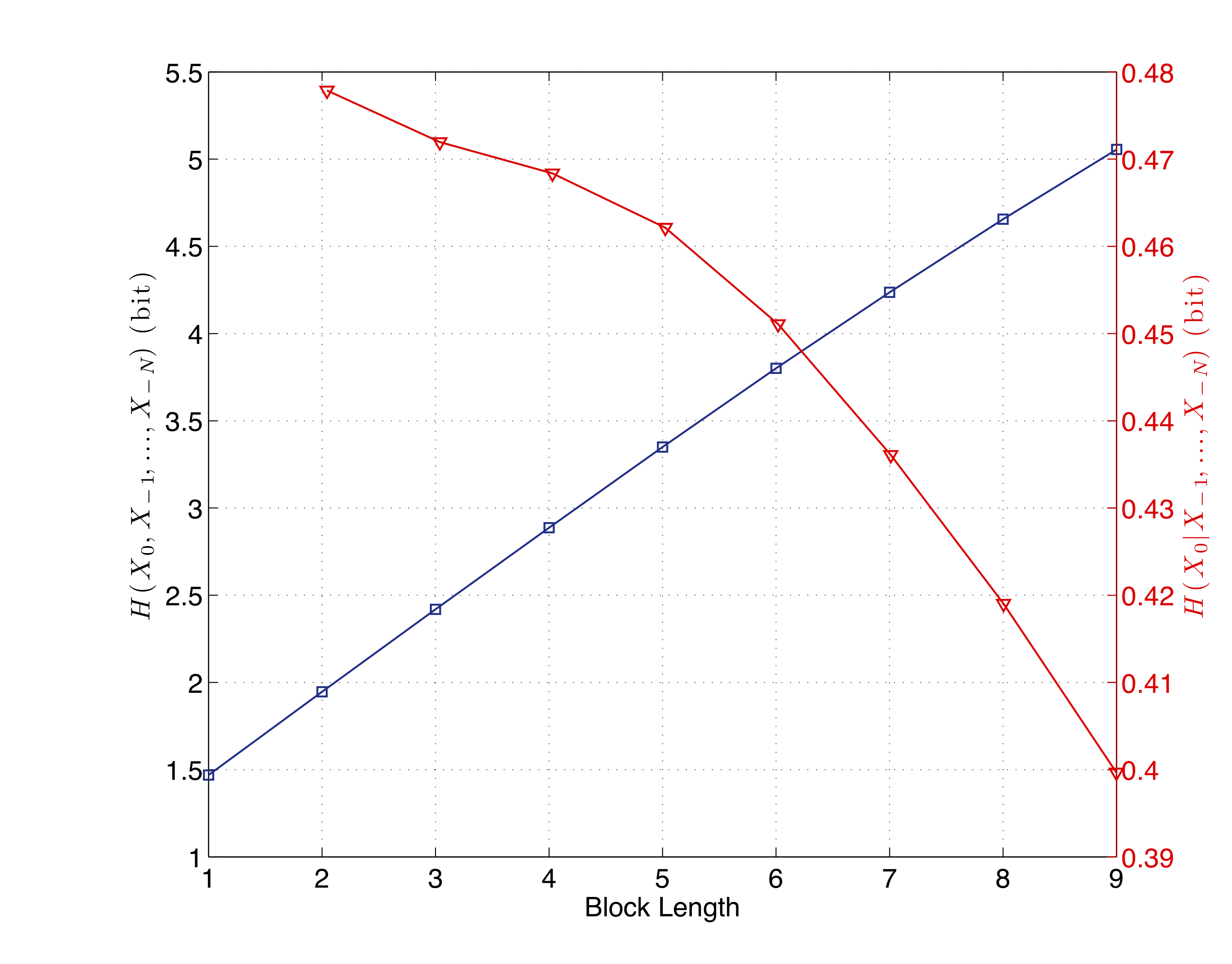}
\end{minipage}
\hspace{0.1cm}
\begin{minipage}[b]{0.5\linewidth}
\centering
   \includegraphics[width=2.95in]{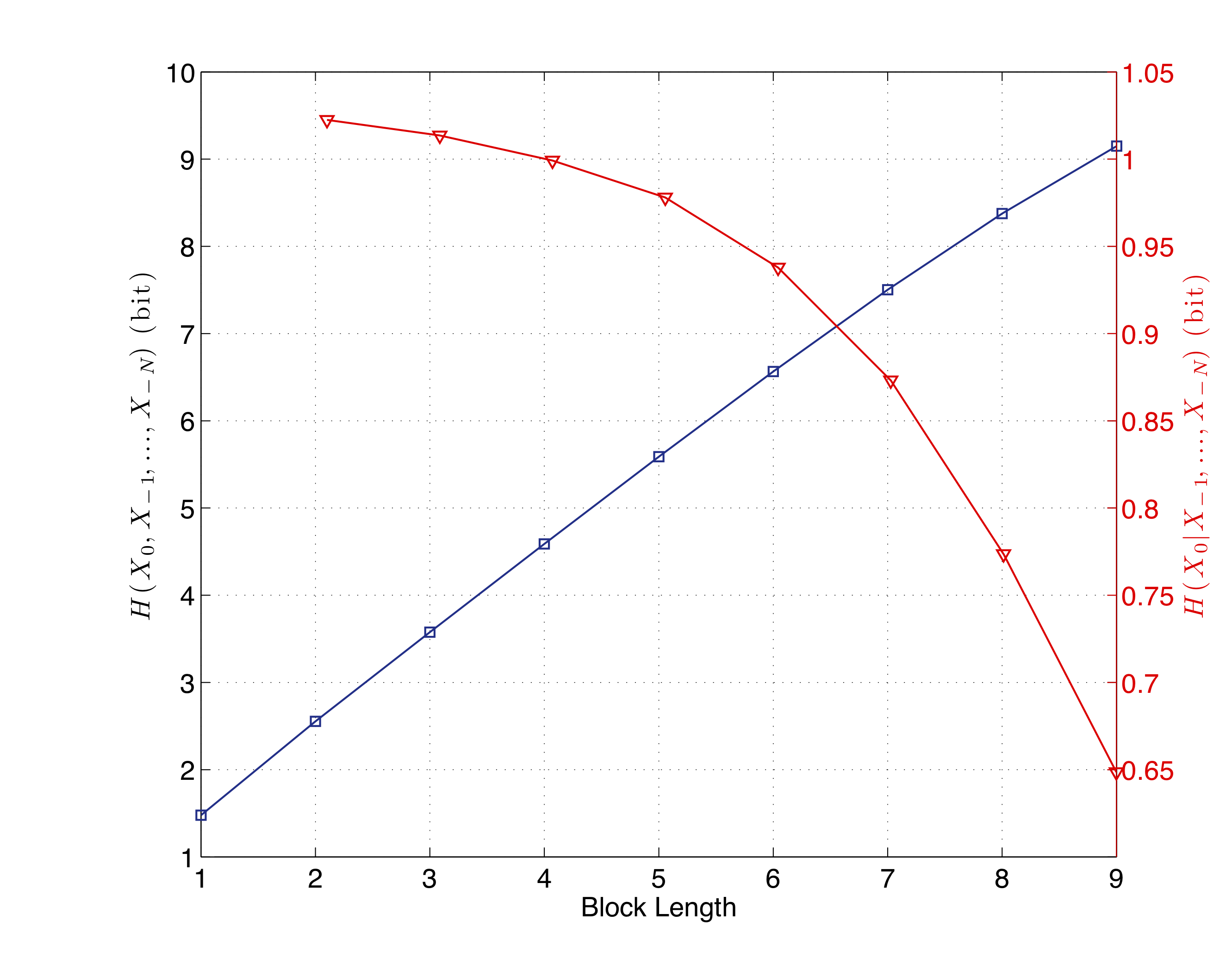}
\end{minipage}
\caption{Block Entropy EUR/USD}
\begin{quote}
The figure shows the unconditional $H(X_{0}, X_{-1} , \ldots X_{-(m+1)}  ) $ and conditional  block entropies $H(X_{0}\vert X_{-1} , \ldots X_{-(m+1)}  ) $ for history lengths of $m=1$ to $m=9$ on the discretized time series of EUR/USD. Two discretization schemes are used. On the left  $q_1=0.05$ and on the right $q_1 = 0.15$.
\end{quote}
\label{fig_blockeurusd}
\end{figure}

Transfer entropy, as defined in equation (\ref{eq_transferconditional2}), uses entropy that is conditional on blocks of history  to determine if processes influence each other. In the remainder of this section, we discuss some structural features of $H(X_{0}\vert X_{-1} , \ldots X_{-(m+1)}  ) $ of the exchange rate processes defined by the discretization. 
In Figures \ref{fig_blockeurusd}  the conditional block entropies and block entropies of the daily EUR/USD draw states are shown. The left figure is for a partition with quantiles $q_1= 0.05$, $q_2= 0.95$ focussing on the large draws. The right figure is for a more equiprobable partition, with quantiles $q_1=0.15$, $q_2=0.85$. The absolute levels are differ by a factor of two, which is explained by the entropies of the symbol distribution, which are $\hat{H}_{q_1=0.05}(\text{EUR/USD}) = 0.968$ and $\hat{H}_{q_1 = 0.15}(\text{EUR/USD}) = 1.48$ (see Table \ref{T:symbolentropy}). 

The conditional entropy for the case $q_1=0.05$ drops slower than that for  $q_1 = 0.15$. This is consistent with the finding that the length of draws is $E[l_D] \approx 2$ (see Table \ref{T:basicstats} the lines for $E[l_{d}],E[l_{u}]$) but conditional on the draw being in the higher quantiles $q=0.05,0.95$ the average length of EUR/USD drawdowns can be estimated as $E[l_D \vert D < q(D,0.1)] = 4.03 \pm 1.65$ and for drawups as $E[l_U \vert U >q(U,0.9)] = 4.21 \pm 1.93$.  These results taken together, show that the bulk of large draws have a length not exceeding $4.03 + 1.65 \approx 5.68$ $4.21+1.93 \approx 6.14$. Therefore, we would expect to see the conditional block entropy to reach a plateau at lengths of over $6.14$.
We confined the discussion here to the daily EUR/USD series as the picture is very similar for the GBP/USD series and for both hourly series.

\FloatBarrier
\section{Empirical Results}
\label{sec_results}
\subsection{Dependency between Draws in EUR/USD and GBP/USD}
\label{subsec_depanalysis}
\begin{figure}[ht]
\begin{minipage}[b]{0.5\linewidth}
\centering
\includegraphics[width=2.95in]{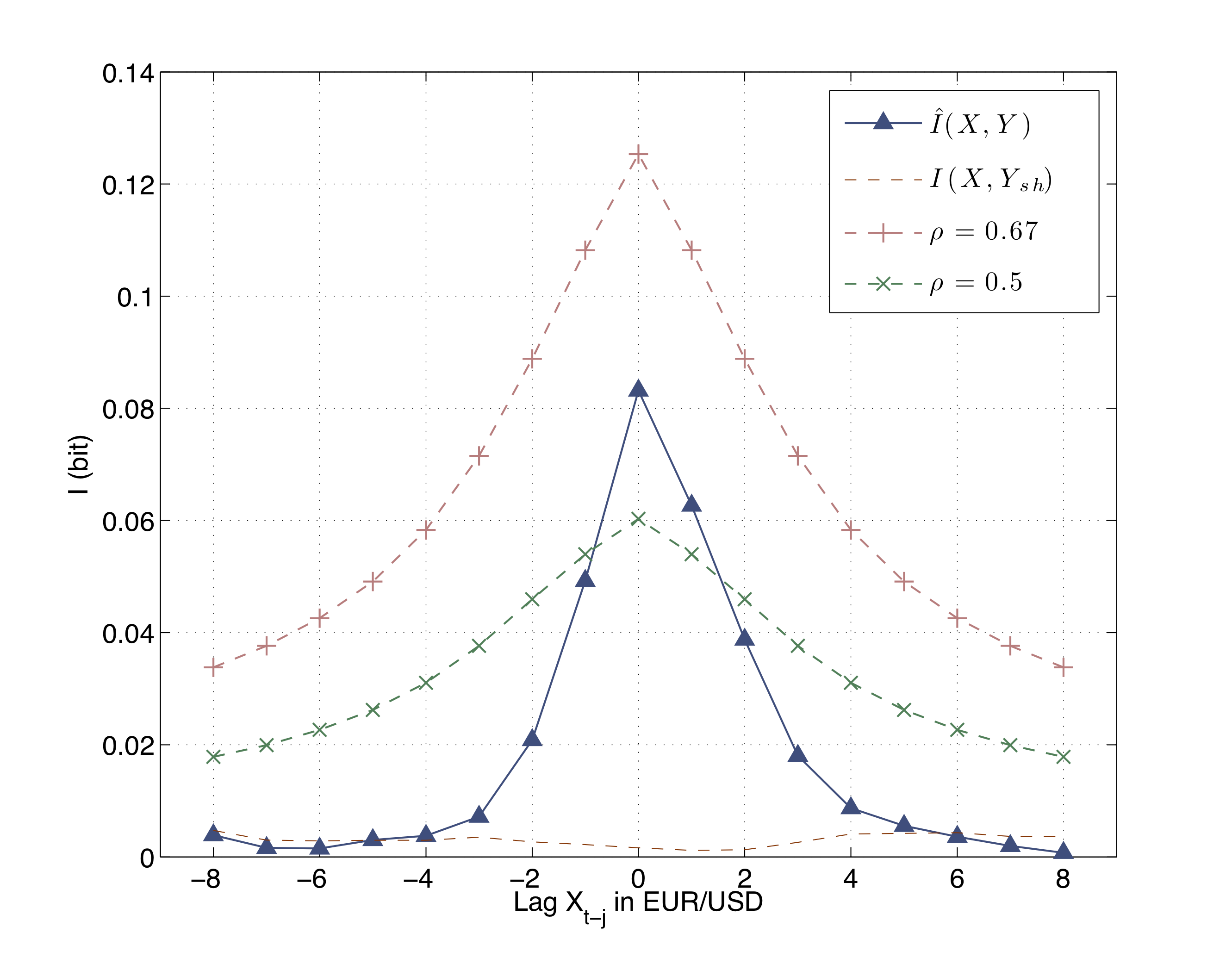}
\end{minipage}
\hspace{0.1cm}
\begin{minipage}[b]{0.5\linewidth}
\centering
   \includegraphics[width=2.95in]{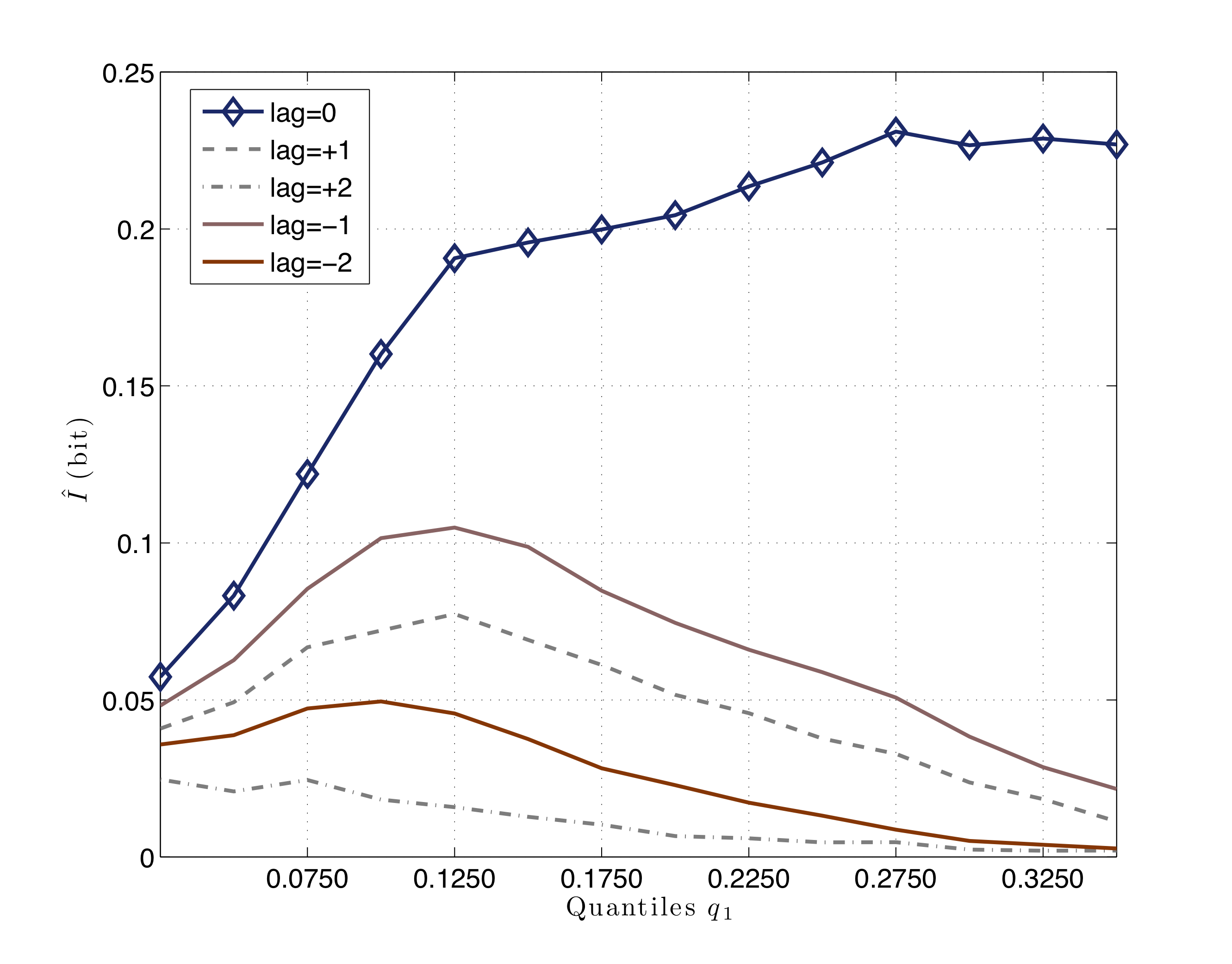}
\end{minipage}
\caption{Mutual Information (daily returns)}
\begin{quote}
The left figure shows $\hat{I}(\text{EUR/USD}_t,\text{GBP/USD}_{t + \tau})$ for lags $\tau = -8, \ldots +8$ and simulated Gaussian processes. On the right figure $\hat{I}(\text{EUR/USD}_{t + \tau},\text{GBP/USD}_t)$ for lags $\tau = -2, \ldots +2$ and various discretizations $q_1$ is shown. The specification is as in Table \ref{t_MIdailydraws}.
\end{quote}
\label{fig_miq0109}
\end{figure}

We estimated the mutual information in equation (\ref{eq_mutualinfo2}), with $\tau = -8, \ldots 8$ for daily and hourly exchange rate returns using the discretization scheme in (\ref{eq_discreteddmag}). With a sample size of around $3,000$ for the daily series and $70,000$ for the hourly series, the na\"{i}ve estimator in (\ref{eq_naiveestimator2}) with $M=3$ (for the three letters '0', '1', '2') is used to estimate the mutual information $I$. As a robustness check against the threshold choice ($q_1 = 0.05$, $q_2 =  1 - q_1 = 0.95$) for the discretization scheme, we have also calculated mutual information based  on other threshold choices. The results for the daily exchange rate returns are reported in Table \ref{t_MIdailydraws} for ($q_1= 0.045,0.050,0.055$) and plotted in Figure \ref{fig_miq0109}.
It is clear from Table \ref{t_MIdailydraws} and Figure \ref{fig_miq0109} that the mutual information is the highest at $\tau = 0$ and drops off sharply for $\tau \neq 0$. The mutual information for $\tau < 3$ is greater than $\tau > 0$ for the same $\vert \tau \vert$, indicating that the EUR/USD is leading with respect to GBP/USD.
\begin{table}[t]
\centering
{\scriptsize
\begin{tabular}{ c  c  c  c  c  c  c }
\hline
 & \multicolumn{3}{c}{$\hat{I}(X_t,Y_{t-\tau})$} & \multicolumn{3}{c}{$\hat{I}(X_t,Y_{t-\tau})$, $q=0.050$}\\ 
 &&&&&& \\
lag $\tau$ & q=0.045 & q=0.050 & q=0.055 & $\mu(\hat{I}(X,Y_{\text{shuffled}}) \pm \hat{\sigma}$ & $\mu(\hat{I}(X_B,Y_B)\vert_{\rho = 0.67} \pm \hat{\sigma}$ & $\mu(\hat{I}(X_B,Y_B)\vert_{\rho = 0.5} \pm \hat{\sigma}$ \\
 \hline
  &&&&&& \\
+4 & 0.00842 & 0.00871 & 0.00973 & 0.00289 $\pm$ 0.00203 & 0.05831 $\pm$ 0.04773 & 0.03105 $\pm$ 0.02221 \\
+3 & 0.01703 & 0.01806 & 0.02037 & 0.00288 $\pm$ 0.00203 & 0.07152 $\pm$ 0.04389 & 0.03766 $\pm$ 0.01956 \\
+2 & 0.03545 & 0.03876 & 0.04383 & 0.00289 $\pm$ 0.00203 & 0.08884 $\pm$ 0.03524 & 0.04595 $\pm$ 0.01427 \\
+1 & 0.05595 & 0.06266 & 0.07181 & 0.00286 $\pm$ 0.00198 & 0.10822 $\pm$ 0.02336 & 0.05398 $\pm$ 0.00901 \\
0 & 0.07187 & \textbf{0.0832} & 0.09617 & 0.00283 $\pm$ 0.00194 & 0.12536 $\pm$ 0.01992 & 0.06029 $\pm$ 0.00897 \\
-1 & 0.04303 & 0.04926 & 0.05633 & 0.00275 $\pm$ 0.00192 & 0.10822 $\pm$ 0.02336 & 0.05398 $\pm$ 0.00901 \\
-2 & 0.01879 & 0.02087 & 0.02278 & 0.00273 $\pm$ 0.00191 & 0.08884 $\pm$ 0.03524 & 0.04595 $\pm$ 0.01427 \\
-3 & 0.00657 & 0.00718 & 0.00771 & 0.00273 $\pm$ 0.00189 & 0.07152 $\pm$ 0.04389 & 0.03766 $\pm$ 0.01956 \\
-4 & 0.00415 & 0.00377 & 0.00398 & 0.00272 $\pm$ 0.00188 & 0.05831 $\pm$ 0.04773 & 0.03105 $\pm$ 0.02221 \\
 \hline
\end{tabular}}
\caption{Mutual Information (daily returns)}
\begin{quote}
In the first three columns, the estimated mutual information $\hat{I}(\text{EUR/USD}_t,\text{GBP/USD}_{t + \tau})$ on the discretized time series of the daily return series is reported. The discretization scheme (\ref{eq_discreteddmag}) has been used with quantile $q=0.050$ and as a robustness check with quantiles $q=0.045,0.055$. In the last three columns, the results of numerical simulations of three processes is shown. $\mu(\hat{I}(X,Y_{\text{shuffled}})$ shows the value, where one discretized time series is shuffled, destroying any correlation or cross correlation information. In columns $\mu(\hat{I}(X_B,Y_B)\vert_{\rho=0.67, 0.5}$, the mutual information of correlated processes, with correation $\rho=0.67, 0.5$ and using the discretization scheme (\ref{eq_discreteddmag}) is calculated.
\end{quote}
\label{t_MIdailydraws}
\end{table}

To understand how sensitive the mutual information estimates are as regards to the particular choice of partition, various quantiles were chosen for which the mutual information was estimated. In columns two and four in Table \ref{t_MIdailydraws}, we report the estimated mutual information for the quantile sets ($q_1=0.045$, $q_2=0.955$) and ($q_1=0.055$, $q_2=0.945$). In the right part of Figure \ref{fig_miq0109} we report the mutual information estimates for the lags $\tau = -2, -1, 0, +1, +2$ and quantiles ranging from $q_1=0.025$ to $q_1=0.35$, which approximates to the equiprobable case. The curves representing the mutual information estimates are continuous, with no spikes at particular quantiles.  This gives some assurance that the choice of quantile does not alter the conclusion substantially. It is also clear from the right part of Figure  \ref{fig_miq0109} that mutual information is the highest for lag $\tau = 0$ and substantially lower for all other lags. The increase in information is higher between $q_1=0.055$ and $q_1= 0.125$ than for other quantile values.

Next, we compare the empirical estimates with simulated results produced by two correlated Gaussian processes.
For each simulated time series, draw states were calculated based on the discretization scheme (\ref{eq_discreteddmag}) and the na\"{i}ve estimator was applied. The correlation factor used in one simulation set is the estimated unconditional correlation of the two return series EUR/USD and GBP/USD which is $\hat{\rho} = 0.673$ (see Table \ref{T:basicstats}). This allows us to quantitatively describe the exclusive influence of correlation on the draw dependence in the time series under the simplifying assumption of Gaussian returns.
The results are depicted in Figure \ref{fig_miq0109} and in the sixth column of Table \ref{t_MIdailydraws}.  The mutual information for the simulated draw states is consistently and markedly higher than that for the empirical series at all lags. The lower mutual information of the pair of empirical draw state series indicates that large draws in the empirical series show a higher degree of independence than what would be expected from a pair of correlated Gaussian processes with the same correlation coefficient. The difference between simulated draws and empirical draws is expected to be even larger if further cross correlation relationships are considered in the simulations.

A further pair of correlated Gaussian processes were simulated with a lower correlation of $\hat{\rho} = 0.5$. The specific factor has been set after experimenting with different correlations, with the aim of having one lower bound on the mutual information at lag $\tau = 0$.
The results of this simulation are also shown in Figure \ref{fig_miq0109}, and the seventh column of Table \ref{t_MIdailydraws}. At lag $0$, the mutual information is lower than that for the empirical series. Also, in this set-up, the mutual information drops much slower for lags $\tau \neq 0$ than the case for the empirical series. 
At lags $\tau \in \{0,1\}$, the mutual information of the simulated pair is higher than the corresponding mutual information of the empirical series. 

To quantify the error when estimating entropy (mutual information), we re-estimate the mutual information of the two given time series, but with one underlying return series being shuffled. 
The result of the shuffle is to strip off any correlation and cross-correlation between the series. Furthermore, any auto-correlation information is also destroyed by the shuffling, leading to a different draw distribution. 
This is done iteratively, generating a sample of shuffled series. The discretization procedure is applied to all shuffled series, and the mutual information is calculated
 and presented in the figure as $I(X_{\text{shuffled}},Y)$. The mean and standard error of the shuffled sample are calculated. The mean is shown in the left diagram in Figure \ref{fig_miq0109} along with the estimated mutual information of the empirical time series. The numerical values for lags $\tau = -4, \ldots +4$ are reported in Table \ref{t_MIdailydraws}. From Figure \ref{fig_miq0109}, one can observe that the mutual information of the shuffled series is very low. Only for lag $\vert \tau \vert \geq 6$ does the mutual information $\hat{I}$ of the empirical series drop below the level of mutual information contained in shuffled series. 

We repeat the procedure above for the hourly exchange rates and report the results in Table \ref{t_EURUSDMIhourly}.
The findings for the hourly relationship are similar to that reported above for the daily returns, in that mutual information, $I$, peaked at $\tau = 0$, and dropped off for $\tau \neq 0$. The $I$ at $\tau < 0$ is marginally higher than the equivalent $\vert \tau \vert$ for $\tau > 0$.
\begin{table}[t]
\centering
{\scriptsize
\begin{tabular}{ c  c  c  c  c  }
\hline
 & \multicolumn{3}{c}{$\hat{I}(X_t,Y_{t-\tau})$} & $\hat{I}(X_t,Y_{t-\tau})$, $q=0.050$\\ 
  &&&& \\

lag $\tau$ & $q_1=0.045$ & $q_1=0.050$ & $q_1=0.055$ & $\mu(\hat{I}(X,Y_{\text{shuffled}}) \pm \hat{\sigma}$ $[10^{-3}]$ \\
\hline
+5 & 0.0054 & 0.00535 & 0.00538 & 0.11361 $\pm$ 0.081 \\
+4 & 0.01163 & 0.01157 & 0.01131 & 0.1123 $\pm$ 0.079 \\
+3 & 0.02479 & 0.02441 & 0.02409 & 0.11351 $\pm$ 0.08 \\
+2 & 0.05141 & 0.05184 & 0.05274 & 0.11517 $\pm$ 0.081 \\
+1 & 0.09124 & 0.09409 & 0.09865 & 0.11619 $\pm$ 0.083 \\
0 & 0.13581 & 0.14376 & 0.15415 & 0.11131 $\pm$ 0.079 \\
-1 & 0.08763 & 0.09058 & 0.09585 & 0.1114 $\pm$ 0.078 \\
-2 & 0.04873 & 0.04992 & 0.05228 & 0.11103 $\pm$ 0.077 \\
-3 & 0.0234 & 0.02408 & 0.02529 & 0.1116 $\pm$ 0.077 \\
-4 & 0.01134 & 0.01182 & 0.01255 & 0.11147 $\pm$ 0.079 \\
-5 & 0.00598 & 0.00633 & 0.00663 & 0.11182 $\pm$ 0.081 \\
\hline
\end{tabular}}
\caption{Mutual Information of hourly returns}
\begin{quote}
In the first three columns, the estimated mutual information $\hat{I}(\text{EUR/USD}_t,\text{GBP/USD}_{t + \tau})$ on the discretized hourly return series is reported. The discretization scheme (\ref{eq_discreteddmag}) has been used with quantile $q=0.050$ and as a robustness check with quantiles $q=0.045,0.055$. $\mu(\hat{I}(X,Y_{\text{shuffled}})$ shows the value, where one discretized time series is shuffled, destroying any correlation or cross correlation relations. 
\end{quote}
\label{t_EURUSDMIhourly}

\end{table}
As in the case of the daily returns, we test the sensitivity of our results to the quantile choice.  
Table \ref{t_EURUSDMIhourly} shows the results for the quantiles $q_1 = 0.045, 0.050, 0.055$ for different lags $\tau = -5, \ldots +5$. The estimated $\hat{I}$ for the hourly series contains higher levels of information than that for the daily draw state series at $\tau = 0$ and other short $\vert \tau \vert$. So there is a stronger dependency among the hourly returns at contemporaneous term and at short lags.

\FloatBarrier
 \subsection{Information Flows between Draws in EUR/USD and GBP/USD}
 \label{subsec_infoflow}
Transfer entropy can be expressed as the difference in two block entropies, as shown in equation (\ref{eq_transferconditional2})). 
Here, we use the three letter partition defined in equation (\ref{eq_discreteddmag}) and the Grassberger estimator $\hat{H}_{\psi}$ in equation (\ref{eq_grassberger}) for the estimation of transfer entropy. 

Since the transfer entropy is not zero for small sample, we follow \citeN{Marschinski:2002dq}  to estimate the mean and standard error by the bootstrapping method, involving reshuffling the predictor series, and destroying all its time series pattern. This process is repeated enough times, after which a mean, $\mu({\hat{T_{sh}}})$, and a standard deviation, $\sigma$, is calculated, and the adjusted \textit{Effective Transfer Entropy} (ET) is calculated as:

\begin{equation} \label{eq_ete_paper1}
ET_{Y \rightarrow X} (m,l) \equiv T_{Y \rightarrow X} (m,l) - \mu({\hat{T_{sh}}}) 
\end{equation}

\begin{figure}[ht]
\begin{minipage}[b]{0.5\linewidth}
\centering
\includegraphics[width=2.95in]{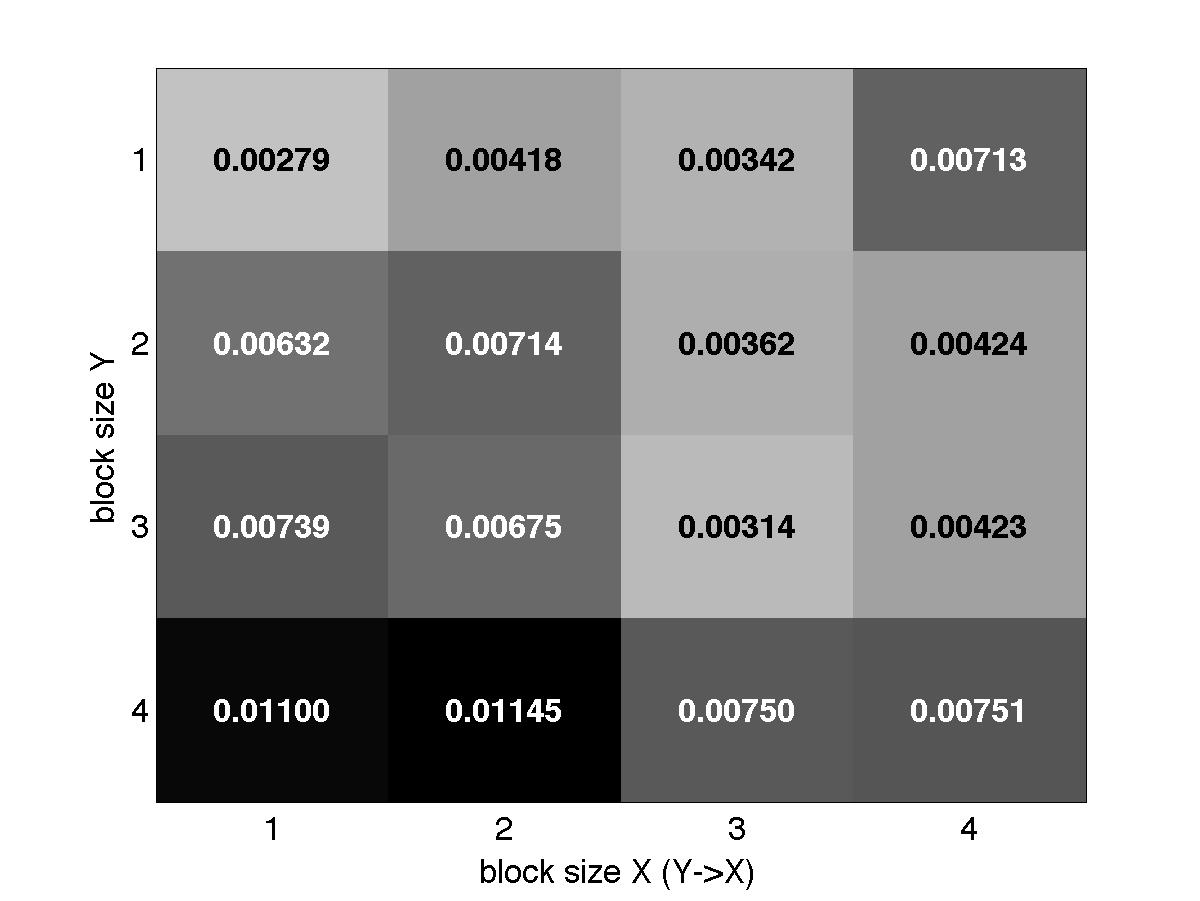}
\end{minipage}
\hspace{0.1cm}
\begin{minipage}[b]{0.5\linewidth}
\centering
   \includegraphics[width=2.95in]{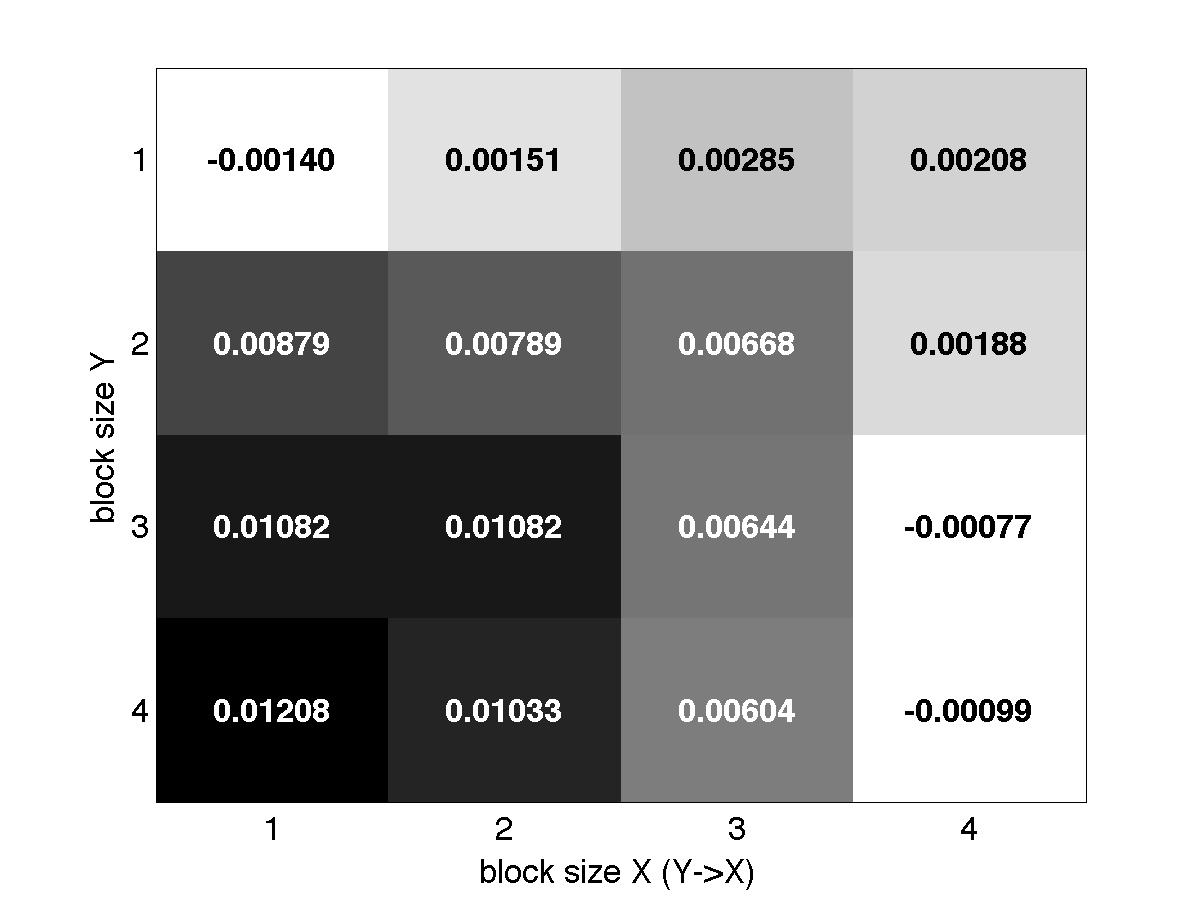}
\end{minipage}
\caption{$\hat{ET}_{\text{EUR} \rightarrow \text{GBP}}$ and $\hat{ET}_{\text{GBP} \rightarrow \text{EUR}}$ (daily returns)}
\begin{quote}
On the discretized series of daily prices, effective transfer entropy, as specified in equation (\ref{eq_ete_paper1}), is shown for block lengths of one to four days' history. The left figure shows the estimated effective transfer entropy $\hat{ET}_{\text{EUR/USD} \rightarrow \text{GBP/USD}} (m,l)$. In the right figure, the information flow, as measured by the effective transfer entropy in the other direction, GBP/USD to EUR/USD, is shown. 
\end{quote}
\label{fig_eteurusd}
\end{figure}

For $ET_{Y \rightarrow X} (m,l)$, $m$ refers to the length of own past history in $X$, and $l$ to the length of past history of $Y$, used in the prediction of $X$. The findings of $ET$ for $m=1, \ldots 4$ and $l=1, \ldots 4$ are presented in Figure \ref{fig_eteurusd}.
The detailed  statistics for the more significant cases (i.e. represented by the darkest block) are presented in Tables \ref{t_TEUSDGBPUSDd} and \ref{t_TGBPUSDEURUSDd}. From Figure \ref{fig_eteurusd}, we note that for a fixed $m$, the $ET$ generally increases as $l$ increases. On the other hand, for the same $l$, $ET$ may decrease as $m$ increases if $X$ can be forecast more efficiently using its own past. 

For the results presented in Figure \ref{fig_eteurusd} and Tables \ref{t_TEUSDGBPUSDd} and \ref{t_TGBPUSDEURUSDd}, we conclude that the maximum $\hat{ET}_{\text{EUR/USD} \rightarrow \text{GBP/USD}}$, is at $m=2$ and $l=4$, whereas for $\hat{ET}_{\text{GBP/USD} \rightarrow \text{EUR/USD}}$, $m=1$ and $l=4$. From the amount of $\hat{ET}$ for the two exchange rates, there is slightly more information flow from GBP/USD to EUR/USD than the other way around.
Note that the negative values of effective transfer entropy in Figure \ref{fig_eteurusd} are due to the fact that the estimated transfer entropy resulting from the shuffling is higher than that of the original time series. More noise is introduced as one increases $m$ and $l$.
When comparing the $ET$ estimates with the mutual information, $\hat{I}$, reported in the previous section, we notice that the amount of information gain is much lower. 

\begin{table}[t]
\centering
{\scriptsize
\begin{tabular}{ c  c  c  c  c  c  c }
\hline
 \multicolumn{2}{c}{History} & \multicolumn{5}{c}{Transfer Entropy} \\ 
 &&&&&& \\
m & l & $\hat{T}_{\text{naive}}$ & $\hat{T}$ & $\mu({\hat{T_{sh}}}) \pm \sigma$  & $\hat{ET}$ & REA \\
$\text{GBP/USD} $ & $\text{EUR/USD}$ & [$10^{-3}$]& [$10^{-3}$]&  [$10^{-3}$] & [$10^{-3}$]  &  [\%]  \\ 
\hline
1 & 4 & 40.774 & 26.176 & 15.18 $ \pm $ 3.973 & 10.996 & 2.291 $\pm$ 0.828  \\
2 & 4 & 45.569 & 31.391 & 19.941 $ \pm $ 4.118 & 11.451 & 2.518 $\pm$ 0.906  \\
3 & 4 & 54.445 & 39.991 & 32.488 $ \pm $ 5.116 & 7.504 & 1.666 $\pm$ 1.136  \\
4 & 4 & 68.163 & 56.329 & 48.815 $ \pm $ 6.376 & 7.515 & 1.683 $\pm$ 1.428  \\
\hline
\end{tabular}}
\caption{Transfer entropy $\text{EUR/USD} \rightarrow \text{GBP/USD}$}
\begin{quote}
The information flow between draws from EUR/USD to GBP/USD is estimated using daily prices.These are calculated for block lengths of up to four days history.
Transfer Entropy is estimated with the na\"{i}ve estimator $\hat{T}_{\text{naive}}$, and the Grassberger estimator $\hat{T}$ are reported. With the predictor series EUR/USD reshuffled repeatedly, $\mu({\hat{T_{sh}}}) \pm \sigma$ reports the bias due to small sample effects. Effective Transfer Entropy $\hat{ET}$ and the relative explanation added (REA) is reported in the last columns.
\end{quote}
\label{t_TEUSDGBPUSDd}
\end{table}

\begin{table}[t]
\centering
{\scriptsize
\begin{tabular}{ c  c  c  c  c  c }
\hline
 \multicolumn{2}{c}{History} & \multicolumn{4}{c}{Transfer Entropy} \\ 
 &&&&& \\
m & l & $\hat{T}$ (bit) & $\mu({\hat{T_{sh}}}) \pm \sigma$ (bit) & $\hat{ET}$ (bit) & REA [\%]  \\
$\text{EUR/USD} $ & $\text{GBP/USD}$ & [$10^{-3}$]&  [$10^{-3}$] & [$10^{-3}$]  & \\ 
\hline
1 & 3 & 18.701 & 7.88 $\pm$ 3.019 & 10.821 & 2.158 $\pm $ 0.602  \\
1 & 4 & 27.073 & 14.992 $\pm$ 3.972 & \textbf{12.081} & \textbf{2.409} $\pm $ 0.792  \\
2 & 1 & 2.963 & 1.458 $\pm$ 1.369 & 1.506 & 0.315 $\pm $ 0.287  \\
2 & 2 & 13.611 & 5.721 $\pm$ 2.527 & 7.891 & 1.649 $\pm $ 0.529  \\
2 & 3 & 22.188 & 11.363 $\pm$ 3.34 & 10.825 & 2.262 $\pm $ 0.698  \\
2 & 4 & 29.685 & 19.356 $\pm$ 4.277 & 10.329 & 2.159 $\pm $ 0.894  \\
\hline
\end{tabular}}
\caption{Transfer entropy $\text{GBP/USD} \rightarrow \text{EUR/USD}$}
\begin{quote}
The information flow between draws from GBP/USD to EUR/USD is estimated using daily returns. These are calculated for block lengths of up to four days' history.
$\hat{T}$ is the Grassberger estimator for Transfer Entropy. With the predictor series EUR/USD reshuffled repeatedly, $\mu({\hat{T_{sh}}}) \pm \sigma$ reports the bias due to small sample effects. Effective Transfer Entropy $\hat{ET}$ and the relative explanation added (REA) is reported in the last columns.
\end{quote}
\label{t_TGBPUSDEURUSDd}
\end{table}

To understand the relative information gain due to the transfer entropy we follow \citeN{Marschinski:2002dq} by relating this information gain to the entropy of the series and produce the REA (\textit{relative explanation added}) measure below:
\begin{align}
REA (m,l) & \equiv \frac{ET_{Y \rightarrow X} (m,l)}{H(X_{0}\vert X_{-1} , \ldots X_{-(m)}  ) } \\
 &=  \frac{H(X_{0}\vert X_{-1} , \ldots X_{-m} , Y_{-1} , \ldots Y_{-l}) }{H(X_{0}\vert X_{-1} , \ldots X_{-m}  ) } - 1 \notag
\end{align}
Here the information flow from $Y$ to $X$, measured by the transfer entropy, is related to the total flow of information measured in $X$ alone, based on the conditional block entropy of $X$. Put differently, it measures the additional information gain by observing the history of $X$ and $Y$ when already observing $X$. 
The results in Tables \ref{t_TEUSDGBPUSDd} and \ref{t_TGBPUSDEURUSDd} suggest that the amount of information gain from the predictive exchange rate is about 2.5\% for both exchange rates. 
For a comparison, the REA for using DAX to predict the Dow Jones Industrial, as estimated by \citeN{Marschinski:2002dq}, using the one-minute tick data, is only 0.4\% for a three symbol discretization. 

Finally, we can conclude that the maximum information gain from the given draw states of a series is to use one or two days of own history ($m=1,2$) and up to four days of the predictor's history ($l=4$). Using longer own history beyond two days may contain more information but also induces more noise. 

\begin{figure}[ht]
\centering
\includegraphics[width=4.5in]{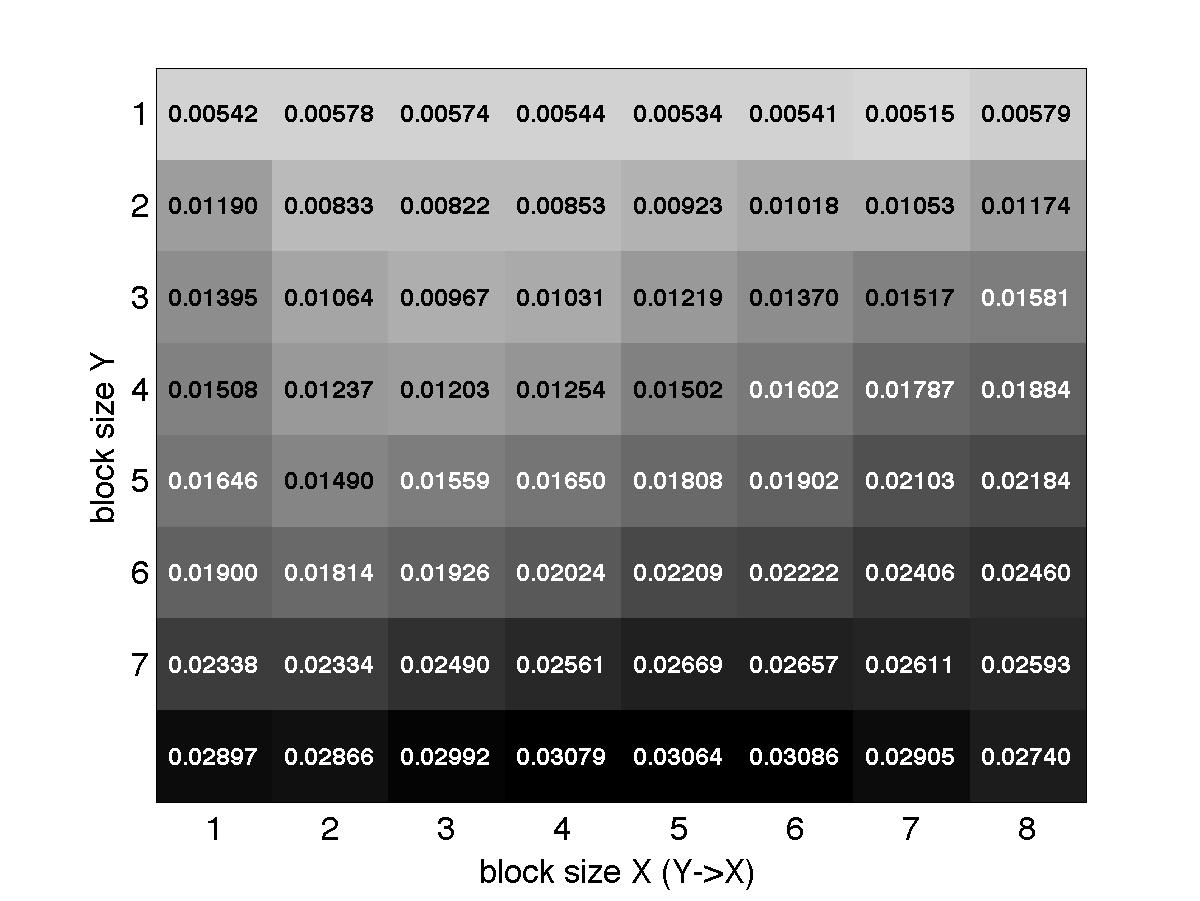}
      \caption{Draws Information Flow from hourly EUR/USD to GBP/USD \label{fig_eteurusdH}}
 \begin{quote}
Effective transfer entropy as specified in equation (\ref{eq_ete_paper1}) is shown for  block lengths of one to eight hours history. 
The figure shows the estimated effective transfer entropy $\hat{ET}_{\text{EUR/USD} \rightarrow \text{GBP/USD}} (m,l)$. 
\end{quote}
\end{figure}

\begin{figure}[ht]
\centering
\includegraphics[width=4.5in]{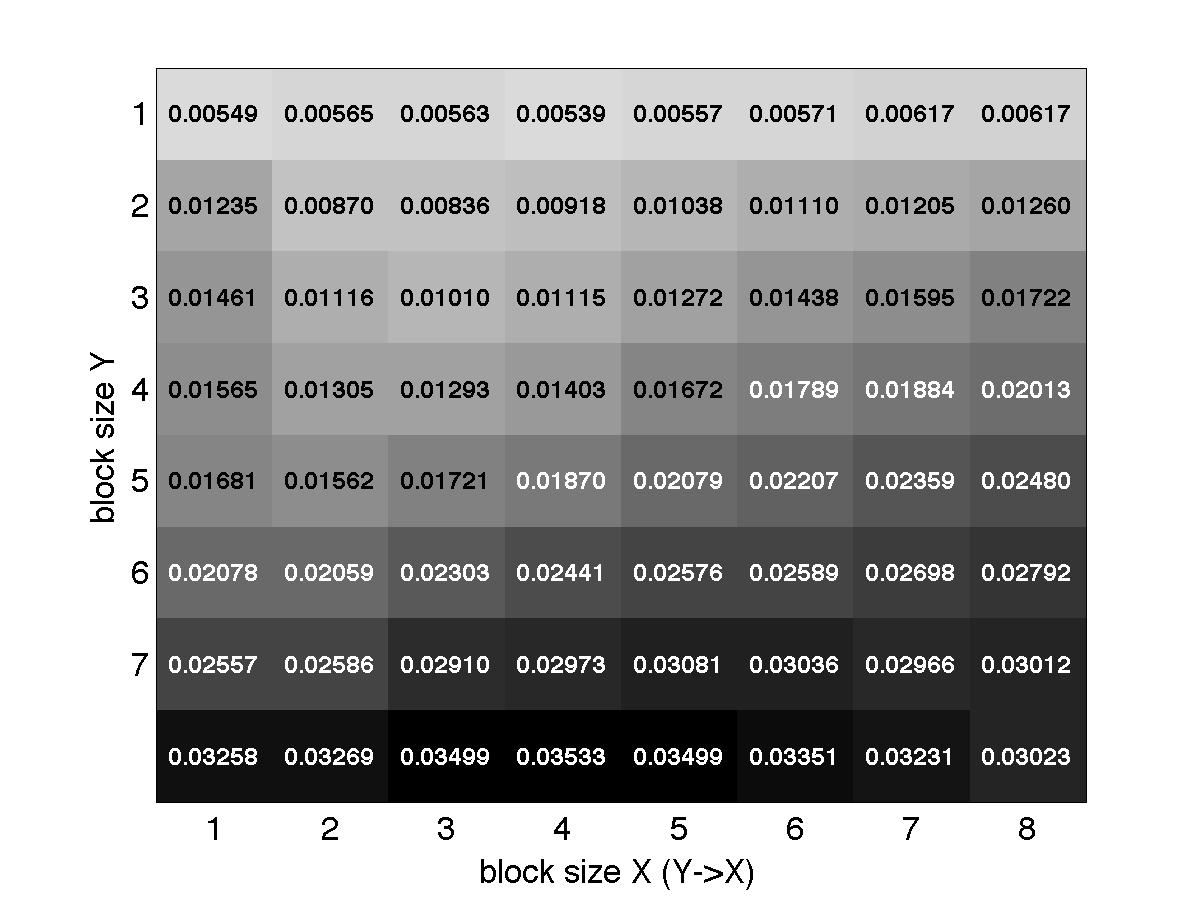}
      \caption{Draws Information Flow from hourly GBP/USD to EUR/USD \label{fig_etgbpusdH}}
\begin{quote}
Effective transfer entropy as specified in equation (\ref{eq_ete_paper1}) is shown for  block lengths of one to eight hours history. 
The figure shows the estimated effective transfer entropy $\hat{ET}_{\text{GBP/USD} \rightarrow \text{EUR/USD}} (m,l)$. 
\end{quote}
\end{figure}

We now repeat the estimation using the hourly returns of EUR/USD and GBP/USD. The $\hat{ET}$ estimated for $m=1,\ldots , 8$ and $l=1, \ldots 8$ are presented in Figures \ref{fig_eteurusdH} and \ref{fig_etgbpusdH}. The levels of $\hat{ET}$ are higher than those for the daily returns. The statistics for the more significant results are summarised in Table \ref{t_TEEURGBPh} and Table \ref{t_TEGBPEURh}.

\begin{table}[t]
\centering
{\scriptsize
\begin{tabular}{ c  c  c  c  c  c }
\hline
 \multicolumn{2}{c}{History} & \multicolumn{4}{c}{Transfer Entropy} \\ 
 &&&&& \\
m & l & $\hat{T}$ (bit) & $\mu({\hat{T_{sh}}}) \pm \sigma$ (bit) & $\hat{ET}$ (bit) & REA \\
$\text{GBP/USD} $ & $\text{EUR/USD}$ &[$10^{-3}$]  &[$10^{-3}$]  & [$10^{-3}$] & [\%]  \\ 
\hline
4 & 8 & 66.219 & 35.427 $ \pm $ 1.253 & 30.792 & 6.849 $\pm$ 0.279  \\
5 & 8 & 77.439 & 46.802 $ \pm $ 1.467 & 30.637 & 6.858 $\pm$ 0.329  \\
6 & 8 & 88.898 & 58.035 $ \pm $ 1.612 & \textbf{30.863} & \textbf{6.963} $\pm$ 0.364  \\
7 & 8 & 96.159 & 67.108 $ \pm $ 1.764 & 29.052 & 6.631 $\pm$ 0.403  \\
8 & 7 & 88.994 & 63.064 $ \pm $ 1.672 & 25.931 & 6.015 $\pm$ 0.388  \\
8 & 8 & 100.936 & 73.534 $ \pm $ 1.806 & 27.402 & 6.356 $\pm$ 0.419  \\
\hline
\end{tabular}}
\caption{Transfer entropy $\text{EUR/USD} \rightarrow \text{GBP/USD}$ (hourly prices)}
\begin{quote}
The information flow between draws from EUR/USD to GBP/USD is estimated using hourly prices.These are calculated for block lengths of up to eight hours' history.
$\hat{T}$ is the Grassberger estimator for Transfer Entropy. With the predictor series EUR/USD reshuffled repeatedly, $\mu({\hat{T_{sh}}}) \pm \sigma$ reports the bias due to small sample effects. Effective Transfer Entropy $\hat{ET}$ and the relative explanation added (REA) is reported in the last columns.
\end{quote}

\label{t_TEEURGBPh}
\end{table}

\begin{table}[t]
\centering
{\scriptsize
\begin{tabular}{ c  c  c  c  c  c }
\hline
 \multicolumn{2}{c}{History} & \multicolumn{4}{c}{Transfer Entropy} \\ 
 &&&&& \\
m & l & $\hat{T}$ (bit) & $\mu({\hat{T_{sh}}}) \pm \sigma$ (bit) & $\hat{ET}$ (bit) & REA \\
$\text{EUR/USD} $ & $\text{GBP/USD}$ &[$10^{-3}$]  &[$10^{-3}$]  & [$10^{-3}$] & [\%]  \\ 
\hline
4 & 8 & 69.081 & 33.756 $\pm$ 1.18 & 35.326 & 7.828 $ \pm$ 0.25  \\	
5 & 8 & 79.416 & 44.428 $\pm$ 1.358 & 34.988 & 7.782 $ \pm$ 0.288  \\	
6 & 8 & 88.707 & 55.197 $\pm$ 1.54 & 33.51 & 7.48 $ \pm$ 0.327  \\	
7 & 8 & 97.141 & 64.83 $\pm$ 1.677 & 32.311 & 7.279 $ \pm$ 0.356  \\	
8 & 8 & 101.737 & 71.508 $\pm$ 1.738 & 30.229 & 6.912 $ \pm$ 0.369  \\	
\hline
\end{tabular}}
\caption{Transfer entropy $\text{GBP/USD} \rightarrow \text{EUR/USD}$ (hourly prices)}
\begin{quote}
The information flow between draws from GBP/USD to EUR/USD is estimated using hourly prices.These are calculated for block lengths of up to eight hours history. 
$\hat{T}$ is the Grassberger estimator for Transfer Entropy. With the predictor series GBP/USD reshuffled repeatedly, $\mu({\hat{T_{sh}}}) \pm \sigma$ reports the bias due to the small sample effect. Effective Transfer Entropy $\hat{ET}$ and the relative explanation added (REA) is reported in the last columns.
\end{quote}

\label{t_TEGBPEURh}
\end{table}
The maximum $ET$  $0.0353$ bits  for the hourly pair is reached in the direction from GBP/USD, considering a history of four hours of EUR/USD and eight hours of GBP/USD, with $REA = 7.828 \pm 0.25$. So marginally more information flows from GBP/USD to EUR/USD than the other way around.
It is clear from the results that the greatest information gain is when $l=8$, i.e. up to eight hours of the past history of the predictive
variable are used. The amount of own history needed is less clear cut and the difference in $\hat{ET}$ is small. Based on the  maximum $\hat{ET}$, optimal $m=6$ for GBP/USD and $m=4$ for EUR/USD.

Based on $REA$ the predictive variable contribute another 7\% to 8\% of forecasting power compared to using only own past history. Given the importance of the EUR/USD pair, its  liquidity and volume, the result of a marginally higher information flow from GBP/USD to EUR/USD than the other way around, seems a little surprising. 
However, we should note that the difference in information for every pair of history blocks $(m,l)$  is very small for the daily return series and is within one standard error $\hat{\sigma}(T_{sh})$ (see the fifth column in Tables \ref{t_TEUSDGBPUSDd} and \ref{t_TGBPUSDEURUSDd}). Moreover, the draws considered are large in magnitude (0.05, 0.95 quantiles of the respective drawdown and drawup distributions), so they belong to a part of the return/drawstate distribution where dependency between the two series could be different from that for the main part of the distribution, hence the result may not necessary match common observations.
An impact analysis by \citeN{Omrane:2009wh} suggests that in the very short term (in hours), positive macroeconomic news announcements in the US increase the GBP/USD volatility more strongly than that of EUR/USD. Moreover, during the crisis of 2007, in the foreign exchange markets, \citeN{Melvin:2009db} found evidence that the volatility of bid-ask spreads for spot trades increased for all major currency pairs, and the increase in the GBP/USD pair was the most dramatic ($+5,000 \%$ increase in average bid-ask spread).

We saw that transfer entropy is able to detect information flows between the draw state of the FX series. The level of information flow is three times greater for the hourly data than for the daily data.
Using relative explanation added (REA), relating the information gained to the entropy of the block history, we were able to show that this information gain is considerable, especially when comparing the results to findings reported elsewhere in the literature. 
\FloatBarrier

\section{Conclusion}
\label{sec_conclusion}
Using entropy measures, this chapter investigates the dependence and spillover relationship between the drawups and drawdowns of EUR/USD and GBP/USD. For the daily series, we found evidence of dependence among the largest draws (i.e. 5\% and 95\% quantiles), but it is not as strong as that expected from the correlated Gaussian series at the same level of correlation. There is a dependence between lead or lagged values of these draws. Similar findings were obtained from the hourly data, although the dependence is much stronger in the hourly data. Next we used transfer entropy to examine the spillover and lead-lag information flow between the drawup/drawdown of the two exchange rates. Such information flow is indeed detectable in both daily and hourly data. The amount of information transferred is considerably higher for the hourly than the daily data. The data demands in estimating the transfer entropy measure are considerable. Given the limited number of observations in the daily series, the lead-lag terms considered was restricted to four by four (days). For the hourly data we were able to extend our analysis to eight by eight (hours). Both daily and hourly series show clear evidence of information flowing from EUR/USD to GBP/USD and in the reverse direction. Robustness tests, using effective transfer entropy $ET$, show that the information measurable is not due to noise. By using the quantity 'relative explanation added' (REA), which relates the information gain to the entropy of the process and comparing our results to documented cases in the literature (see  \citeN{Marschinski:2002dq}, \citeN{Dimpfl:2011tp},  \citeN{Kwon:2008vl}), we conclude that there is a measurable information transfer between the two exchange rates that is potentially useful for forecasting and risk management. Many questions remain open, and these will be addressed in future analyses. For the moment, we do not have a model for the data generating process to understand how strong the draw state dependence between two time series is. The other immediate improvement planned is to employ the bootstrapping method from \citeN{Horowitz:2003hu} in order to make inferences. One such solution is a standard block bootstrap. However, as has been outlined by \citeN{Lahiri:1999wq}, standard methods such as non-overlapping-block bootstraps, moving-block bootstraps or bootstraps with random block length produce biased estimates. \citeN{Horowitz:2003hu} introduced a bootstrapping method based on the transition probabilities of the underlying Markov process. In this the process, $Y$ is simulated based on the calculated transition probabilities, thereby destroying the dependencies between $Y$ and $X$, but retaining the dynamics of the series $Y$. Transfer entropy is then estimated again using the simulated time series. Repeating this procedure yields the distribution of the transfer entropy estimate under the null hypothesis of no information flow. The third extension which we plan to follow is the use of Renyi's entropy instead of Shannon's entropies in calculating transfer entropy. The Renyi entropy lends itself readily to the analysis of a tail event, as shown by \citeN{Jizba:2011dq}, and would allow for a more robust way to examine different parts of the draw tail distributions.

\chapter{Information Flows between FX Volatility Regimes}\label{chap_paper2}
\section*{\centering \large Abstract}

We use state space models of volatility to investigate volatility spillovers between exchange rates. Our use of entropy-related measures in the investigation of dependencies of two state space series is novel. A set of five daily exchange rates from emerging and developed economies against the dollar from 1999-2012 is used. 
We find that among the currency pairs, the co-movement of EUR/USD and CHF/USD volatility states show the strongest observed relationship. With the use of transfer entropy, we find evidence for information flows between the volatility state of AUD, CAD and BRL. We can further measure an information flow from EUR/USD to GBP/USD, which indicates a causal volatility spillover relationship.

\clearpage
\section{Introduction}
Relations between large drawdowns and drawups of two currency pairs were the focus of the previous chapter. Tumultuous periods, which in most cases are the market regimes generating large draws, are the topic of this study. Using hidden Markov models to identify volatile periods, we investigate volatility transmissions between foreign exchange rates 
applying the information theoretic tools that we used in the previous chapter.

History provides many examples of financial crises and how turbulence in financial markets spread across markets within a short time. Evidence for volatility transmissions has been found in many markets. \citeN{Edwards:2000ja} analysed weekly stock market data for a group of Latin American countries with a switching GARCH model (SWARCH). A univariate version of the SWARCH model is used to identify breakpoints in the ARCH model of conditional variance and highly volatile periods. The multivariate SWARCH is used to find evidence of volatility co-movements across countries. Volatility spillover effects have also been identified between different asset classes. \citeN{Baba:2008ba} found that during the second half of 2007, turmoil in the money markets spilled over to the FX swap markets and the cross-currency basis swap market. The mechanism underlying the spilling over was that at the time, the FX swap market was increasingly used by financial institutions to cover US dollar funding shortages.

One of the first studies on volatility spillover focussing on foreign exchange markets was \citeN{Engle:1990go}. Using the Japanese yen and US dollar exchange rate, they found evidence of intraday volatility spillover across markets. \citeN{Hong:2001ci} investigated the German mark and Japanese yen denominated in US dollars and found evidence of unidirectional volatility spillover from the German mark to the Japanese yen. On the other hand, \citeN{Baillie:1989uf} did not find significant evidence of volatility spillover between the daily nominal US dollar rates of British pound, German mark, Swiss franc and yen in 1980-1985.

From a methodological point of view, the extent to which the approaches identify spillover, co-movement or interdependence vary considerably. \citeN{Engle:1990go} employed a GARCH model in a vector autoregression to test if conditional variances were affected by the squared innovations (i.e. news, information) in other markets. \citeN{Cheung:1996fg} used the cross-correlation function between squared residuals, which were standardized by their individual conditional variance estimators, as a test for volatility spillover. Empirical studies \citeN{Poon:2003ww} have shown that GARCH-estimated volatilities appear to be too persistent following large shocks. In general, the standard single-state GARCH model does not provide for the different speeds of volatility adjustment in different states. \citeN{Hamilton:1989jg}  introduced a model in which changes in regimes are governed by a Markovian hidden state process. It gives more flexibility to the different adjustment speeds of volatility, and the persistence of the estimated volatility for the different states are very different from that estimated using the standard single-state GARCH. Later, 
\citeN{Hamilton:1994bq} adapted this model and introduced the SWARCH model, which they applied to identify volatility regimes in stock returns. SWARCH models has a Markov-modulated GARCH process, where the conditional variance is state dependent. \citeN{Engle:1990go} used this model in their analysis of volatility spillover. \citeN{Biaikowski:2005kr} used a bivariate Markov switching process, where each state corresponds to a bivariate normal distribution. They find evidence of feedback spillovers (based on the Granger causality test) between the Japanese and Hong Kong equity indices during the Asian crisis in 1997. 

Volatility spillover is a situation whereby a change in volatility in one market is related to a change in volatility in another market. When the change in volatility regime of two markets is contemporaneous, it is termed a co-movement, or interdependent when changes in two markets at different periods are related. In the literature, the term spillover is sometimes specifically reserved for causality in variance. When drawing conclusions from our analysis, we will highlight which of the specific situations we refer to.

In this study, we follow \citeN{Hamilton:1989jg} in identifying volatility regimes among a set of currencies quoted against the US dollar. We use squared currency returns as a proxy for daily volatility. A two-state hidden Markov model is defined, with which we model high and normal volatility states. Following the procedures in Chapter \ref{chap_paper1}, we use the entropy framework with mutual information and transfer entropy to evaluate various interdependency hypotheses. The estimated state process is a discrete Markov process, which lends itself to the analysis with information theoretic tools. We find evidence of various volatility regime relationships between the currency pairs in our sample. Among the European currencies (EUR/USD, GBP/USD, CHF/USD), there are volatility regime co-movement relationships that are measurable. CAD/USD and AUD/USD exhibit a volatility co-movement relationship and, to a lesser extent, interdependency with time lag. We also find evidence for information flows between the volatility states of currency pairs. Most notable is an information flow from EUR/USD to GBP/USD, which indicates a causal volatility spillover relationship, in line with the findings in \citeN{Inagaki:2006kj}, who reported evidence for such a unidirectional volatility spillover from EUR to GBP between 1999 and 2004. Our work showcases also the usefulness of the concepts of mutual information and transfer entropy when applying them to analyse volatility spillover relationships.

 \FloatBarrier
\section{Hidden Markov Models}
\subsection{Foundations}
\label{subsec_hmm}
A Hidden Markov Model (HMM) consists of a random process  $\{y_1, \ldots , y_n\}$, where the probability distribution depends on the realisation of a 'hidden' markov chain  $\{q_1, \ldots , q_n\}$ with finite state space $I = \{1, \ldots N\}$. The distribution of $\{y_t\}$ depends on the state $q_t$, so $P(y_t \vert q_t)$. A  Hidden Markov Model is denoted by $\theta = \{P(q_1), P(q_{t+1} \vert q_{t}), P(y_t\vert q_{t}) \}$ :
\begin{itemize}
\item  $P(q_1)$ is an initial state probability.
\item $a_{ij} \equiv P(q_{t+1}=i \vert q_{t} = j)$ for $1 \leq i,j \leq N$ are the transition probabilities between the states.
\item $P(y_t\vert q_{t})$ are the emission probabilities. 
\end{itemize}
There are no restrictions as to the choice of the emission probability distributions. In many  applications, normal distributions, a mixture of normal distributions and Poisson distributions have been used (\citeN{Zucchini:2009tn}).
\begin{figure}[ht]
\centering
\includegraphics[scale=0.25]{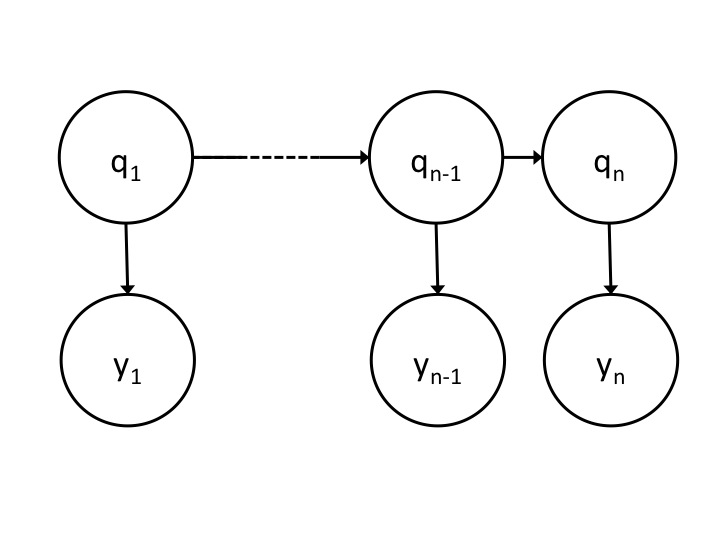}
\caption{A Hidden Markov Model, where $\{q_t\}$ represents the hidden states, $\{y_t\}$ the emissions and the arrows describe dependence relations.}
\label{fig:hmm}
\end{figure}
The set-up of a Hidden Markov model is depicted in Figure \ref{fig:hmm}. Each node in the figure represents the value of a random variable. A line connecting two nodes represents dependence and the absence of a connecting line implies conditional independence. 
The hidden state $q_{t+1}$ at time $t+1$ depends only on $q_t$ in the previous state as shown in equation (\ref{eq_statedyn}), and the emission at time $t+1$ depends only on $q_{t+1}$ at time $t+1$ as shown in equation (\ref{eq_statedep}):
\begin{eqnarray}
P(q_{t+1} \vert q_{1}, \ldots q_{t},  y_{1}, \ldots y_{t})  = P(q_{t+1} \vert q_{t}) \label{eq_statedyn}\\
P(y_{t+1} \vert q_{1}, \ldots q_{t},  y_{1}, \ldots y_{t}) = P(y_{t+1} \vert q_{t+1}) \label{eq_statedep}
\end{eqnarray}
Equation (\ref{eq_statedyn}) implies that the Markov process $\{q_1, \ldots , q_n\}$ is fully determined by the initial state probability $P(q_1)$ and the transition probabilities $P(q_t \vert q_{t-1})$. These relations allow for various conditional independence relations that are used in the estimation of HMM parameters. See \citeN{Barber:2012vq} for a more general discussion on the conditional independence relations in graphical models and the d-separation feature.

 \FloatBarrier
\subsection{HMM Estimation}
\label{sec_hmmest}
We will explain in this section two central methods for estimating the HMM models. The Expectation Maximisation Algorithm (EM), otherwise known as the Baum-Welch Algorithm, enables us to find the maximum likelihood estimates of the HMM model parameters when we cannot observe all the data (\citeN{Barber:2012vq}). There is also the Viterbi algorithm that provides the inference of the most likely state sequence of the hidden states, given the observations.

First of all, we note that there are two ways of indexing a time series, depending on the context. Starting at time $t_1$ and ending at time $t_2$, a sequence $\{y_t \}$ is denoted $y_{t_1}^{t_2} = y_{t_1}, y_{t_1 +1}  \ldots, y_{t_2} $. For a particular reference point in time with the observation denoted by $y_0$, we look back with $ y_{-n},\ldots , y_{-1}, y_0 $ and forward with $y_0, y_{1}, \ldots y_{n}$.

The forward-backward algorithm is used to calculate the conditional probability $P(q_{t}, y_1^T)$ of a hidden state at a specific time, given all the observations in the time series $\{ y_t \}_{1 \leq t \leq T}$. 
The forward algorithm is concerned with the \underline{joint} probability of $P(q_{t}, y_1^t)$ for all $t \in \{1, \ldots T \}$ and state space $I = \{1, \ldots , N\}$. In the backward algorithm the \underline{conditional} probability $P(y_{t+1}^T \vert q_t) $ is computed for all $t \in \{1, \ldots T \}$.
Using the backward and the forward algorithm together, the conditional probability $P(q_{t} \vert y_1^T)$ can be computed, assuming that the emission probabilities $P(y_t\vert q_{t})$, transition probabilities $P(q_{t+1} \vert q_{t})$ and initial distribution $P(q_1)$ are known. This follows from the independence relationship in equations (\ref{eq_statedyn}) and  (\ref{eq_statedep}) as shown below:
\begin{eqnarray}
P(q_{t} \vert y_1^T)  \propto  P(q_{t}, y_1^T) &=& P(q_{t}, y_1^t,y_{t+1}^T) \nonumber \\  
 & = & P(y_{t+1}^T \vert q_t,y_1^t) \cdot P(q_t,y_1^t) \nonumber \\ 
  & = & P(y_{t+1}^T \vert q_t) \cdot P(q_t,y_1^t) \nonumber \\
  & = & \beta_t (q_t) \cdot \alpha_t (q_t) \label{eq:forwardbackward} 
\end{eqnarray}
For the second last line in equation (\ref{eq:forwardbackward}) above, the first item $P(y_{t+1}^T \vert q_t)$ is calculated using the backward algorithm and the second item $P(q_t,y_1^t)$ is calculated using the forward algorithm. The terms of the last line in equation (\ref{eq:forwardbackward}) above are explained below.

The probabilities $ P(q_t,y_1^t)\equiv \alpha_t (q_t) $ for a given set of parameters  $\theta = \{P(q_1), P(q_{t+1} \vert q_{t}), P(y_t\vert q_{t}) \}$ obey a recursive relationship that can be calculated using the independence relations (\ref{eq_statedyn}) and (\ref{eq_statedep}) as follows:
\begin{eqnarray}
P(q_{t}, y_1^t)  &=& \sum_{q_{t-1} = 1}^N P(q_{t}, q_{t-1},y_1^t) \nonumber \\ \nonumber
&=& \sum_{q_{t-1} = 1}^N P(y_{t} \vert q_{t},q_{t-1},y_1^{t-1}) P(q_{t} \vert q_{t-1},y_1^{t-1}) P(q_{t-1},y_1^{t-1})\\ \nonumber
&=&  \sum_{q_{t-1} = 1}^N P(y_{t} \vert q_{t}) P(q_{t} \vert q_{t-1}) P(q_{t-1},y_1^{t-1})\\ \nonumber
&=& \sum_{q_{t-1} = 1}^N P(y_{t} \vert q_{t}) P(q_{t} \vert q_{t-1}) \cdot \alpha_{t-1}(q_{t-1}) \\
 &=& \alpha_t (q_t) \label{eq_forward}
\end{eqnarray}
For a given set of parameters $\theta = \{P(q_1), P(q_{t+1} \vert q_{t}), P(y_t\vert q_{t}) \}$,  the probabilities $P(y_{t+1}^T \vert q_t) \equiv \beta_{t}(q_{t})$ are similarly calculated based on a backward algorithm as follows:
\begin{eqnarray}
 P(y_{t+1}^T \vert q_t)  &=& \sum_{q_{t+1} = 1}^N P(y_{t+1}^T,q_{t+1} \vert q_t) \nonumber  \\ \nonumber
&=& \sum_{q_{t+1} = 1}^N P(y_{t+2}^T \vert q_{t+1},q_{t},y_{t+1}) P(y_{t+1} \vert q_{t+1},q_t) P(q_{t+1} \vert q_t)\\ \nonumber
&=& \sum_{q_{t+1} = 1}^N P(y_{t+2}^T \vert q_{t+1}) P(y_{t+1} \vert q_{t+1}) P(q_{t+1} \vert q_t)\\ \nonumber
&=& \sum_{q_{t+1} = 1}^N \beta_{t+1}(q_{t+1}) \cdot P(y_{t+1} \vert q_{t+1}) \cdot P(q_{t+1} \vert q_t)\\ 
& = & \beta_t (q_t) 
\label{eq_backward}
\end{eqnarray}

Both algorithms have a computational complexity of order $O (N^2 \cdot T)$. We explain this briefly for the backward algorithm. 
In equation (\ref{eq_backward}), $\beta_t (q_t)$ is a sum over all states $1, \ldots , N$ repeated at each point in time $ t \in \{1, \ldots , T\}$ and for each of the possible states at time $t$, giving $N^2 \cdot T$ calculations in total. In comparison, if $P(q_{t}, y_1^t)$ is summed over all combinations of state sequences as follows:
\begin{equation}
P(q_{t}, y_1^t) = \sum_{q_{1} = 1}^N  \cdots \sum_{q_{t} = 1}^N P(q_{t}, q_{t-1},y_1^t),
\end{equation}
the calculation involves all possible state paths from time $1$ to time $t$ and the joint probability of a certain state. This is repeated for all $ t \in \{1, \ldots , T\}$ giving $N^T \cdot T$ calculations in total.

The parameters $\theta$ in a HMM model are estimated based on the \textit{Expectation Maximisation Algorithm} below: 
\begin{equation} \label{eq_mlparameter}
\theta = \text{arg} \max_{\theta} P(y_1, \ldots ,y_T \vert \theta)
\end{equation} 
The algorithm proceeds iteratively in two steps. First, the estimation step computes the conditional expectations of the missing data given the observations and the current estimate of $\theta$. Next, the maximisation step maximizes the log likelihood with respect to $\theta$ for the complete data set, including the hidden states.

\begin{eqnarray}
\xi_{ij} (t) &=& P(q_t = i , q_{t+1}=j \vert y_1^T, \theta ) \\
 & = & \frac{\alpha_t (q_t = i) \cdot a_{ij} \cdot P(y_{t+1} \vert q_{t+1}=j) \cdot \beta_{t+1}(q_{t+1}=j)   }{P(y_1^T \vert \theta)} \\
 & = & \frac{\alpha_t (q_t = i) \cdot a_{ij} \cdot P(y_{t+1} \vert q_{t+1}=j) \cdot \beta_{t+1}(q_{t+1}=j)   }{\sum_{1 \leq t \leq T}\beta_t (q_t) \cdot \alpha_t (q_t)}
\end{eqnarray}

Finally, the estimated likelihood is used to classify the state, specifically given the estimated parameters ($\hat{\theta}$) from equation (\ref{eq_mlparameter}):
\begin{equation} \label{eq_mlviterbi}
q_1^T = \text{arg} \max_{q_1^T} P(y_1^T, q_1^T \vert \hat{\theta})
\end{equation} 

\FloatBarrier
\section{Data}
\label{sec_data}
The data set consists of log returns of Bloomberg daily fixing for AUD (Australian Dollar), BRL (Brazilian Real), CAD (Candadian Dollar), CHF (Swiss Franc), EUR (European Monetary Union currency unit) and the GBP (British Pound) quoted against the US dollar for the period from January 1, 1999 to March 11, 2012.\footnote{In the text, we will most of the times refer to the currency pairs AUD/USD, BRL/USD, CAD/USD, CHF/USD, EUR/USD and GBP/USD by their base currency AUD, BRL, CAD, EUR, GBP only.} There are 3,180 data points for each currency pair.
\begin{figure}[ht]
\centering
\includegraphics[width=400pt]{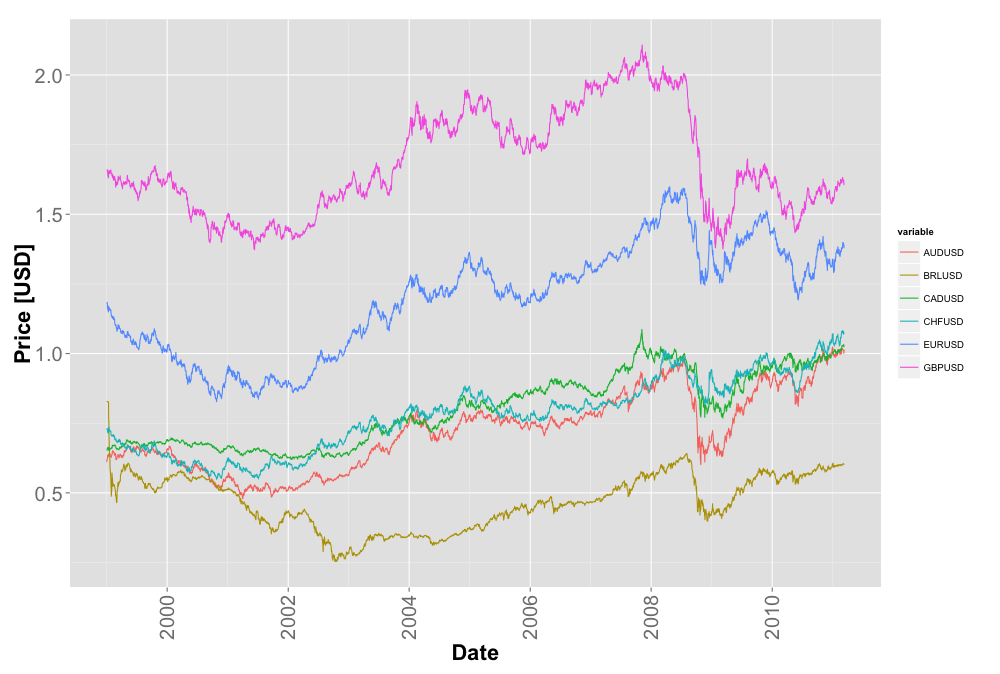}
\caption{Currency Time Series}
\begin{quote}
All the exchange rates in the sample period from January 1999 to March 2012 are quoted against US dollar.
\end{quote}
\label{fig:allccytimeseries}
\end{figure}
The sample period includes several market regimes and significant economic events, such as the post dot.com crash, the financial crisis of 2007, and the beginning of the European debt crisis in 2010. Table \ref{T:ccycorrelation} shows the unconditional correlation of all currency pairs. The correlation between the EUR and CHF is the highest with $\hat{\rho} = 0.875$. Higher correlations are measured between the European currencies, EUR, GBP and CHF, with AUD and CAD also exhibiting a higher correlation. BRL has the weakest correlations with all the other currency in the sample.

\begin{table}[t]
\centering
{\small
\begin{tabular}{ c c  c  c  c c  c }
\hline
 & AUD & BRL & CAD & CHF & EUR & GBP \\
 \hline
AUD &  &  &  &  & &  \\
BRL & 0.333 &  &  & &  &  \\
CAD & 0.623 & 0.306 &  &  &  &  \\
CHF & 0.395 & 0.075 & 0.303 &  &  &  \\
EUR & 0.543 & 0.175 & 0.428 & 0.875 &  & \\
GBP & 0.505 & 0.176 & 0.416 & 0.57 & 0.651 &  \\
\hline 
\end{tabular}}
\caption{Correlation Matrix}
\begin{quote}
Matrix of unconditional correlations between the daily returns of exchange rates in the sample. 
\end{quote}
 \label{T:ccycorrelation}
\end{table}

The return statistics for the sample are summarized in Table \ref{T:basicstats1}.
\begin{table}[t]
\centering
{\small
\begin{tabular}{ l  c  c  c c  c c }
\hline
 & AUD & BRL & CAD & CHF & EUR & GBP\\
 \hline 
 \multicolumn{7}{l}{\textit{Panel A: Quantiles} } \\ 
 Min $[10^{-3}]$ & -72.8 & -99.76 & -33.08 & -29.76 & -25.22 & -34.65 \\
 q(0.1) $[10^{-3}]$ & -9.38 & -11.47 & -6.16 & -8.24 & -7.94 & -6.83 \\
q(0.25) $[10^{-3}]$ & -3.98 & -4.61 & -2.95 & -3.91 & -3.86 & -3.3 \\
q(0.5) $[10^{-3}]$ & 0.56 & 0 & 0.19 & 0.09 & 0.09 & 0.07 \\
q(0.75) $[10^{-3}]$ & 4.88 & 5.13 & 3.37 & 4.18 & 3.8 & 3.37 \\
q(0.9) $[10^{-3}]$ & 9.02 & 10.98 & 6.61 & 8.75 & 8.08 & 7.08 \\
Max $[10^{-3}]$ & 82.47 & 103.44 & 39.51 & 46.92 & 34.51 & 29.26 \\
 &&&&&& \\
 \multicolumn{7}{l}{\textit{Panel B: Moments} } \\
$\mu$ $[10^{-3}]$ & 0.16 & -0.11 & 0.15 & 0.13 & 0.06 & -0.02 \\
$\sigma$ $[10^{-3}]$ & 8.69 & 11.61 & 5.89 & 6.91 & 6.59 & 5.88 \\
$\mu_2$ $[10^{-5}]$ & 7.54 & 13.48 & 3.47 & 4.78 & 4.34 & 3.45 \\
$\mu_3 /\sigma^3$ & -0.42 & -0.36 & -0.15 & 0.14 & 0.06 & -0.28 \\
$\mu_4 / \sigma^4 - 3$ & 9.97 & 12.17 & 3.36 & 1.62 & 1.19 & 2.29 \\ 

\hline 
\end{tabular}}
\caption{Summary Statistics}
\begin{quote}
This table shows the basic statistics for exchange rate returns in the sample. \textit{Panel A} shows the various quantiles of the return distribution, including the maximum and minimum. \textit{Panel B} shows the first and second moments, the standard deviation ($\sigma$), the standardized skewness ($\frac{\mu_3}{\sigma^3}$) and the excess kurtosis ($\frac{\mu_3}{\sigma^3}-3$).
\end{quote}
 \label{T:basicstats1}
\end{table}
From Table \ref{T:basicstats1}, it can be seen that AUD and BRL have daily volatility at 0.87\% and 1.16\% respectively; the highest volatility in the sample. These two currencies also stand out in terms of extreme returns (min, max), and the kurtosis indicates a significant fat tailedness. The other currencies have comparable return distributions, as measured by the quantiles and moments, with EUR/USD being the world's most important currency pair, having the lowest volatility and the least fat tails.

 \FloatBarrier
\section{HMM Estimation Results}
\label{subsec_hmmestim}
Here we are interested in market states associated with higher or lower volatility. As a proxy for volatility, we use the daily return squares $r_t^2$ and estimate the Hidden Markov Model with two volatility states $q_t = 1,2$. 
The state dependent emission distributions are assumed to be normally distributed as follows: 
\begin{equation}
P(y_t \vert q_t = i) \simeq \mathcal{N} (\mu_i , \sigma_i) \qquad i=1,2
\label{eq_hmmemission}
\end{equation}
where $y_t = \{r^2_t\}$ for $1 \leq t \leq T$. The HMM model is estimated via the maximum likelihood method described in section \ref{sec_hmmest}. The estimated parameters are then used in the Viterbi Algorithm (see section \ref{sec_hmmest}) to identify the most likely hidden state sequence. 

The estimated HMM parameters (transition matrix $a_{ij}$, emission distribution parameter $\mu_i$ and $\sigma_i$) for the two hidden states are reported in Table \ref{t_HMM}. Next to the estimated parameters are the standard errors, derived from the asymptotic covariance matrix of estimates, which is computed by finite difference approximations. 
\renewcommand{\arraystretch}{1.2}
\begin{table}[t]
\centering
{\small
\begin{tabular}{ c  c  c  c }
\hline
& \textit{Transition Matrix} &  \multicolumn{2}{c}{\textit{Emission Distribution}} \\ 
&&&\\
& $\hat{a}_{ij} = p(q_{t+1}=j \vert q_{t}=i )$ & $\mathcal{N}_1 \binom{\mu}{\sigma^2}$ $[10^{-5}]$ & $\mathcal{N}_2 \binom{\mu}{\sigma^2}$ $[10^{-5}]$ \\[0.2cm]
\hline
&&&\\
AUD &
$\begin{pmatrix}
0.318_{(0.024)} & 0.683_{(0.024)} \\
0.146_{(0.008)} & 0.855_{(0.008)} \\
 \end{pmatrix}$
 & 
\begin{tabular}{c}
$31.85_{(2.4)}$ \\
$0.04_{(0.01)}$ \\
\end{tabular} 
& 
\begin{tabular}{c}
$2.34_{(0.06)}$ \\
$0.01_{(0.01)}$ \\
\end{tabular} \\
&&&\\
BRL & 

$\begin{pmatrix}
0.452_{(0.023)} & 0.549_{(0.023)} \\
0.142_{(0.008)} & 0.859_{(0.008)} \\
 \end{pmatrix}$
 & 
\begin{tabular}{c}
$54.97_{(4.03)}$ \\
$0.11_{(0.01)}$ \\
\end{tabular} 
& 
\begin{tabular}{c}
$2.74_{(0.08)}$ \\
$0.01_{(0.01)}$ \\
\end{tabular} \\
&&&\\

CHF & 

$\begin{pmatrix}
0.368_{(0.018)} & 0.633_{(0.018)} \\
0.374_{(0.013)} & 0.627_{(0.013)} \\
 \end{pmatrix}$
 & 
\begin{tabular}{c}
$11.2_{(0.39)}$ \\
$0.01_{(0.01)}$ \\
\end{tabular} 
& 
\begin{tabular}{c}
$0.99_{(0.03)}$ \\
$0.01_{(0.01)}$ \\
\end{tabular} \\
&&&\\

CAD & 
$\begin{pmatrix}
0.341_{(0.021)} & 0.66_{(0.021)} \\
0.21_{(0.01)} & 0.791_{(0.01)} \\
 \end{pmatrix}$
 & 
\begin{tabular}{c}
$11.5_{(0.51)}$ \\
$0.01_{(0.01)}$ \\
\end{tabular} 
& 
\begin{tabular}{c}
$0.92_{(0.03)}$ \\
$0.01_{(0.01)}$ \\
\end{tabular} \\
&&&\\
GBP & 
$\begin{pmatrix}
0.362_{(0.02)} & 0.639_{(0.02)} \\
0.272_{(0.012)} & 0.729_{(0.012)} \\
 \end{pmatrix}$
 & 
\begin{tabular}{c}
$9.48_{(0.37)}$ \\
$0.01_{(0.01)}$ \\
\end{tabular} 
& 
\begin{tabular}{c}
$0.89_{(0.03)}$ \\
$0.01_{(0.01)}$ \\
\end{tabular} \\
&&&\\

EUR & 
$\begin{pmatrix}
0.369_{(0.018)} & 0.632_{(0.018)} \\
0.364_{(0.013)} & 0.637_{(0.013)} \\
 \end{pmatrix}$
 & 
\begin{tabular}{c}
$10.29_{(0.33)}$ \\
$0.01_{(0.01)}$ \\
\end{tabular} 
& 
\begin{tabular}{c}
$0.91_{(0.03)}$ \\
$0.01_{(0.01)}$ \\
\end{tabular} \\
&&&\\
\hline

\end{tabular}}
\caption{Hidden Markov Model Estimates}
\begin{quote}
The estimated transition matrices $\hat{a}_{ij}=p(q_{t+1}=j \vert q_{t}=i )$ for the Markov hidden states are reported with the standard error derived from asymptotic covariance matrix of estimates.  In the right column, the estimated state dependent emission distributions for $r_t^2$ are reported, with standard error in brackets, for which we assume a normal distribution $(\mathcal{N}_i$, state $i=1,2$). All parameter estimates are statistically significant at the 99\% confidence level.
\end{quote}
\label{t_HMM}
\end{table}
It is clear from Table \ref{t_HMM} that the volatility level in state 1 is much higher than in state 2. The mean of the state dependent squared return distribution for the 'excited' state ranges between 9.48 $[10^{-5}]$ for GBP to 54.97 $[10^{-5}]$ for BRL. The 'normal' state $r_t^2$ distribution has a range that is lower than for the 'excited' state, in between 0.89 $[10^{-5}]$ for GBP to 2.74 $[10^{-5}]$ for BRL. AUD and BRL have volatility levels distinct from the group, CHF, EUR, CAD and GBP.
Of all six currency pairs, BRL is the most persistent and most likely to remain in the same state (high or low volatility) for a long period, whereas AUD tends to stay much longer in the low volatility state than all other other currencies. 
In terms of the dynamics of the underlying Markov series, the majority of the currencies have a probability $p(q_{t+1}=1 \vert q_{t}=1 )$ staying in the excited state lies between 0.318 (AUD) to 0.369 (EUR), with the exception of  BRL at 0.452. The probability to remain in the 'normal' state is higher than that for the 'excited' state for all currencies. 

\begin{figure}[ht]
\centering
\includegraphics[width=4.5in]{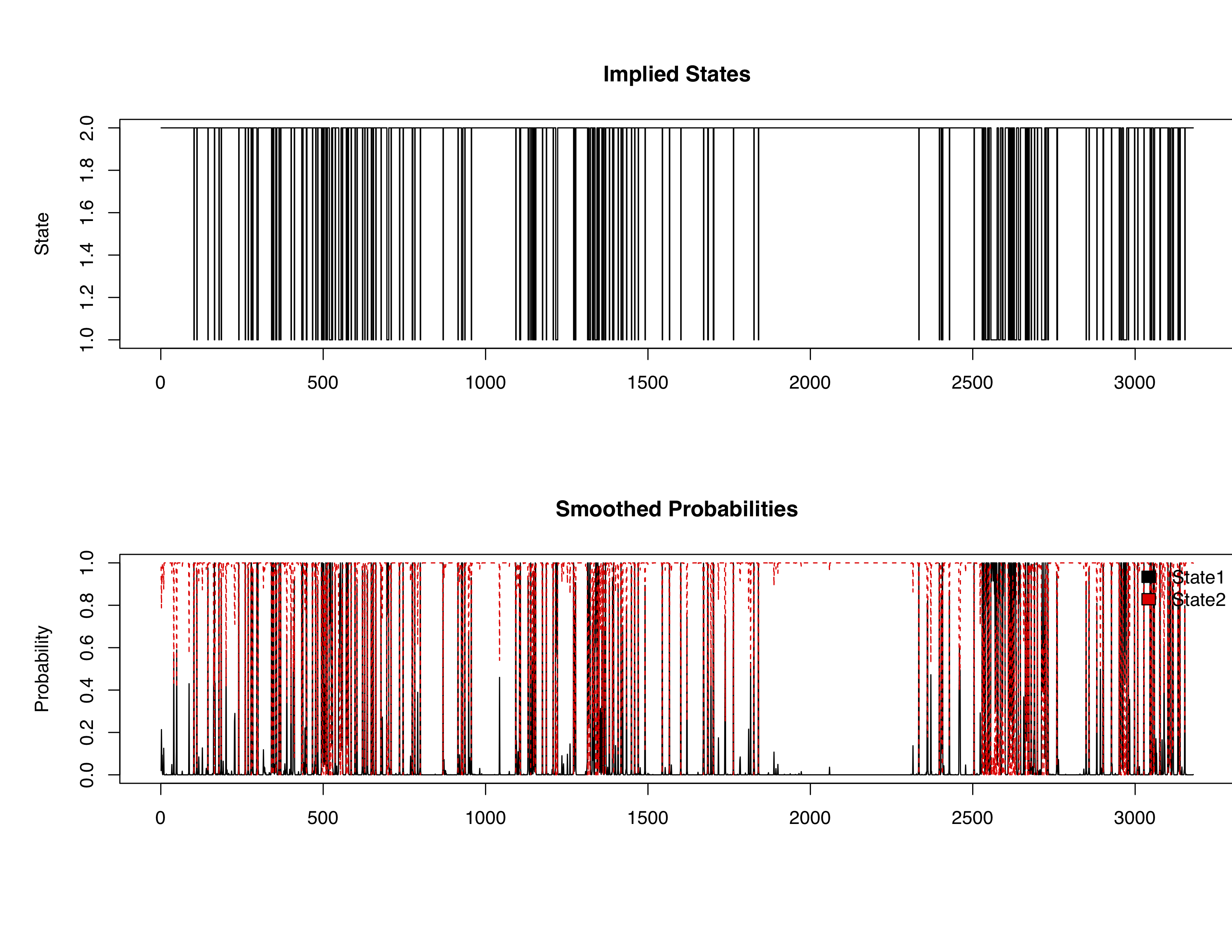}
      \caption{State Sequence for the daily EUR/USD exchange rate \label{fig_statesequenceEUR}}
\begin{quote}
The upper graph shows the most likely volatility state sequence as determined by the Viterbi Algorithm for the daily EUR/USD exchange rate. The low volatility state and high volatility state are represented by $\{1,2\}$. In the lower graph, the smoothed probabilities of the two states are shown. 
\end{quote}
\end{figure}

The optimal state sequence associated with a given observation sequence is determined via the Viterbi Algorithm, which is based on dynamic programming methods. The top picture in Figure \ref{fig_statesequenceEUR} shows the derived state sequence for the EUR time series. The lower picture in Figure \ref{fig_statesequenceEUR} depicts the associated probability to be in a specific state. The diagram allows for the 'eyeballing' of the goodness of fit for the states. 
It is clear that there are periods when the identification of the state is less certain than other periods. In the interest of brevity, we omit the diagnostic diagrams for the other currencies.

\begin{figure}[ht]
\begin{minipage}[b]{0.5\linewidth}
\centering
\includegraphics[width=2.95in]{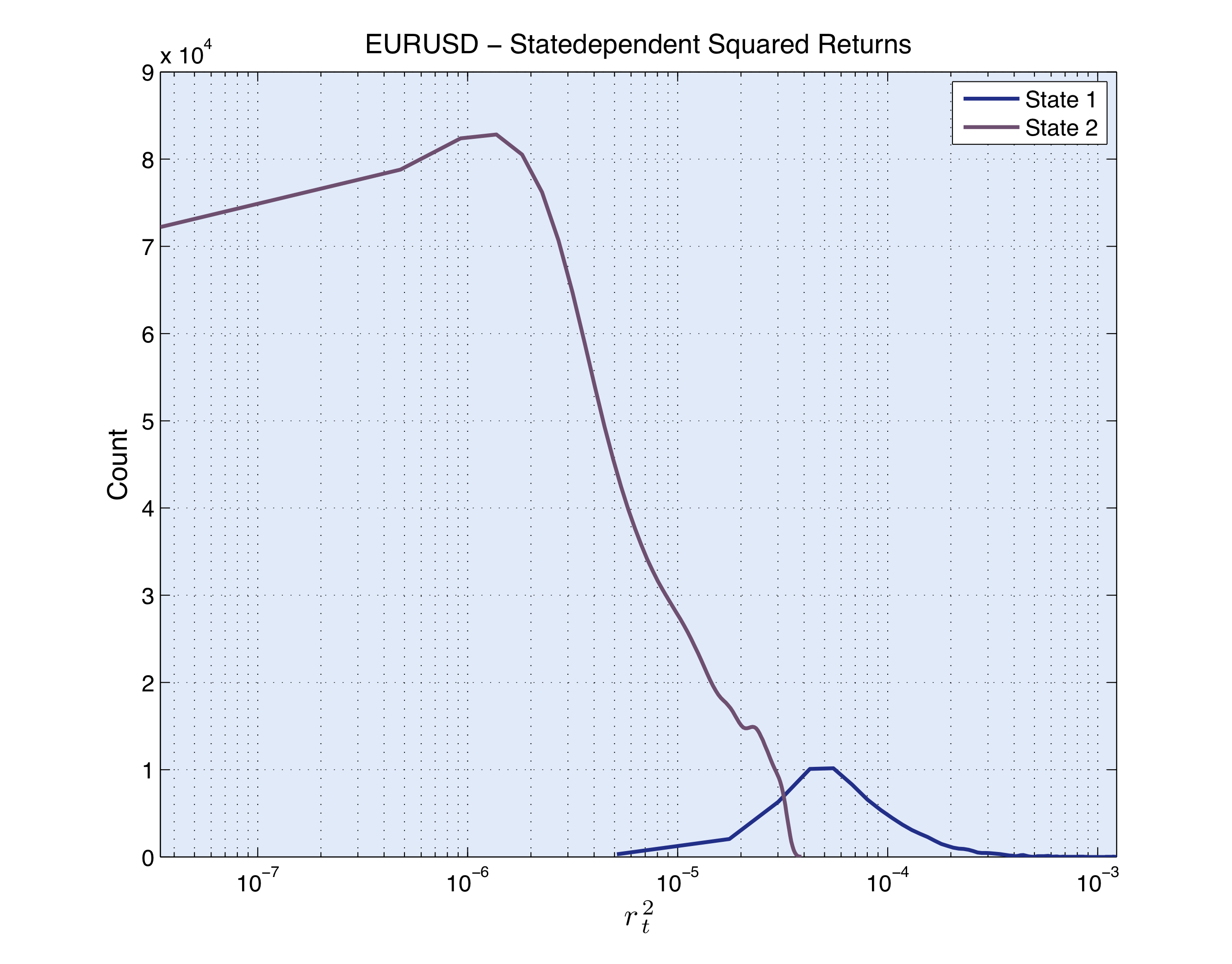}
\end{minipage}
\hspace{0.1cm}
\begin{minipage}[b]{0.5\linewidth}
\centering
   \includegraphics[width=2.95in]{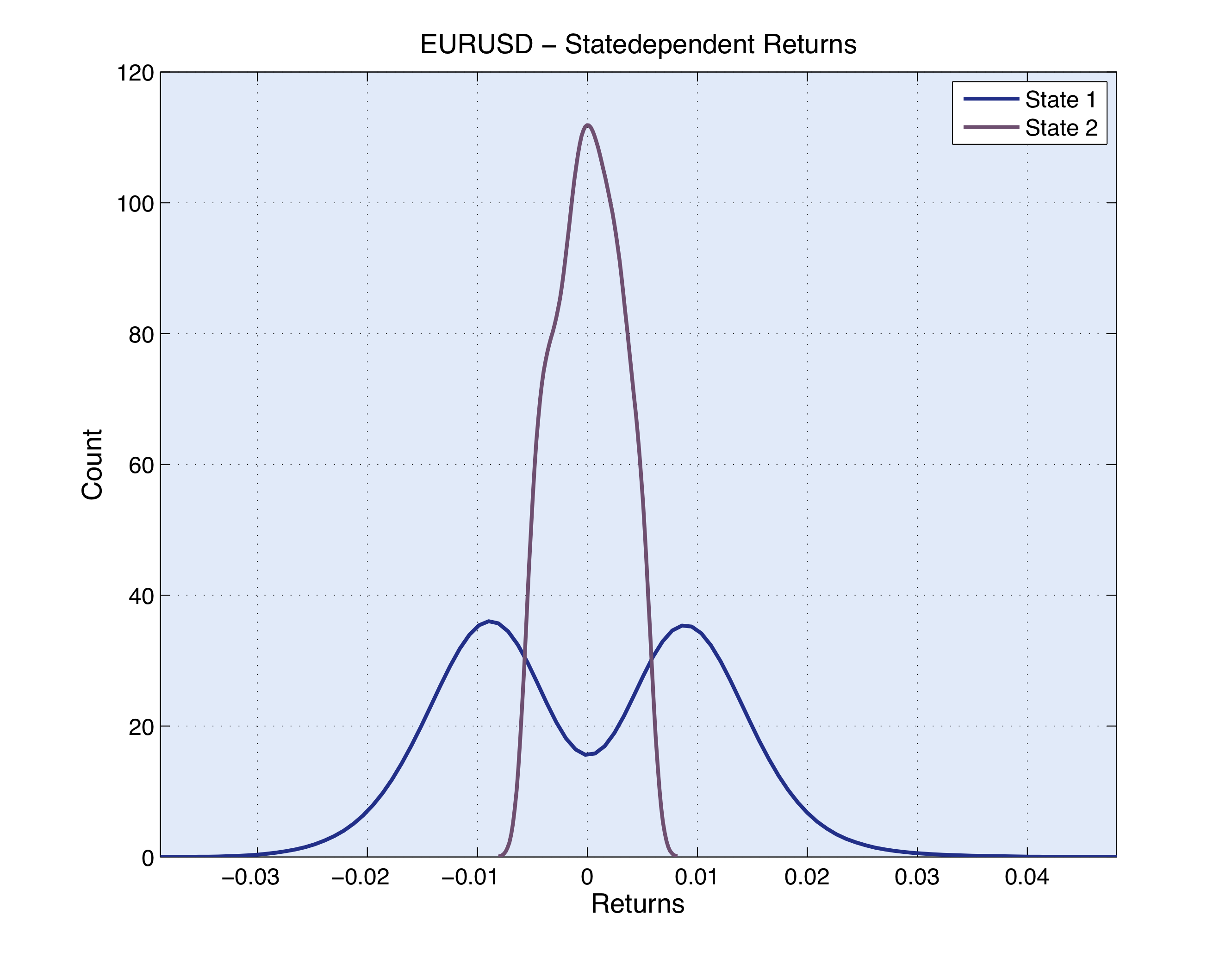}
\end{minipage}
\caption{EUR/USD state dependent return distributions}
\begin{quote}
The left figure shows the state dependent squared returns $r_t^2$ distribution. The right figure shows the return distribution.
\end{quote}
\label{fig_statertns}
\end{figure}

Figure \ref{fig_statertns} plots the kernel density estimates for $r_t$ and $r_t^2$ distributions of daily EUR/USD. The results are very similar for all the other exchange rates and, hence, they are not presented here. Figure \ref{fig_statertns} shows on the right picture that long tails, both negative and positive, are captured within state 1, the 'excited' market state. The smaller positive and negative returns are in state 2, the 'normal' market state. 
Figure \ref{fig_statertns} shows a small overlap of the emission distributions of the two volatility states and a smaller density count for the 'excited' state . 

In Appendix  \ref{sec_allccystateseries} the volatility state sequences for all currencies are shown over time, with the high volatility states marked in red. The financial crisis of 2008 to 2009 is clearly identifiable as a period when all currency pairs in the sample were in volatile periods. Other financial markets events are also visible. The crash of May 2010, the start of the financial crisis in August 2007 and other well known volatile periods can be observed. The period from August 2006 to May 2007 is an extended period of tranquillity, in which all currency pairs in the sample had only a few, short-lived volatile episodes.

 \FloatBarrier
\section{Dependence between Hidden States}
In this section, we apply the Shannon entropy to measure the dependencies between the series. We use the mutual information in section \ref{subsec_hmm_mutualinfo} for measuring the dependence between the states of two series and transfer entropy in section \ref{subsec_transferentropy} to measure the information flow between the states of two series. 

We first analyse the entropy and information content for each state series individually. Understanding the entropy of the state sequences, especially the entropy of blocks of the process, are helpful as these form the basic input in the calculation of mutual information and transfer entropy. They also shed light on how much information is contained in the past, given the current state of events, can be used to predict the future of the state sequence. In section \ref{subsec_entropyblock}, we introduced various measures to determine the information content of series of random variables, such as the entropy rate $h_{\infty} = \lim_{n \rightarrow \infty} H(X_{0}\vert X_{-1} , \ldots X_{-(n+1)}  ) $, which quantifies the average information needed to predict a future observation given the past. 
One of the assumptions of the HMM model is that the hidden states form a Markov process. Although the hidden state sequence is, generally, not a Markov process,\footnote{\citeN{Spreij:2001ka} proved the conditions underwhich the hidden state sequence is a Markov process.} it is interesting to use the entropy measurements of the estimated state sequence to test for the Markov property. 

As the Markov chain is irreducible and aperiodic, the distribution of the $n$th random variable tends to a stationary distribution as $n \rightarrow \infty$. In section \ref{subsec_entropyblock}, we derived the stationary distribution and the corresponding entropy for a Markov process. Given the estimated transition matrices, we can calculate the stationary distribution of the Markov process. In panel A of Table \ref{t_entropyrates}, the probabilities of the states for the different currency pairs are listed.  
The relative frequency of the 'excited' state $\pi_1$ is lower than that of the 'normal' state $\pi_2$ for all currencies. AUD and BRL have the lowest $\pi_1$ while EUR has the highest  $\pi_1$.  AUD and BRL also have the highest excess kurtosis in the sample with EUR having the lowest excess kurtosis (see Table \ref{T:basicstats1}). In our set-up, we stipulated a simple model for squared returns consists of two states with normal distributions. In the case of AUD and BRL, in which the tails are heavier, the 'excited' state's mean of the emission distribution is drawn more to the tails, capturing the more extreme and less frequent volatility regimes. 

The entropy and the entropy rate of the Markov process can be derived from the stationary distribution (see section \ref{subsec_entropyblock}). Panel B of Table \ref{t_entropyrates} lists the entropy of the Markov process $H(X)$, given the transition probabilities and assuming a stationary distribution. $\hat{H}(X)$ is the entropy as measured from the state sequence derived through the Viterbi Algorithm. $H(X)$ is greater than $\hat{H}(X)$ for all currencies. So the estimated state sequence is more regular and has a lower entropy than its stationary Markovian counterpart. One possible explanation for this is the 'small sample effect', which we mentioned in Chapter \ref{chap_entropy}. The calculated stationary distribution is the theoretical entropy, which does not suffer the bias of the empirical estimate. Another possible explanation for the difference between the two figures is that the estimated state sequence is in fact not a Markov process.

The HMM state series, estimated in the previous section, are assumed to be Markov processes. This implies that the future is independent from the past, given the present. This conditional independence from past histories longer than one day, implies that the entropy of the series can be simplified $H(X_{n}\vert X_{n-1} , \ldots X_{1}  ) = H(X_{n}\vert X_{n-1} )$. 
In Panel C of Table \ref{t_entropyrates}, the entropy rates of the hidden Markov processes $h_{\infty}$, assuming stationarity, are listed. The estimated entropy rates $\hat{h}_{\infty} = \hat{H}(X_0 \vert X_{-1})$  are slightly lower than the stationary distribution entropy rates,  $H(X)$. This is to be expected given that stationarity is a limiting case for the process for which we only have a finite sample. Also, small sample bias, again, plays a part for the underestimation (\citeN{2003physics...7138G}). This small sample effect can also be observed when estimating entropies for longer histories. 
\begin{table}[!h]
\centering
{\small
\begin{tabular}{ c  c  c  c  c  c  c  c }
\hline
 & AUD & BRL & CAD & CHF & EUR & GBP \\
 \hline  
 \multicolumn{7}{l}{\textit{Panel A: Frequencies of Stationary Distribution} } \\
$\pi_1$ & 0.177 & 0.206 & 0.241 & 0.372 & 0.366 & 0.299 \\
$\pi_2$ & 0.824 & 0.795 & 0.76 & 0.629 & 0.635 & 0.702 \\
 \hline 
 \multicolumn{7}{l}{\textit{Panel B: Process Entropy} } \\
$H(X)$ & 0.6715 & 0.733 & 0.7964 & 0.9515 & 0.9471 & 0.8789 \\
$\hat{H}(X)$ & 0.6363 & 0.7049 & 0.7613 & 0.9253 & 0.9198 & 0.8406 \\
   \hline
    \multicolumn{7}{l}{\textit{Panel C: Entropy Rates} } \\
$h_{\infty}$ & 0.6526 & 0.6723 & 0.7846 & 0.9515 & 0.9471 & 0.8732 \\
$\hat{H}(X_2 \vert X_1)$ & 0.6199 & 0.6535 & 0.7517 & 0.9252 & 0.9197 & 0.8351 \\
  \hline
\end{tabular}}
\caption{Entropy rate $h_{\infty}$ and  sample $\hat{H}(X_2 \vert X_1)$ in (bit) for the stationary distribution of HMM process.}
\label{t_entropyrates}
\end{table}
The highest rate associated with CHF is $0.9515$ bit, and is close to the maximum of $1$. Only EUR has a similar high entropy rate. This relates to the observation that for both currency pairs, the stationary distribution is more equally weighted and hence the states are less predictable.

\subsection{Hidden State Dependence}
\label{subsec_hmm_mutualinfo}

In Chapter \ref{chap_entropy}, we introduced mutual information $I(X,Y)_{\tau} = H(Y) - H(Y \vert X_{\tau})$ as a measure of the dependency between two random variables $X$ and $Y$. It measures the average gain in information about $Y$ from observing $X_{\tau}$. We have seen in Chapter \ref{chap_entropy} that $I(X,Y) = I(Y,X)$, so $X$ provides the same amount of information on $Y$ as $Y$ does on $X$. It is in this sense not a directional measure. 

We estimate mutual information for lags $\tau = 0,1,-1$, to analyse volatility co-movement ($\tau=0$), and volatility spillovers ($\tau = \pm 1$). The estimates of finite samples are biased \citeN{2003physics...7138G}. These biases are related to sample size, the size of the state space and the entropy rate of the underlying process. Here, we follow the approach in Chapter \ref{chap_paper2} to quantifying this bias by calculating the mutual information with one process being reshuffled, and thereby destroying all of its time series patterns. With sufficient repetitions of this reshuffling process, we generate a sample for which the mutual information includes the noise due to small sample. As an estimate of the noise, we take the 99\% quantile of the measured mutual information of the reshuffled series $q_{0.95}(\hat{I}(X_{sh},Y))$ and report them in brackets in the table. Panels \textit{A}, \textit{B} and \textit{C} in Table  \ref{t_mitau}, show the results for the different lags $\tau = 0,-1, +1$.
\begin{table}[t]
\centering
{\scriptsize
\begin{tabular}{ c  c  c  c  c  c  c }
\hline
 & $\text{AUD}_{t+\tau}$ & $\text{BRL}_{t+\tau}$ & $\text{EUR}_{t+\tau}$ & $\text{GBP}_{t+\tau}$ & $\text{CAD}_{t+\tau}$ & $\text{CHF}_{t+\tau}$ \\
\hline
 \multicolumn{7}{l}{\textit{Panel A: $\tau = 0$ } } \\
   &  &  &  &  &  & \\
$\text{AUD}_{t}$ &  & 16.611 (1.442)  & 44.18 (1.512)  & 43.649 (1.523)  & 63.445 (1.537)  & 32.566 (1.532)  \\
 $\text{BRL}_{t}$&  &  & 6.082 (1.484)  & 4.979 (1.502)  & 10.154 (1.506)  & 1.785 (1.555)  \\
$\text{EUR}_{t}$ &  &  &  & 74.079 (1.474)  & 16.987 (1.468)  & 305.02 (1.5)  \\
$\text{GBP}_{t}$ &  &  &  &  & 25.378 (1.536)  & 60.996 (1.504)  \\
$\text{CAD}_{t}$ &  &  &  &  &  & 12.19 (1.475)  \\
$\text{CHF}_{t}$ &  &  &  &  &  &  \\
 \hline
 \multicolumn{7}{l}{\textit{Panel B: $\tau = -1$ } } \\
   &  &  &  &  &  & \\
$\text{AUD}_{t}$ &  & \textbf{4.578} (1.447)  & \textbf{3.805} (1.526)  & \textbf{5.673} (1.517)  & \textbf{5.118} (1.542)  & 0.437 (1.526)  \\
 $\text{BRL}_{t}$ &  &  & 0.035 (1.526)  & \textbf{1.663} (1.509)  & \textbf{2.589} (1.512)  & 0.001 (1.548)  \\
$\text{EUR}_{t}$ &  &  &  & 1.127 (1.47)  & \textbf{2.289} (1.476)  & 0.532 (1.498)  \\
$\text{GBP}_{t}$ &  &  &  &  & \textbf{6.417} (1.53)  & 0.048 (1.505)  \\
$\text{CAD}_{t}$ &  &  &  &  &  & 0.162 (1.509)  \\
$\text{CHF}_{t}$ &  &  &  &  &  &  \\
 \hline
  \multicolumn{7}{l}{\textit{Panel C: $\tau = +1$ } } \\
  &  &  &  &  &  & \\
$\text{AUD}_{t}$ &  & \textbf{3.947} (1.568)  & 0.839 (1.529)  & \textbf{7.525} (1.517)  & \textbf{7.305} (1.521)  & 1.027 (1.526)  \\
 $\text{BRL}_{t}$ &  &  & 0.136 (1.512)  & 0.664 (1.509)  & \textbf{3.872} (1.489)  & 0.191 (1.526)  \\
$\text{EUR}_{t}$ &  &  &  & 0.292 (1.47)  & 0.173 (1.488)  & 0.11 (1.522)  \\
$\text{GBP}_{t}$ &  &  &  &  & \textbf{5.197} (1.5)  & 0.07 (1.495)  \\
$\text{CAD}_{t}$ &  &  &  &  &  & 0.552 (1.509)  \\
$\text{CHF}_{t}$&  &  &  &  &  &  \\
 \hline
\end{tabular}}
 \caption{Volatility States Co-Movement and Interdependence}
 \begin{quote}
This table shows the estimates for Mutual Information $\hat{I}(X,Y)_{\tau} $ $[10^{-3}]$ (in bits) for the hidden state sequences with different time lags. Volatility co-movement ($\tau =0$) is shown in panel \textit{A} and volatility states interdependence relations ($\tau =  \pm 1$) are shown in panels \textit{B} and \textit{C}. The error estimates are shown in brackets. These are produced by generating reshuffled series of one of the series. The 99\% quantile is reported $q_{0.99}(\hat{I}(X_{sh},Y))$ $[10^{-3}]$. In panel \textit{A}, all measured mutual information $\hat{I}(X,Y)$ are significant. The significant measurements in panels \textit{B} and \textit{C} are shown in bold.
\end{quote}
 \label{t_mitau}
 \end{table}

We have seen in section \ref{subsec_paper2info} that mutual information measures the divergence of two random variables from being independent.
In this particular setting, with binary valued series, mutual information is a summation of four terms, representing the possible combinations of values in the two random variables $X_{\tau},Y$: $I(X;Y)_{\tau}  = - \sum_{x_{\tau} ,y \in \{1,2 \}} p(x_{\tau},y) \cdot log_2 \frac{p(x_{\tau} ,y)}{p(x_{\tau} )p(y)}$.  Any statistical significant measurement of $\hat{I}(X,Y)_{\tau} > 0$, implies that the random variables are more often in the same state, high volatility or low volatility, and at the specified time (lagged by $\tau$ in the case of $X$), than if those variables are independent. 

For all currency pairs, the mutual information peaks at lag $\tau = 0$, where all estimated values are above the 99\% quantile of 'noise'. The strongest observed relationship is noted between EUR and CHF, with a mutual information of $305.02 [10^{-3}]$ bits. This is approximately five times stronger than that for any of the other currencies. The next strongest co-movement relationship is that between GBP and EUR with $74.079 [10^{-3}]$ bits. The connection between GBP and CHF with $\hat{I} = 60.996 [10^{-3}]$ bits completes the triangle strong link of European currencies in the sample. The other strong co-movement relationship outside this triangle of european group is that between AUD and CAD with $63.445 [10^{-3}]$ bits measured. 


For all currency pairs, mutual information dropped sharply when $\tau \ne 0$. This means that a change in volatility regime in one series has the strongest influence on the volatility regime of another series on the same day. Panels B and C show that at $\tau \pm 1$, only 13 of the 30 pairs have mutual information statistically significant at the 99\% confidence level. The strongest two-way leag-lag connections are AUD-GBP, AUD-CAD, AUD-BRL, CAD-GBP and, to a lesser extent, CAD-BRL. The weaker one-way connections include ${\text{AUD}_{0}}$-${\text{EUR}_{-1}}$, ${\text{BRL}_{0}}$-${\text{GBP}_{-1}}$ and ${\text{EUR}_{0}}$-${\text{CAD}_{-1}}$.

\FloatBarrier
\subsection{Hidden State Information Flow}
\label{subsec_transferentropy}
While mutual information quantifies the deviation from $X$ and $Y$ being independent, transfer entropy quantifies the deviation from $X$ being determined by its own history only (via conditional probabilities). Unlike mutual information, transfer entropy $T_{Y \rightarrow X} (m,l)$, introduced in Chapter \ref{chap_entropy} and defined below, is not symmetric and takes into account only the statistical dependencies originating from $Y$ and not those from a common signal. 
\begin{equation*}
 T_{Y \rightarrow X} (m,l) = H(X_{0} | X_{-1} , \ldots X_{-m}  ) - H(X_{0} | X_{-1} , \ldots X_{-m} , Y_{-1} , \ldots Y_{m-l}) 
 \end{equation*}
To measure any dependence of volatility states estimated by the HMM-model, the transfer entropy between the hidden state sequence of two series is estimated. The transfer entropy $T_{Y \rightarrow X}$ from $Y$ to $X$ is the degree to which $Y$ makes less uncertain the future of $X$ beyond the degree that $X$ is already giving information on resolving the future of $X$. In this context, a detected information flow from $Y$ to $X$ indicates that past volatility states in $Y$ (up to specified length $l$) contain information, that makes the next day's volatility state in $X$ more predictable. We will in the present case use the most likely hidden states determined by the Viterbi algorithm. 

We use the two-letter alphabet for the volatility state and the Grassberger estimator $\hat{H}_{\psi}$ in equation (\ref{eq_grassberger}) for the estimation of transfer entropy. 
To calculate the estimation error, we follow \citeN{Marschinski:2002dq}  (as in Chapter \ref{chap_paper1}) and estimate the mean and standard error by the bootstrapping method, involving reshuffling the predictor series, destroying all its time series pattern. This process is repeated enough times, after which a mean, $\mu({\hat{T}_{sh}})$, and a standard deviation, $\sigma$ is calculated. The effective transfer entropy (ET) (see Chapter \ref{chap_entropy} for details) is formed by deducting the mean of the shuffling simulations from the estimated transfer entropy, $ET_{Y \rightarrow X} (m,l) \equiv T_{Y \rightarrow X} (m,l) - \mu({\hat{T}_{sh}}) $.
For $ET_{Y \rightarrow X} (m,l)$, $m$ refers to the amount of own past history of $X$ included in the prediction, whereas $l$ refer to the amount of past history of $Y$ used in predicting $X$. 
Also, as in Chapter \ref{chap_paper1}, we use the REA (\textit{relative explanation added}) measure from \citeN{Marschinski:2002dq} to guage the relative information gain due to the transfer entropy. REA defined below relates the information gain from TE to the entropy of the series:
\begin{equation*}
REA (m,l) = \frac{ET_{Y \rightarrow X} (m,l)}{H(X_{0}\vert X_{-1} , \ldots X_{-(m)}  ) } 
\end{equation*}

\begin{table}[t]
\centering
{\scriptsize
\begin{tabular}{ l  c  c  c  c  c  c }
\hline
\multicolumn{1}{c}{$Y \rightarrow X$}  & \multicolumn{2}{l}{Block} &  \multicolumn{4}{c}{Transfer Entropy } \\
   &  &  &  &  &  & \\
$\hat{ET}_{Y \rightarrow X}(m,l)$ & m & l & $\hat{T}$ (bit) & $\mu({\hat{T_{sh}}}) \pm \sigma$ (bit) & $\hat{ET}$ (bit) & REA \\
& &  &[$10^{-3}$]  &[$10^{-3}$]  & [$10^{-3}$] & [\%]  \\ 
\hline
\textit{Panel A: AUD} $\rightarrow X$    &  &  &  &  &  & \\
${\text{AUD} \rightarrow \text{BRL}}$ & 1 & 4 & 10.192 & 1.118 $ \pm $ 1.81 & 9.074 & 1.388 $\pm$ 0.277  \\
${\textbf{AUD} \rightarrow \textbf{CAD}}$ & 4 & 4 & 38.986 & 20.042 $ \pm $ 4.756 & \bf{18.945} & 2.609 $\pm$ 0.655  \\
${\text{AUD} \rightarrow \text{CHF}}$ & 1 & 2 & 4.354 & 0.007 $ \pm $ 0.782 & 4.348 & 0.47 $\pm$ 0.085  \\
${\text{AUD} \rightarrow \text{EUR}}$ & 1 & 4 & 6.701 & 0.811 $ \pm $ 1.878 & 5.891 & 0.641 $\pm$ 0.205  \\
${\text{AUD} \rightarrow \text{GBP}}$ & 1 & 4 & 15.203 & 0.953 $ \pm $ 1.927 & 14.251 & 1.706 $\pm$ 0.231  \\
\hline
\textit{Panel B: BRL}  $\rightarrow X$   &  &  &  &  &  & \\
${\text{BRL} \rightarrow \text{AUD}}$ & 4 & 4 & 27.868 & 21.749 $ \pm $ 4.747 & 6.119 & 1.015 $\pm$ 0.788  \\
${\text{BRL} \rightarrow \text{CHF}}$ &  &  &  &   &  &     \\
${\textbf{BRL} \rightarrow \textbf{CAD}}$ & 4 & 4 & 42.537 & 21.016 $ \pm $ 5.026 & \bf{21.521} & 2.963 $\pm$ 0.692  \\
${\text{BRL} \rightarrow \text{EUR}}$ & 1 & 4 & 2.314 & 0.47 $ \pm $ 1.92 & 1.844 & 0.201 $\pm$ 0.209  \\
${\text{BRL} \rightarrow \text{GBP}}$ & 4 & 4 & 25.327 & 20.143 $ \pm $ 5.04 & 5.184 & 0.62 $\pm$ 0.603  \\
\hline
\textit{Panel C: EUR}  $\rightarrow X$   &  &  &  &  &  & \\
${\text{EUR} \rightarrow \text{AUD}}$ & 3 & 4 & 16.152 & 7.569 $ \pm $ 4.105 & 8.584 & 1.413 $\pm$ 0.676  \\
${\text{EUR} \rightarrow \text{BRL}}$ & 1 & 3 & 2.468 & 0.007 $ \pm $ 1.22 & 2.461 & 0.377 $\pm$ 0.187  \\
${\text{EUR} \rightarrow \text{CAD}}$ & 1 & 4 & 6.311 & 0.049 $ \pm $ 1.81 & 6.263 & 0.833 $\pm$ 0.241  \\
${\text{EUR} \rightarrow \text{CHF}}$ & 4 & 4 & 21.894 & 12.486 $ \pm $ 5.885 & 9.408 & 1.017 $\pm$ 0.637  \\
${\textbf{EUR} \rightarrow \textbf{GBP}}$ & 4 & 4 & 32.756 & 16.867 $ \pm $ 5.883 & \bf{15.889} & 1.901 $\pm$ 0.704  \\
\hline
\textit{Panel D: CHF}  $\rightarrow X$   &  &  &  &  &  & \\
${\text{CHF} \rightarrow \text{AUD}}$ & 3 & 2 & 1.201 & 0.045 $ \pm $ 1.668 & 1.157 & 0.191 $\pm$ 0.275  \\
${\text{CHF} \rightarrow \text{BRL}}$ & 2 & 4 & 7.522 & 0.566 $ \pm $ 2.745 & 6.956 & 1.099 $\pm$ 0.434  \\
${\text{CHF} \rightarrow \text{CAD}}$ & 4 & 4 & 25.756 & 19.925 $ \pm $ 5.876 & 5.832 & 0.803 $\pm$ 0.809  \\
${\textbf{CHF} \rightarrow \textbf{EUR}}$ & 4 & 4 & 31.099 & 12.061 $ \pm $ 5.951 & \bf{19.038} & 2.081 $\pm$ 0.651  \\
${\text{CHF} \rightarrow \text{GBP}}$ & 4 & 4 & 27.589 & 16.597 $ \pm $ 5.804 & 10.993 & 1.315 $\pm$ 0.695  \\
\hline
\textit{Panel E: CAD}  $\rightarrow X$   &  &  &  &  &  & \\
${\text{CAD} \rightarrow \text{AUD}}$ & 4 & 4 & 35.408 & 22.94 $ \pm $ 5.049 & 12.469 & 2.069 $\pm$ 0.838  \\
${\textbf{CAD} \rightarrow \textbf{BRL}}$ & 4 & 4 & 38.384 & 22.719 $ \pm $ 5.138 & \bf{15.666} & 2.542 $\pm$ 0.834  \\
${\text{CAD} \rightarrow \text{CHF}}$ & 1 & 4 & 2.674 & 0.254 $ \pm $ 1.845 & 2.421 & 0.262 $\pm$ 0.2  \\
${\text{CAD} \rightarrow \text{EUR}}$ & 4 & 4 & 31.312 & 18.243 $ \pm $ 5.384 & 13.069 & 1.428 $\pm$ 0.589  \\
${\text{CAD} \rightarrow \text{GBP}}$ & 1 & 4 & 9.957 & 0.377 $ \pm $ 1.904 & 9.581 & 1.147 $\pm$ 0.228  \\
\hline
\textit{Panel F: GBP}  $\rightarrow X$   &  &  &  &  &  & \\
${\text{GBP} \rightarrow \text{AUD}}$ & 1 & 4 & 11.395 & 0.328 $ \pm $ 1.959 & 11.068 & 1.785 $\pm$ 0.316  \\
${\text{GBP} \rightarrow \text{BRL}}$ & 4 & 4 & 29.678 & 23.055 $ \pm $ 5.459 & 6.624 & 1.075 $\pm$ 0.886  \\
${\textbf{GBP} \rightarrow \textbf{CAD}}$ & 1 & 4 & 11.946 & 0.169 $ \pm $ 1.881 & \bf{11.778} & 1.566 $\pm$ 0.251  \\
${\text{GBP} \rightarrow \text{CHF}}$ & 1 & 4 & 4.083 & 0.044 $ \pm $ 1.818 & 4.04 & 0.437 $\pm$ 0.197  \\
${\text{GBP} \rightarrow \text{EUR}}$ & 1 & 2 & 1.586 & 0.019 $ \pm $ 0.794 & 1.568 & 0.171 $\pm$ 0.087  \\
\hline
\end{tabular}}
\caption{Volatility States Information Flows}
\begin{quote}
This table presents the results on information flows between the hidden volatility state series. All combinations of currency pairs are listed by the 'source' process. For all currency pairs, the 
combination of blocks $(m,l)$ from one to four days past history is selected with the highest $\widehat{ET}$. For each flow $Y \rightarrow X$, $m$ blocks of past states in $X$ are taken into account and $l$ past states in $Y$. 
Transfer Entropy $\hat{T}$, effective TE $\hat{ET}$ and the relative explanation added REA is shown.
\end{quote}
\label{t_TEallccy}
\end{table}
Table \ref{t_TEallccy} shows that the largest information flow, as measured by the effective transfer entropy $ET$, is that between BRL and CAD with $\hat{ET}_{\text{BRL} \rightarrow \text{CAD} }(4,4) = 21.521$  [$10^{-3}$]. This information flow is also the strongest in terms of REA. It adds REA = 2.963 [\%] to the information already contained in the history of volatility states of CAD itself.  
This compares well with the REA = 1.32\% reported in \citeN{Marschinski:2002dq} for an information flow from the Dow Jones to the DAX equity index based on one-minute intraday returns over the period from May 2000 to June 2001.

The currency pairs BRL, CAD with AUD form a triple group producing the strongest information flow compared with the whole sample.  
A direct linkage between the currency pairs can be clearly seen for AUD and CAD, which have a higher correlation $\hat{\rho} = 0.623$ (Table \ref{T:ccycorrelation}) and for which we measured a high mutual information $\hat{I}(\text{AUD},\text{CAD})_{\tau = 0} = 63.445 $ (Table \ref{t_mitau}). Transfer Entropy picks up a strong information flow between BRL and CAD, with ${\text{BRL} \rightarrow \text{CAD}}$ being stronger than ${\text{CAD} \rightarrow \text{BRL}}$ with $\hat{ET}_{\text{CAD} \rightarrow \text{BRL} }(4,4) = 15.666$  [$10^{-3}$]. So, although the co-movement between the volatility states of BRL and CAD measured with Mutual Information is weak, we can detect with transfer entropy a strong flow, pointing to BRL volatility influencing CAD volatility.

The strong information flow in Europe is among CHF, EUR and GBP with $\hat{ET}_{\text{CHF} \rightarrow \text{EUR} }(4,4) = 19.038$  [$10^{-3}$] and $\hat{ET}_{\text{EUR} \rightarrow \text{GBP} }(4,4) = 15.889$  [$10^{-3}$]. 
\citeN{Inagaki:2006kj} examines the currency pairs GBP and EUR for volatility spillover in the 1999-2004 period. Consistent with our results, the author found support for a unidirectional volatility spillover from EUR to GBP, with the EUR having a one-sided impact on GBP. 

In the previous chapter, we investigated the relationship between daily and hourly draws in GBP and EUR. The transfer entropy measured between volatility states $\hat{ET}_{\text{EUR} \rightarrow \text{GBP} }(4,4) = 15.889$  [$10^{-3}$] is comparable to our results from chapter \ref{chap_paper1}. For draws on daily returns we estimated $\hat{ET}_{\text{EUR} \rightarrow \text{GBP} }(2,4) = 11.451$  [$10^{-3}$], with a shorter history of GBP of two days. 
This corresponds to findings in \citeN{Rebonato:2009eu}, which points to a link between large draws and excited volatility states. 
\citeN{Rebonato:2006xs} conjectured that there are at least two regimes ('normal' and 'excited') in the US\$  interest rate market. Following this insight, \citeN{Rebonato:2009eu} used a hidden Markov model for interest rates. The authors show that large draws are primarily from excited states, especially if the draws have been short. The excited states exhibit positive auto-correlation (where consecutive bursts exist), and in the 'normal' and 'quiet' states, returns have negative auto-correlation, with high fractions of reversals in the normal state. In Appendix \ref{sec_stateduration} we compare the duration of the series, staying in the same volatility states, and draws in our data, to highlight the connection between large draws and high volatility states.

 \FloatBarrier
 \section{Conclusion}
In this chapter, we follow \citeN{Hamilton:1989jg} to model volatility regimes for a set of currencies quoted against the US dollar. A two state hidden Markov model was defined, with which we modelled high and normal volatility states. Following Chapter \ref{chap_paper1}, we used mutual information and transfer entropy to evaluate various volatility regime relationships between the currency pairs. 
We found volatility regime co-movement relationships among the European currencies (EUR/USD, GBP/USD, CHF/USD). CAD/USD and AUD/USD exhibit a volatility co-movement relationship and, to a lesser extent, interdependency with a time lag. 
We also find evidence for information flows from EUR/USD to GBP/USD, which indicates a causal volatility spillover relationship. This finding is in line with findings in \citeN{Inagaki:2006kj}, reported a unidirectional volatility spillover from EUR to GBP in the period 1999-2004. The work showcases also the usefulness of the concepts of mutual information and transfer entropy when applying them to analyse volatility co-movements and spillover relationships among volatility states.

\FloatBarrier
\section{Appendix - State Duration} \label{sec_stateduration}
In this appendix, we define drawdowns (drawups) and draws, as in Chapter \ref{chap_paper1}.  All draw statistics for the sample are summarized in Table \ref{T:basicstats2}.
\begin{table}[t]
\centering
{\small
\begin{tabular}{ l  c  c  c c  c c }
\hline
 & AUD & BRL & CAD & CHF & EUR & GBP\\
 \hline 
 \multicolumn{7}{l}{\textit{Panel A: Draws} } \\
$E[D]/E[d] = E[l_{d}]$ & 1.827 & 2.036 & 1.874 & 1.84 & 1.877 & 1.958 \\
$E[U]/E[u] = E[l_{u}]$ & 2.12 & 2.113 & 1.984 & 1.894 & 1.932 & 2.008 \\
$E[D]$ & 0.012 & 0.016 & 0.008 & 0.01 & 0.01 & 0.009 \\
$\sigma(D)$ & 0.014 & 0.023 & 0.009 & 0.009 & 0.01 & 0.01 \\
$E[U]$ & 0.013 & 0.016 & 0.009 & 0.01 & 0.01 & 0.009 \\
$\sigma(U)$ & 0.013 & 0.017 & 0.009 & 0.01 & 0.01 & 0.009 \\[-1.8ex] 

&&&&&& \\
 \hline 
 \multicolumn{7}{l}{\textit{Panel B: Draw Distribution} } \\
$\hat{z}(D)$ & 0.993 & 0.859 & 1.026 & 1.163 & 1.072 & 1.03 \\
$\sigma_{\hat{z}}(D)$ & 0.023 & 0.02 & 0.023 & 0.025 & 0.023 & 0.024 \\
$\hat{\chi}(D)$ [$10^{-3}$] & 11.6 & 14.4 & 8.1 & 10.1 & 9.7 & 8.9 \\
$\sigma_{\hat{\chi}}(D)$ [$10^{-3}$] & 0.4 & 0.5 & 0.3 & 0.3 & 0.3 & 0.3 \\
$\hat{z}(U)$ & 1.104 & 1.013 & 1.087 & 1.088 & 1.132 & 1.096 \\
$\sigma_{\hat{z}}(U)$ & 0.025 & 0.023 & 0.023 & 0.025 & 0.027 & 0.024 \\
$\hat{\chi}(U)$ [$10^{-3}$] & 12.7 & 15.4 & 8.8 & 10.4 & 10.1 & 9 \\
$\sigma_{\hat{\chi}}(U)$ [$10^{-3}$] & 0.4 & 0.5 & 0.3 & 0.3 & 0.3 & 0.3 \\
\hline 
\end{tabular}}
\caption{Summary Statistics}
\begin{quote}
Statistics of drawdowns and drawups for all sample currencies
\end{quote}
 \label{T:basicstats2}
\end{table}
Table \ref{T:basicstats2} shows $E[D], \sigma[D], E[U]$ and $\sigma[U]$, which are the mean and standard deviations of the drawdowns and drawups. BRL stands out with a higher mean and volatility than the others and an asymmetry between drawdowns and drawups. 
$E[d], \sigma[d], E[u], \sigma[u]$ gives the average daily price drop within a drawdown (drawup) and the standard deviation.  
$E[D]/E[d] = E[l_{d}]$ is the ratio of the average drawdown magnitude to the average daily price drop in a drawdown. In the case of \textit{iid}, Gaussian increments, this equals $2$, a result which is valid even in the case where the time series lacks homoskedasticity. 
A value below 2 represents a thin tail distribution, and a value above 2 represents thick tail distribution. Table \ref{T:basicstats1} shows $E[l_{u}] > E[l_{d}] $ for all currency pairs. Except for AUD and BRL, many currency pairs have $E[l_{d}] $ lower than 2.
Interestingly for interest rates,  this measure is consistently lower than the \textit{iid} Gaussian value of $2$, as shown in  \citeN{Rebonato:2006xs}.

\citeN{Johansen_largestock} show that draws follow a modified exponential distribution $P_{se}(D) = P(D=0) e^{- \left( \frac{\vert D \vert}{\chi} \right)^z}$ where $z>1$ represents lower kurtosis and $z < 1$ higher kurtosis. Table \ref{T:basicstats2} show that most currencies have a $z$ close to 1, with the  exception of BRL/USD drawdowns ($\hat{z=0.859}$) indicating a high kurtosis for drawdowns.

Given the transition probabilities of the hidden state sequence, we can calculate the probability of a sequence of same states with a given length to occur. 
The probability of a sequence of length $n$ to stay in state $i$, $ q_k=i, \ldots , q_{k+n-1}=i$ is $a_{ii}^{n-1} \cdot (1 - a_{ii})$.
The expected duration of a sequence in state $i$ is then $\sum_{n \geq 0} n  \cdot a_{ii}^{n-1} \cdot (1-a_{ii}) = 1/(1-a_{ii})$. 

For a transition probability of $\frac{1}{2}$, which is the example of a fair coin, we have, on average, two consecutive heads or tails being tossed. For the six currency pairs, the expected times of remaining in a certain state are listed in the Table \ref{t_HMMdurat}. 
\begin{table}[!h]
\centering
{\scriptsize
\begin{tabular}{| c | c | c | c | c | c | c |}
\hline
  & AUDUSD & BRLUSD & CADUSD & CHFUSD & GBPUSD & EURUSD \\ 
 \hline
State 1 & 1.466 & 1.824 & 1.517 & 1.582 & 1.567 & 1.584 \\ 
State 2 & 6.859 & 7.048 & 4.781 & 2.681 & 3.689 & 2.751 \\ 
 \hline
\end{tabular}}
\caption{Expected Time State Duration}
\label{t_HMMdurat}
\end{table}

\begin{figure}[ht]
\begin{minipage}[b]{0.5\linewidth}
\centering
   \includegraphics[width=2.9in]{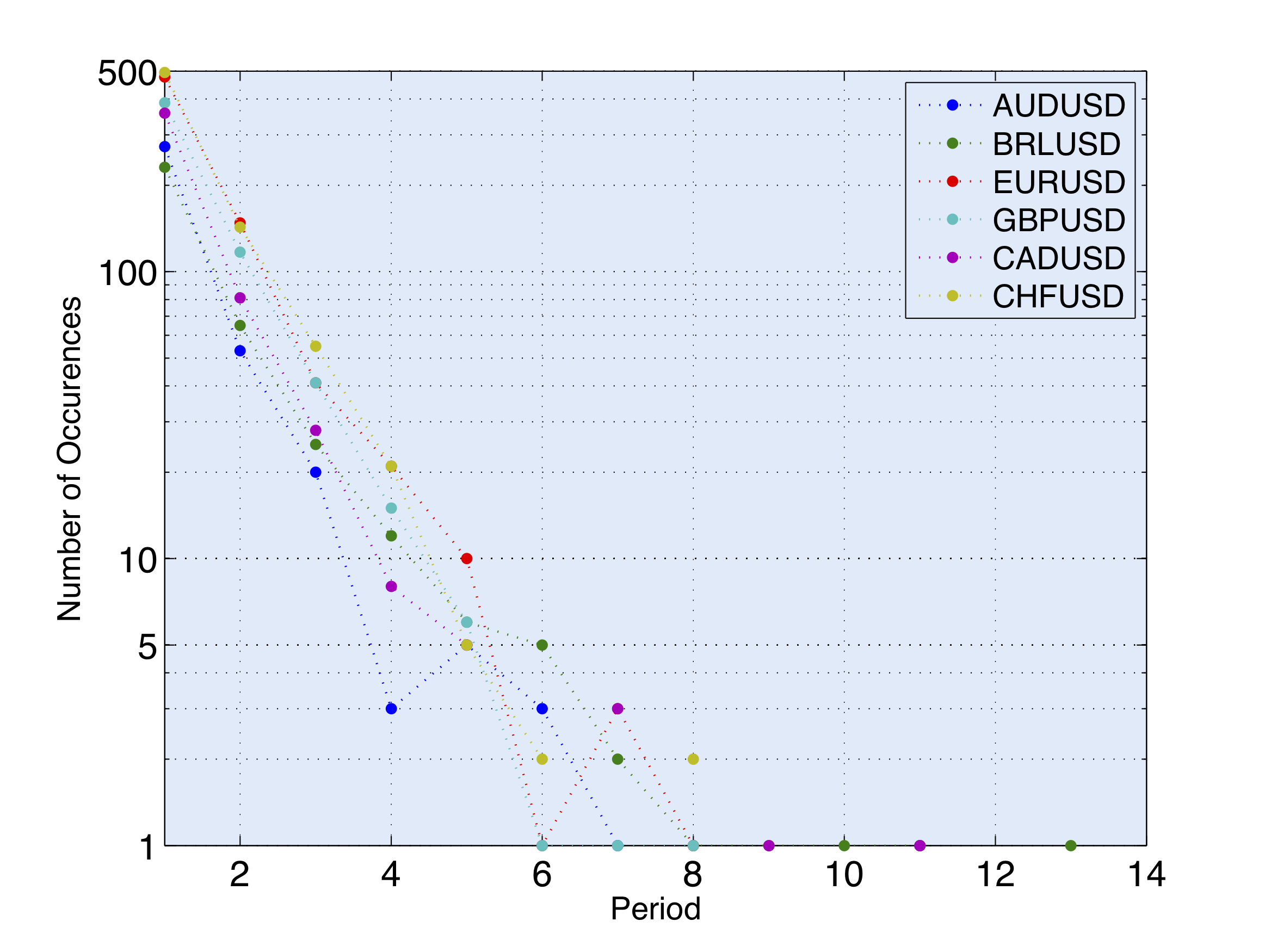}
      \caption{Duration of sequence remaining in the high volatility state 1 \label{fig_statedur1}}
\end{minipage}
\hspace{0.1cm}
\begin{minipage}[b]{0.5\linewidth}
\centering
   \includegraphics[width=2.8in]{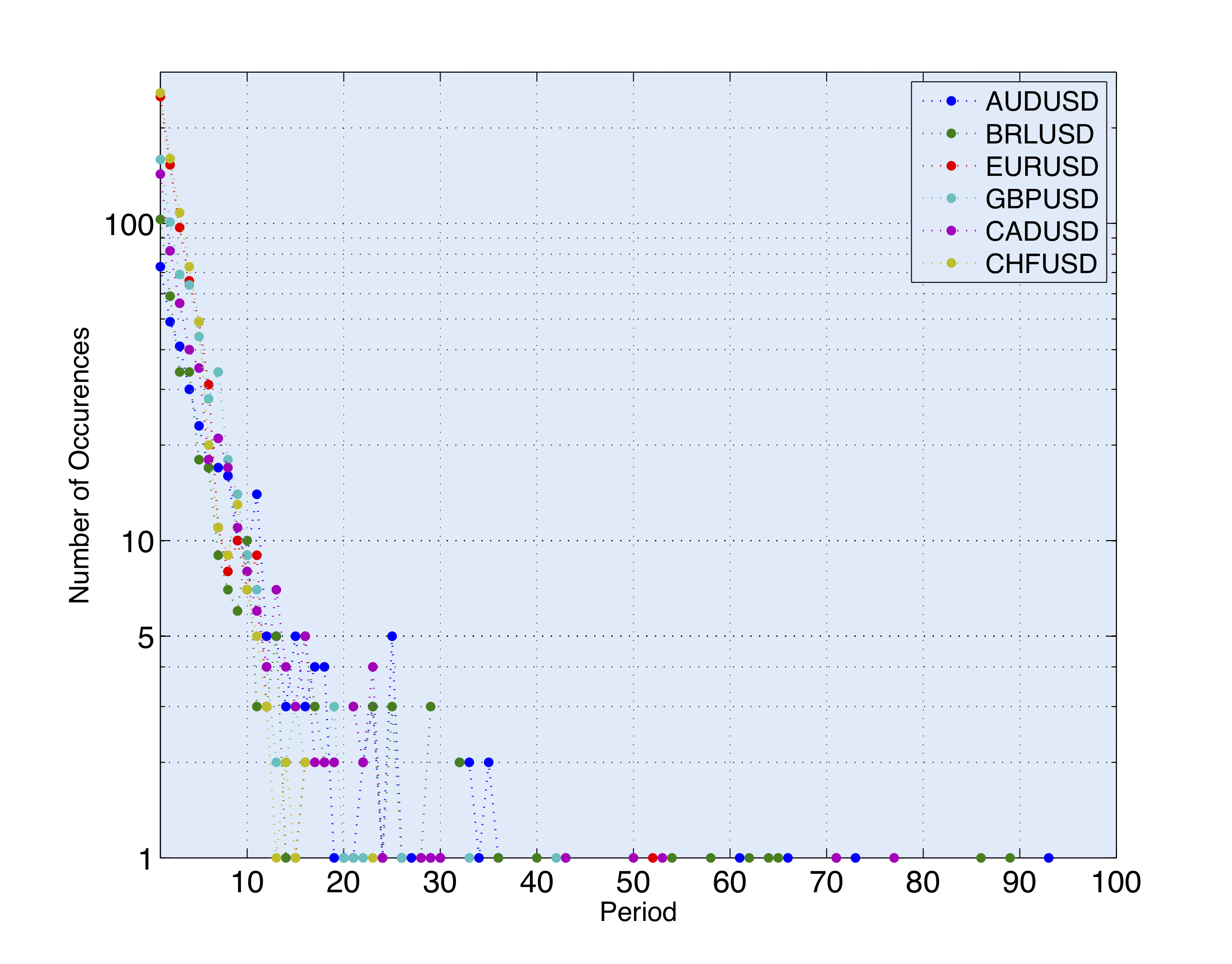}
      \caption{Duration of sequence remaining in the low volatility state 2 \label{fig_statedur2}}
\end{minipage}
\end{figure}

The distribution of state sequences staying in a given state is depicted in Figure \ref{fig_statedur1} for state 1 and in Figure \ref{fig_statedur2} for state 2.
The sequences remaining in one state follow the theoretical predictions. For longer periods there are marked deviations to the theoretical predictions for the underlying Markov sequences. 
The duration of same state sequences relates to the question of whether the HMM model is able to capture large drawdowns and large drawups over different lengths within such a sequences.

\newpage
 \section{Appendix - Currency State Series}
 \label{sec_allccystateseries}

\includegraphics[scale=0.4]{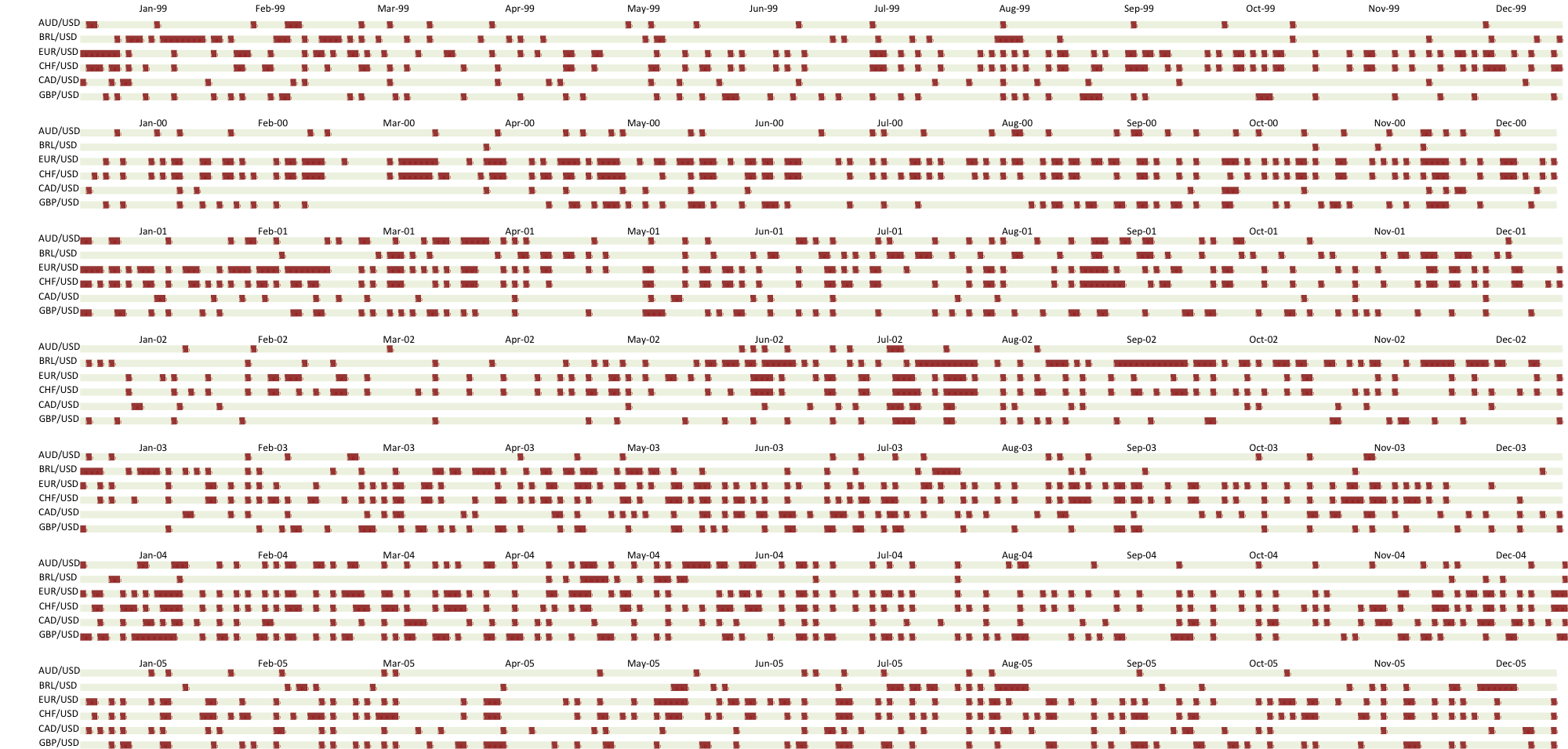}
\includegraphics[scale=0.4]{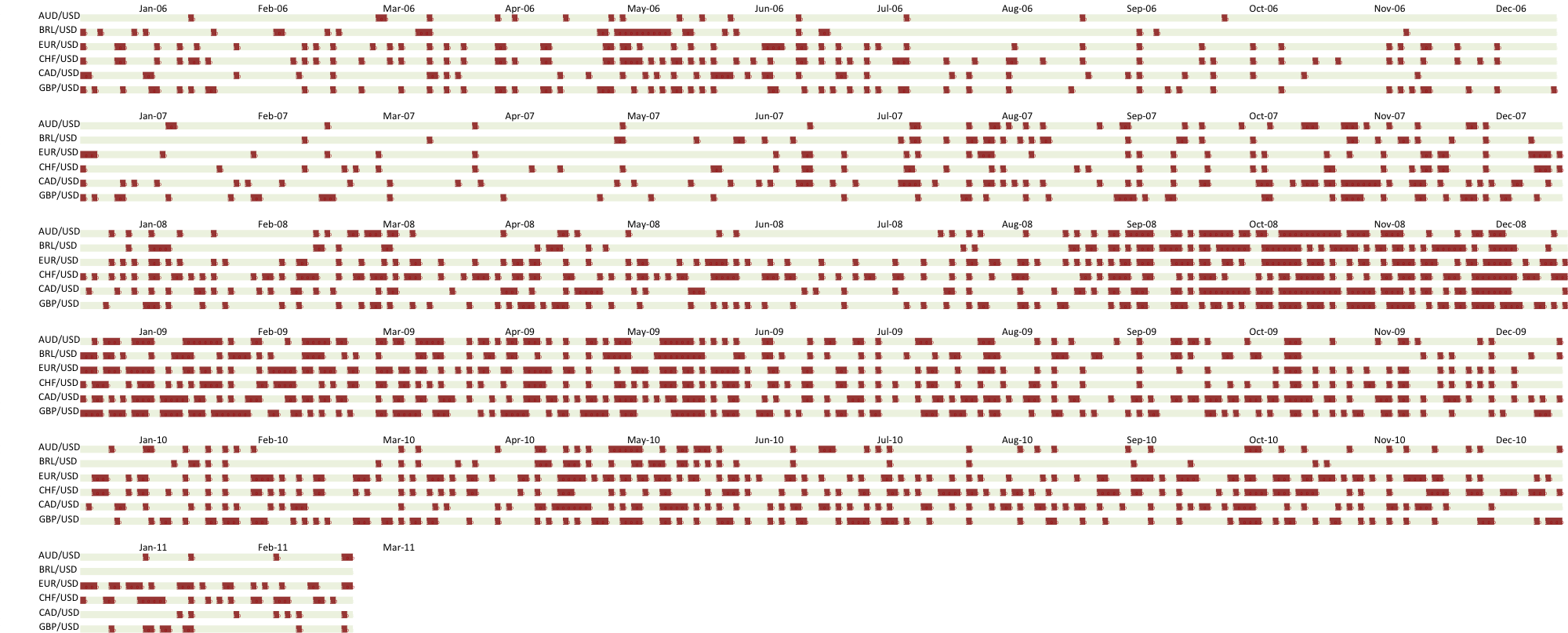}

\chapter{Volatility Timing of Hedge Funds and Entropy}\label{chap_paper3}
\section*{\centering \large Abstract}
This study uses the approximate entropy of S\&P realised volatility to detect changes in volatility regime in order to assess the market volatility timing and investing strategies of US equity focus hedge funds.  A one-factor model is used, conditioned on the entropy of market volatility, to measure the dynamic of hedge funds equity exposure. On a cross section of 1,903 hedge funds, we find that, over the period from 2000 to 2014, these hedge funds adjust their exposures dynamically in response to changes in volatility regime. This adds to the literature on the volatility timing behaviour of hedge fund managers, but using entropy as a model independent measure of volatility regime change.
\clearpage
\section{Introduction}
Hedge fund managers are in general free to change trading strategies, to allocate capital to asset classes and to choose leverage. Hedge fund managers often reallocate their capital dynamically in response to a change in market environment. To pursue their investment goals, hedge fund managers scan the markets for signals and information that can give support to their analyses regarding the current and the future states of the market. From the information gathered, the fund manager forms an expectation of the future states of the market and develops a trading strategy to make profitable use of this knowledge. Absolute and relative price levels, market volatility and trends are information that play an important role in asset allocation and portfolio selection.

A high degree of irregularity and unpredictability in market prices and volatility inhibits the managers' ability to predict future development of the markets. Fund managers are more likely to choose lower levels of equity exposure when the market is in a state of heightened irregularity where no clear view can be formed on expected returns or volatility. In this case, {\it ceteris paribus}, the fund managers may respond by actively reducing the capital allocation of investments in risky assets or delaying further investments in equity markets. Conversely, fund managers will seek more (long or short) exposure to the market in an environment with higher regularity and predictability. 

We extend the market timing model of \citeN{Ferson:1996je}, which assumes funds' equity risk exposure is a linear function of the market risk factors. Various modifications to this baseline model are tested using the approximate entropy (ApEn) of market volatility series as a state variable of the market condition. ApEn captures serial correlation and it is sensitive to changes in market volatility. A drop in entropy could signal volatility regime change. A switch to high volatility state will prompt fund asset allocation response and severely curtail the manager's ability to future market movements.  In response to a change in volatility regime, the manager may increase or reduce equity risk exposure by adjusting the funds' market beta. Volatility is an important determinant for the level of market beta a fund is choosing. For instance, for funds wishing to keep their funds volatility stable over time targeting at a specific Sharpe ratio, the managers are likely to reduce the fund's beta when a switch to higher volatility regime is detected.

Hedge fund managers make good candidates for the study. They are lightly regulated investment managers who have a lot more flexibility in implementing an investment strategy than mutual funds managers. \citeN{Brunnermeier:2004vu} found hedge fund manager are among the most sophisticated investors, who dynamically adjusting their portfolio to changing markets. We use a cross section of hedge funds from Lipper Tass over the sample period from 2000 to 2014, who have a geographic focus on the US equity markets. As a volatility variable, we use the realized volatility based on five-minute price ticks and aggregated to the daily figure, as well as the VIX index for market implied volatility.

The current analysis adds to the market volatility timing research on hedge funds. Research on fund managers' volatility timing ability has been thus far concentrated on mutual funds, whereas our focus is on hedge funds. We will demonstrate the use of ApEn in detecting changes in the time series structure of volatility. Simulations based on popular econometric models of volatility clearly show the usefulness of ApEn measure in distinguishing different volatility patterns and in detecting volatility regime changes. On a sample of 1,903 hedge funds, we find that managers adjust the funds' market beta as volatility is settling into a new regime. Dead funds show no such market beta adjustment skill, differing from live funds where such an adjustment ability was noted. The results are consistent with the findings in \citeN{Busse:1999hi} on mutual funds. The results also show the usefulness of entropy as a model independent measure of changes in volatility regime and patterns.

There is an extensive literature on market timing models and conditional performance evaluation of actively-managed investment vehicles.
\citeN{Treynor:1966ui} were among the first to study whether investment manager were able to profitably time changes in market conditions. The authors formulated a model to test the timing ability, extending the CAPM by a quadratic term to capture the nonlinear exposure to the market, depending on an expected high or low market return. In the sample of 57 mutual funds, the hypothesis of market timing ability was accepted for just one fund. 
\citeN{Henriksson:1981gu} formulated a market timing model where the manager forecasts market excess return over the risk-free rate. Depending on the forecast, the manager adjusts the portfolio's beta. This is modelled as an option payoff on the market return, with the strike price equal to the risk-free rate. On the basis of this model, the authors developed parametric and non-parametric tests for an investment manager's forecasting skill.
\citeN{Ferson:1996je} formulated a model of market timing where the risk exposure evolves linearly with the market observables. In their analysis, various variables that are public information are used as market conditions. Using a monthly data set of mutual funds, the authors found evidence for several variables of a change in risk exposure in response to public information on the economy. Furthermore, they were able to show that an unconditional version of the \citeN{Treynor:1966ui} model and the  \citeN{Henriksson:1981gu} models are misspecified when applied to na\"{i}ve strategies, in which conditional market timing models are an improvement. 
In measuring performance, researchers have attempted to distinguish market timing ability from stock selection skill. These skills are not easy to separate. \citeN{Korajczyk:1986ik} found that if managers trade options or levered securities, evidence for market skill is spurious. The authors construct portfolios that "show artificial timing ability when no true timing ability exists". To test for market timing ability, \citeN{Jiang:2007he} used mutual funds holdings information to avoid spurious findings. On the basis of the holdings based test, the authors were able to find positive market timing ability in US equity mutual funds. Using daily information, \citeN{Bollen:2001jr} showed that mutual funds may possess market timing ability. The higher frequency information adds to the power of the tests used. Unfortunately, higher frequency information on a fund's performance is unavailable for hedge funds. Using modified versions of various market timing models with daily conditioning variables, \citeN{Patton:2013fs} found evidence of varying hedge fund risk exposures across and within months. Market timing ability has also been investigated in relation to fund managers investing in bond markets, as in the study by \citeN{Chen:2010da}.

Timing market volatility has also been actively researched. \citeN{Busse:1999hi} uses daily returns of mutual funds to investigate a fund's ability to time market volatility. His findings suggest that funds decrease market exposure when market volatility is high. The value of this timing ability is in higher Sharpe ratios for successful volatility timing managers. The author further shows that surviving funds show sensitivity to market volatility, whereas non-surviving funds do not show this sensitivity. The economic value of volatility timing for investors has been investigated by \citeN{Fleming:2001jt}. Using conditional analysis on short horizon asset allocation strategies, the authors found volatility timing strategies outperform unconditional static strategies that have the same target return and volatility. 

Various information theoretic quantities have been used to measure and analyse the irregularity, patterns and serial structure of time series. \citeN{1991PNAS...88.2297P} 
used approximate entropy (ApEn) to investigate and classify systems in terms of their complexity. ApEn is a family of measures that is able to differentiate varying degrees of serial correlation within a time series. For specific stochastic processes, such as stationary processes, analytical formulas for the calculation of ApEn are shown (see \citeN{1991PNAS...88.2297P}). In a series of papers, including \citeN{Pincus:1995bb}, \citeN{Pincus:1992fn} and \citeN{1992PNAS...89.4432P}, various statistical qualities of ApEn have been investigated. So far there have been few applications of ApEn to address economics research questions. \citeN{Pincus:2008tb} showcases the usefulness of ApEn as a model independent measure of serial structure and irregularity in financial and economic time series. \citeN{Duan:2010vi} investigate empirical changes in patterns of volatility and the degree of predictability using daily data of equity, foreign exchange, commodity and interest rates. The authors found that ApEn is useful for measuring predictability of a time series and found that longer sample period increases the degree of predictability. The investigated time series exhibited a entropy value different from that of a random walk, from which the authors concluded that the financial time series is to some degree predictable.

 \FloatBarrier
\section{Entropy of Volatility}
\label{sec_apent_ch3}
Various information measures have been introduced to enhance our understanding of stochastic time series. In section \ref{sec_apent}, we introduced  approximate entropy, $\text{ApEn}(m,r,N)(u)$, as a measure for the regularity or dependence structure of a time series. 
Let $u$ be a sequence of numbers $u(1), \ldots u(N)$ of length $N$. Given a non-negative number $m$ with $m \leq N$, we form $m$-block of subsequences $x(i) \equiv (u(i), u(i+1) \ldots , u(i+m-1))$. The distance between two blocks is measured by $d(x(i),x(j)) \equiv \text{max}_{k=1,2, \ldots m} (\vert u(i+k-1) - u(j+k-1) \vert)$. For a given block $x(i)$ we count the fraction of block $x(j)$ that have a distance of less than $r$ to the given block $C^m_i (r) $, to quantify the regularity of a particular pattern.  With $\Phi^m(r) = \frac{1}{N-m+1} \sum_{i=1}^{N-m+1} \text{log} C_i^m (r)$ the approximate entropy $\text{ApEn}(m,r,N)$ is defined as follows.
\begin{align}
\text{ApEn}(m,r,N)(u) &= \Phi^m(r) - \Phi^{m+1}(r) \qquad , \thickspace m\geq 0 \label{eq_apen} \\
\text{ApEn}(0,r,N)(u) &= - \Phi^1(r) 
\end{align}

We proved the following formula in section \ref{sec_apent}, which supports the intuition on ApEn.
\begin{equation}
\label{eq_minusApEn}
- \text{ApEn}(m,r,N)(u) \approx \frac{1}{N-m}  \sum_{i=1}^{N-m} \text{log} \left( \frac{C_i^{m+1} (r)}{C_i^{m} (r)} \right) 
\end{equation}
Equation (\ref{eq_minusApEn}) shows that ApEn measures the logarithmic likelihood that sequences of patterns that are close for $m$ observations will remain close for the next $m$ observations. ApEn reflects persistence, serial correlation and regularity. Lower ApEn values correspond to higher persistence, stronger dependence and predictability, and \textit{vice versa} for higher ApEn values.

Here, we are concerned with the dynamics of volatility. The following sub-sections will showcase the behaviour of approximate entropy of volatility using models that are well known for capturing the salient features of financial markets volatility. 
Although it is possible to deduct analytically tractable expressions for the approximate entropy of specific processes (see equation (\ref{eq_apeniid})), this is not, in general, the case for the random processes we are concerned with in this analysis. Therefore, we will use simulations to examine how the entropy of volatility changes with respect to changing volatility level and persistence.

In line with \citeN{Pincus:2004vi}, the threshold $r$ is set equal to $0.2 \hat{\sigma}$ with $\hat{\sigma}$ being the sample standard deviation of the series being investigated. If the observed series is volatility, then $\hat{\sigma}$ is the volatility of volatility. The chosen threshold value strikes a balance between statistical validity and the amount of data available. A larger $r$ achieves a smaller standard error and a lower bias, but at the expense of losing much information on the data generating process. By normalising $r$ using the estimated volatility $\hat{\sigma}$ of the time series, ApEn becomes a scale and translation invariant measure. This makes ApEn a clean measure for irregularity complementing a measure for variability, such as standard deviation. Constructed in this way, one can compare irregularity in data sets that have very different volatilities.\footnote{Again, it is worth pointing out that the time series in this chapter to which we apply ApEn to are various volatility measures. Since we are interested in the regularity of volatility; all the 'volatility' mentioned here is in fact 'volatility of volatility'.}

\subsection{GARCH(1,1)}
Since the introduction of the General Auto Regressive Conditional Heteroscedacticity $\text{GARCH}(p,q)$ model by \citeN{Baillie:1989uf}, it has become one of the most widely used volatility models for financial time series. The GARCH model is easy to estimate and it captures the salient features of volatility, namely clustering and mean reversion, as well as fat tailed asset returns distribution. The most popular choice of the $\text{GARCH}(p,q)$ specification has been the GARCH(1,1) model, defined as follows: 
\begin{equation}
\sigma_t^2 = \alpha_0 + \alpha_1 \cdot a_{t-1}^2 + \beta_1 \cdot \sigma_{t-1}^2 \label{eq_garch}
\end{equation}
\noindent where $a_t = r_t - \mu_t = \sigma_t \cdot \epsilon_t $ is the stock return innovations at time $t$ for return series $r_t$ with mean $\mu_t$.\footnote{The reader is reminded that that $r_t$ with a time subscript denotes returns, whereas $r$ without a subscript denotes the threshold of the approximate entropy measure.} 

To determine the 'regularity' of volatility from a GARCH(1,1) process, one can calculate  $\text{ApEn}(m=1,r,N)$ of the squared innovations $\{a_t^2\}$. Assuming $\vert a_{t+j-1}^2 - a_{t-1}^2 \vert < r$, we have from equation (\ref{eq_garch}),
\begin{align}
\vert \sigma^2_{t+j}-\sigma^2_{t} \vert &= \vert \alpha_0 + \alpha_1 a_{t+j-1}^2 + \beta_1  \sigma_{t+j-1}^2 - (\alpha_0 + \alpha_1 a_{t-1}^2 + \beta_1  \sigma_{t-1}^2) \vert \nonumber \\
&= \vert \alpha_1 (a_{t+j-1}^2 - a_{t-1}^2) + \beta_1 (\sigma_{t+j-1}^2 - \sigma_{t-1}^2) \vert \nonumber \\
&\leq \alpha_1 \vert a_{t+j-1}^2 - a_{t-1}^2 \vert + \beta_1 \vert \sigma_{t+j-1}^2 - \sigma_{t-1}^2 \vert \nonumber \\
& \leq \alpha_1 r+ \beta_1 \vert \sigma_{t+j-1}^2 - \sigma_{t-1}^2 \vert  \nonumber 
\end{align}
Since $a_{t+j}^2= \sigma^2_{t+j} \epsilon_{t+j}^2$ ,
\begin{align}
\vert \text{E}(a_{t+j}^2) - \text{E}(a_{t}^2)   \vert &= \vert \sigma^2_{t+j}-\sigma^2_{t} \vert \\
&\leq \alpha_1  r+ \beta_1 \vert \sigma_{t+j-1}^2 - \sigma_{t-1}^2 \vert \label{eq_garchapen}
\end{align}

If $\vert \sigma_{t+j-1}^2 - \sigma_{t-1}^2) \vert < r$, then 
$\vert \text{E}(a_{t+j}^2) - \text{E}(a_{t}^2)   \vert \leq \alpha_1  r + \beta_1 r < r$ as  $\alpha_1+\beta_1 < 1$. 
Equation (\ref{eq_garchapen}) provides the conditions for a high likelihood that a succeeding pair of squared innovations to be close, when the preceding squared innovations of a GARCH process are close $\vert a_{t+j-1}^2 - a_{t-1}^2 \vert < r$. If the likelihood is high, then the entropy of $\{a_t^2\}$ is low, which means that the conditional volatility shows more regularity, that is, a higher degree of serial correlation. 

\begin{figure}[ht]
\centering
\includegraphics[width=300pt]{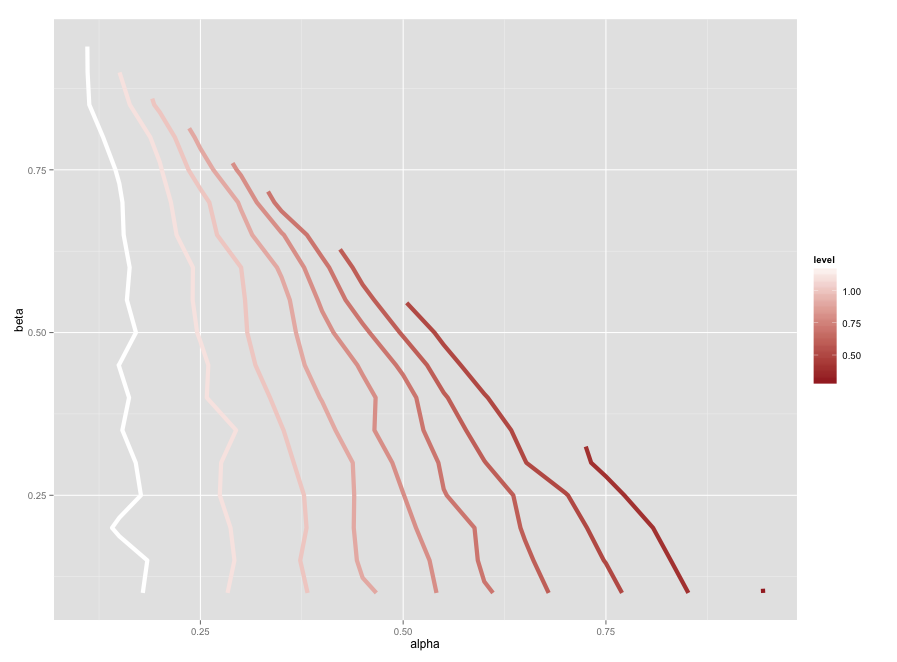}
\caption{ApEn: GARCH(1,1)}
\begin{quote}
$\text{ApEn}(m=1,r=0.2 \hat{\sigma},N=400)$ measured for $r_t^2$, where returns $\{r_t\}$ are generated from GARCH(1,1) process with normal distributed shocks, and varying combination of $\alpha_1$ (x-axis), $\beta_1$ (y-axis) with constraint $\alpha_1 + \beta_1 < 1$. Same levels of ApEn are connected by a contour line and graphed in different colours, where white corresponds to the highest ApEn values measured.
\end{quote}
\label{fig:garchapenall}
\end{figure}

To study how the parameters, $\alpha_1$ and $\beta_1$ impact on ApEn of squared returns,
we ran Monte Carlo simulations on GARCH(1,1) processes with various $\alpha_1$ and $\beta_1$ parameters  to produce a series with 400 time steps. The mean and standard error of the simulated $\text{ApEn}(m=1, r=0.2 \hat \sigma, N=400)$ were recorded. The ApEn values are shown in Figure \ref{fig:garchapenall} in a contour plot for all parameter combinations. Similar levels of ApEn are connected with lines. The upper right-hand corner of the plot is empty as only parameter combinations with $\alpha_1 + \beta_1 < 1$ are considered. 

The ApEn range from a low of 0.29 bit at $(\alpha_1,\beta_1)=(0.9,0.05)$ to a high of 1.25 bit at  $(\alpha_1,\beta_1)=(0.05,0.7)$. The recorded maximum, $1.26 \pm 0.08$ bit, matches that of a Brownian motion using the same ApEn parameters $(m=1, r=0.2 \hat{\sigma}, N=400)$. The lower (higher) ApEn levels are associated with higher (lower) $\alpha_1$, the auto regressive component of the squared shocks. Furthermore, $\alpha_0$, which reflects the overall unconditional variance of the GARCH process in equation (\ref{eq_garchapen}), has no impact on ApEn. 

\subsection{Changing Volatility Level} \label{subsec_volmodelsapen} 

\begin{figure}[ht]
\subfloat[$\sigma_1 =0.01$ up to time $t=500$ followed by $\sigma_2 = 0.05$]{
\begin{minipage}[b]{0.5\linewidth}
\centering
\includegraphics[width=2.8in]{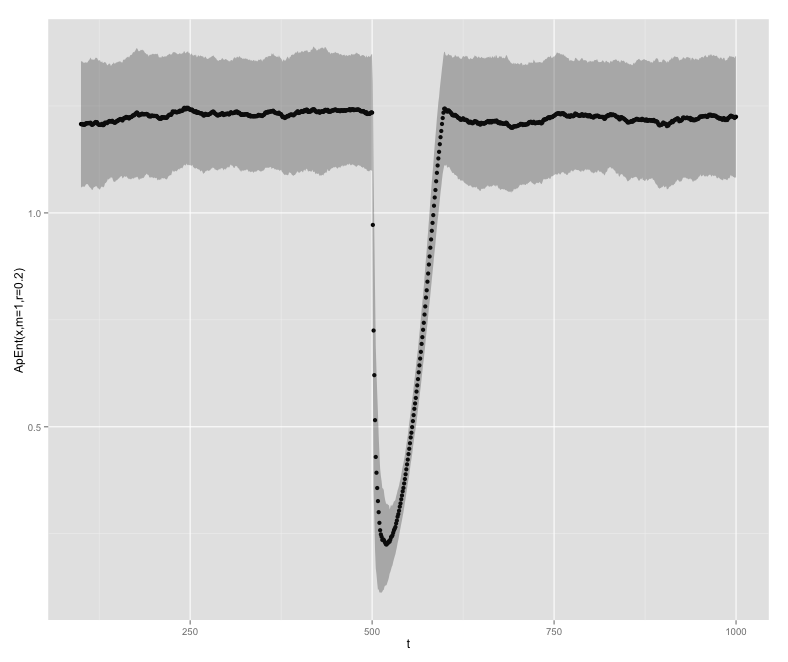}
\end{minipage}
\hspace{0.1cm}
\begin{minipage}[b]{0.5\linewidth}
\centering
   \includegraphics[width=2.55in]{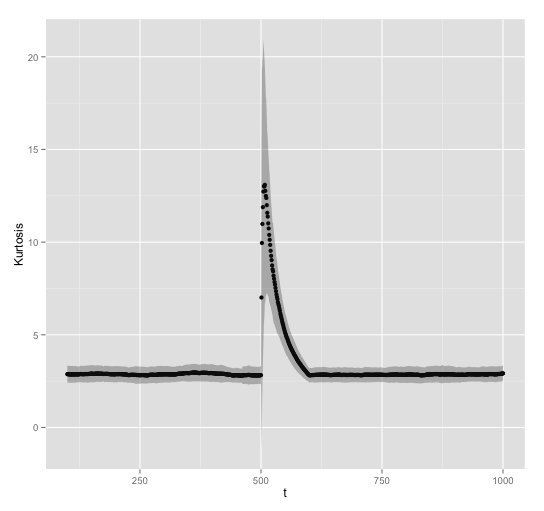}
\end{minipage}}
\hspace{0.0cm}
\subfloat[$\sigma_1 =0.05$ up to time $t=500$ followed by $\sigma_2 = 0.01$]{
\begin{minipage}[b]{0.5\linewidth}
\centering
\includegraphics[width=2.7in]{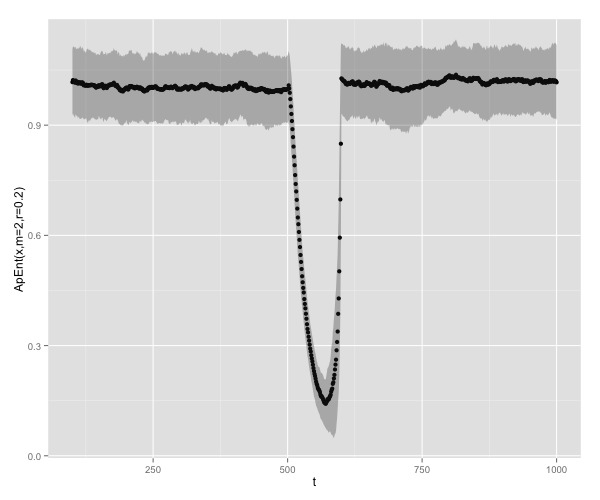}
\end{minipage}
\hspace{0.1cm}
\begin{minipage}[b]{0.5\linewidth}
\centering
   \includegraphics[width=2.7in]{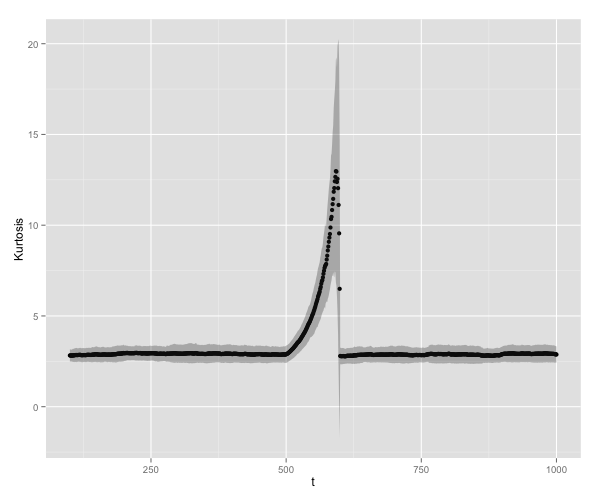}
\end{minipage}}
\caption{ApEn: Volatility regime change}
\begin{quote}
Random process $\{r_t \}$, $t \in \{1, \ldots 1000\}$ of normally distributed shocks, with a change in the distribution at $t=501$ as defined in equation (\ref{eq_volswitch}). The left picture shows $\text{ApEn}(1,r=0.2 \hat{\sigma}, N=100))$ for $\{r_t^2 \}$. The right picture shows the sample kurtosis $\hat{\kappa}=\frac{\mu_4}{\sigma^4}$ over the same rolling window of 100 time steps. 
\end{quote}
\label{fig:volswitchHL}
\end{figure}

We now turn our attention to the ability of ApEn to detect a change in the level of volatility in a time series. By this we mean a jump up or down in the level of volatility in the data generating process, similar to the empirically observed alternating high and low volatility clustering. We perform a simple numerical experiment which gives us insight into the dynamics of entropy when the level of volatility is shifting. A time series of 1000 time steps is simulated based on equation (\ref{eq_volswitch}) below:

\begin{equation}
\label{eq_volswitch}
r_{t} = 
  \begin{cases}
  r_t \in \mathcal{N}(0,\sigma_1^2) & t \leq 500  \\
  r_t \in \mathcal{N}(0,\sigma_2^2) & t > 500 
  \end{cases}
\end{equation}
The data generating process has a predetermined jump from a lower to a higher volatility $\sigma_1 < \sigma_2$ at the middle of the time series. ApEn is measured on the squared returns over the rolling time window: $\text{ApEn}(m=1, r=0.2\hat{\sigma}, N=100)$. As shown in the top panel (i) of Figure \ref{fig:volswitchHL}, for $t \leq 500$, $\text{ApEn}$ over the rolling window is stable at a level of $1.22 \pm 0.009$ bit. This is also the case for $t>601$, when the entire rolling window is in the high volatility state. For $500<t \leq 600$, when the rolling time window includes the point where the switch in volatility occurs, the $\text{ApEn}$ falls to a low of 0.22 bit at $t=520$. It is clear that the threshold $r=0.2\hat{\sigma}$ used in the calculation of $\text{ApEn}$ is sensitive to the sample standard deviation of the series (volatility of squared return in this case), and shifted upwards as the volatility (of volatility) increases. With a higher threshold, the likelihood of the observations (squared returns in this case) to stay close becomes higher for the lower volatility part of the return series, lowering the entropy. The volatility of squared returns is measured by the kurtosis, the fourth moment of the data generating process. In top panel (i) of Figure \ref{fig:volswitchHL}, the (standardised) kurtosis is plotted over time. It follows closely the shape of entropy in response to the step change in volatility: a sharp drop when crossing $t=500$, followed by a slow increase. This is not surprising as, apart from the step change in volatility, the data generating process is a random walk with \textit{iid} white noise. The process is not meant to exhibit any persistence. A similar picture emerges when the switch is from a high volatility state to a low volatility state as shown in the bottom panel, panel (ii), of Figure \ref{fig:volswitchHL}. Unlike the previous case, the increase in ApEn and kurtosis is characterised by a slower change, followed by a sharper reversion to the mean level. 

\begin{figure}[ht]
\centering
\includegraphics[width=400pt]{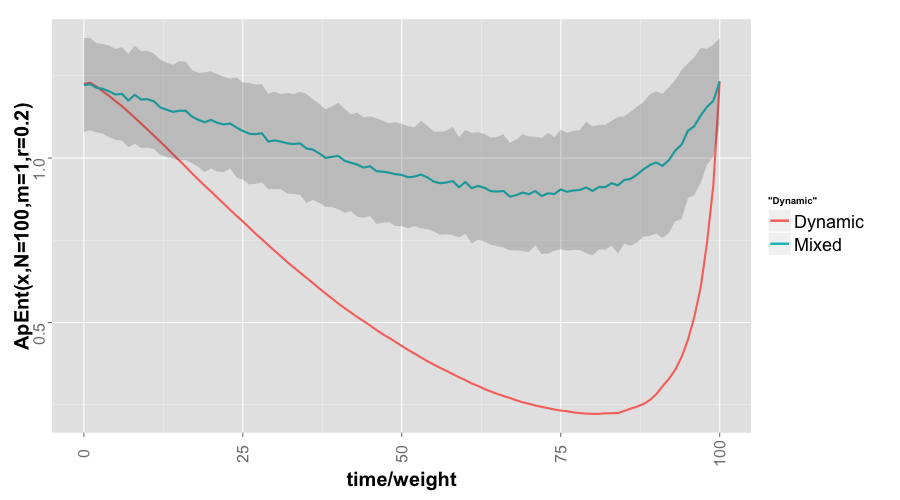}
\caption{ApEn: Regime change versus mixed normal distribution.}
\begin{quote}
Mean of $\text{ApEn}(m=1,N=100, r=0.2 \hat{\sigma})$ from monte carlo simulations of a mixed normal distribution $w_1 \cdot \mathcal{N}(0,\sigma_1^2) + w_2 \cdot \mathcal{N}(0,\sigma_2^2)$ (line 'Mixed') and for the rolling window of a regime change (line 'Dynamic') as in Figure \ref{fig:volswitchHL} (ii). The average of the estimated ApEn over 1000 simulations is shown, with the grey band around the mean ApEn of the mixed distribution showing the standard error. The mixing weights range from $w_2=1$, so drawing only from the high volatility distribution, to $w_1=1$ (all low volatility draws) on the right end. The rolling window of the dynamic regime change simulation runs accordingly from a rolling window with only high volatility squared returns, to one with only low volatility squared returns. 
\end{quote}
\label{fig:apendyn_mix}
\end{figure}

As mentioned above, the change in ApEn during a volatility regime change can be partly attributed to the change in the higher moments of the distribution in the rolling window. To understand the extent to which the serial correlation within the two regimes in this scenario plays a role, we set up an additional simulation, in which we take samples from a mixture of normal distributions. Specifically, we run numerical simulations, where we draw from a mixture normal distribution $w_1 \cdot \mathcal{N}(0,\sigma_1^2) + w_2 \cdot \mathcal{N}(0,\sigma_2^2)$, with varying weights $w_1+w_2=1$. The normal distributions have the same specification as in the previous experiment and the approximate entropy parameters are the same: $m=1, N=100, r= 0.2 \hat{\sigma}$. According to the different weights, a sample path is produced by (uniformly) randomly selecting one of the normal distributions to draw from, thus producing a time series of squared returns which are drawn from the same distributions in the same weighted manner as the volatility regime experiment specified in equation (\ref{eq_volswitch}).

Figure \ref{fig:apendyn_mix} shows both the result from this numerical experiment and the result from the volatility regime change experiment. The rolling time window starts and ends with a sample of 100 data points drawn from either one of the normal distributions. In between, the rolling window moves from the high volatility regime to the low volatility regime. The graph depicting the mixture distribution ('Mixed' in Figure  \ref{fig:apendyn_mix}) starts at the left, with weights from $w_1=0$, $w2=1$ (so selecting only from the higher volatility normal distribution) and ends with a mixture of $w_1=1$, $w_2=0$ at the right end of the figure.
One can see that the ApEn paths follow the same shape, but with the regime change time series 'Dynamic' following a more pronounced path. The lowest entropy in the mixture simulation is $\widehat{\text{ApEn}}=0.882577 \pm 0.1634$ bit, measured at $(w_1=0.67, w_2=0.33)$. 
The main difference between the two time series is the serial correlation structure in the volatility regime change example, which is absent in the simulations from a static mixed distribution. ApEn measures the log likelihood of two squared returns being close, conditional on the preceding two squared returns are also close. Drawn from a random choice of two distributions in the mixed distribution, the likelihood of staying close is only dependent on the shape of the distribution and not from the previous draw. The case of a volatility change has a serial structure when the change is within the rolling window.

\subsection{Markov Switching Volatility}

\citeN{Hamilton:1989jg} introduced a popular stochastic volatility model that allows the variance of returns to switch between discrete levels and the switching is determined by a hidden Markov chain. This specification retains from the GARCH(p,q,) model the important feature of volatility clustering but offers the added flexibility of different volatility persistence levels at different states. For a model with two states, denoted as $h_n = 1,2$, and normally distributed white noise in each state, the return process can be written as 
\begin{equation}
P(r_t \vert h_t = i) \simeq \mathcal{N} (\mu_i , \sigma_i) \qquad i=1,2
\label{eq_hmmemission}
\end{equation}

The switching between the two states is driven by a first order Markov process in which the transition probability between the states, $P(h_{t+1} \vert h_{t} )$, is given by the following matrix, the elements of which are the transition probability $T_{i,j} = P(h_{t+1} = i \vert h_{t}=j )$,
\begin{equation}
T = 
\begin{pmatrix}
P(1 \vert 1)& P(2 \vert 1)\\
P(1 \vert 2)& P(2 \vert 2)
\end{pmatrix}
=
\begin{pmatrix}
p& 1-p\\
1-q& q
\end{pmatrix}
 \label{eq_markovmatrix}
\end{equation}

\begin{figure}[ht]
\centering
\includegraphics[width=300pt]{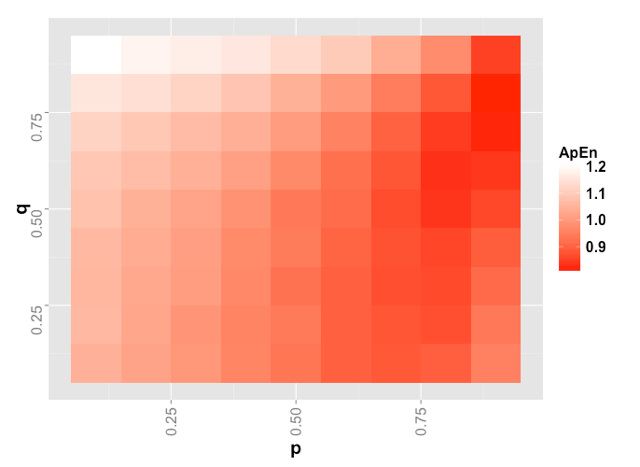}
\caption{ApEn: Markov switching process}
\begin{quote}
Numerical simulation of Markov process with combinations of transitions probabilities $(p,q)$ in equation (\ref{eq_markovmatrix}). For each combination, the mean $\text{ApEn}(1,r=0.2 \hat{\sigma}, N=400)$ of the numerical simulations is reported.
\end{quote}
\label{fig:volswitch}
\end{figure}

We simulate and estimate the impact of transition probabilities on the ApEn measure.  The time series of returns are drawn from a normal distribution in the numerical experiment, with the state dependent volatility controlled by the transition matrix.  In general, if the finite-state Markov chain is irreducible and aperiodic, the stationary distribution is unique. Given any starting distribution, the distribution of $h_t$ tends to the stationary distribution as $t \rightarrow \infty$ which can be easily calculated given the transition matrix. We simulated a sample length of 400 data points using various specifications for the transition matrix in equation  (\ref{eq_markovmatrix}) and non-zero transition probabilities. State 1  is the low volatility state with $\sigma_1=0.1$ and state 2 is the high volatility state with $\sigma_2=0.5$. The estimated $\text{ApEn}(m=1, r=0.2\hat{\sigma}, N=400)$ for the squared returns, presented in Figure \ref{fig:volswitch}, range from a low of $0.8075 \pm 0.0933$  bit at $(p,q)=(0.9.0.8)$ to a high of $ 1.2047 \pm 0.0883$ bit at $(p,q)=(0.1.0.9)$. 

The maximum ApEn of a Markov switching process is associated with a low probability of switching from the high to the low volatility state $(P(1 \vert 2)=1-q=0.1)$ and a high probability of the reverse transition ($P(2 \vert 1)=1-p=0.9$). For larger sample sizes, where the distribution of states tends to the equilibrium, such a combination of transition probabilities is associated with a process that is often in high volatility state $ \frac{P(2\vert 1)}{P(1\vert2)  + P(2 \vert 1)}=0.9$ and seldom in the low volatility state $ \frac{P(1\vert 2)}{P(1\vert2)  + P(2 \vert 1)}=0.1$. Similarly, a low ApEn is to be expected for high probability of low volatility state. 
This is consistent with the pattern of ApEn response to a change in volatility shown above in Figure \ref{fig:volswitchHL}. 

 \FloatBarrier
\section{Data}
Monthly hedge fund data, such as return performances, investment style and assets under management (AuM) have been extracted from the Lipper Tass database for the January 2000 to December 2014 period. The Lipper database constitutes one the most extensive hedge fund data sources and has been widely used in academic research. Since 1994, the database also retains data on funds that went out of business, which we also include in our analysis. The information provided by any hedge fund to Lipper, or any other database, is self-reported, without any legal or regulatory obligation involved. The motivation for self-reporting is purely to get added recognition and legitimacy. As such, it gives rise to various biases in the data (\citeN{Fung:2000fg}). Hence, to correct for the well-documented survivorship and back fill biases, we select all funds with a minimum of 24 months reported performances and delete the first 12 months of fund performances at the time when the fund started self-reporting.

Only hedge funds are selected and not mutual funds. We further narrowed the selection to hedge funds that have the United States as the self-reported geographical area of focus. Volatility timing is most prevalent in equity markets. Consequently, we exclude all fixed income, credit or currency focus funds, which account for 334 of the 2,704 US focus hedge funds. Given the equity focus, the set of factors that are of potential relevance in explaining fund performance is reduced. Our final data set, after adjusting for the aforementioned biases and restrictions, contains 1,903 hedge funds.

\begin{figure}[ht]
\centering
\caption{Hedge Funds: AuM, live funds}
\begin{quote}
Hedge funds from January 2000 to December 2014
\end{quote}
\subfloat[The number of live funds]{
\includegraphics[width=5in]{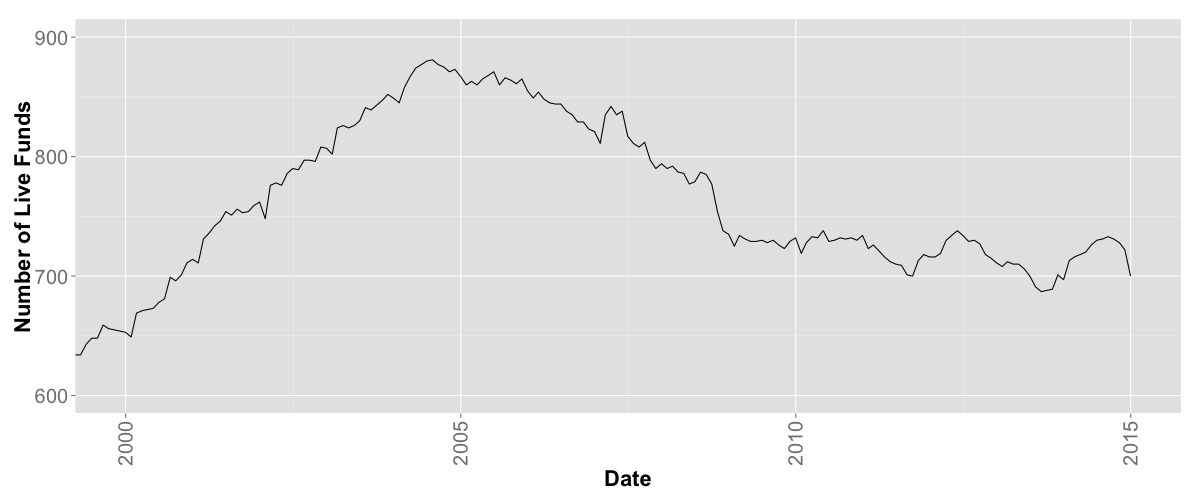}}
\hspace{0.0cm}
\centering
\subfloat[Assets under management (AuM, \$ million)]{
   \includegraphics[width=5in]{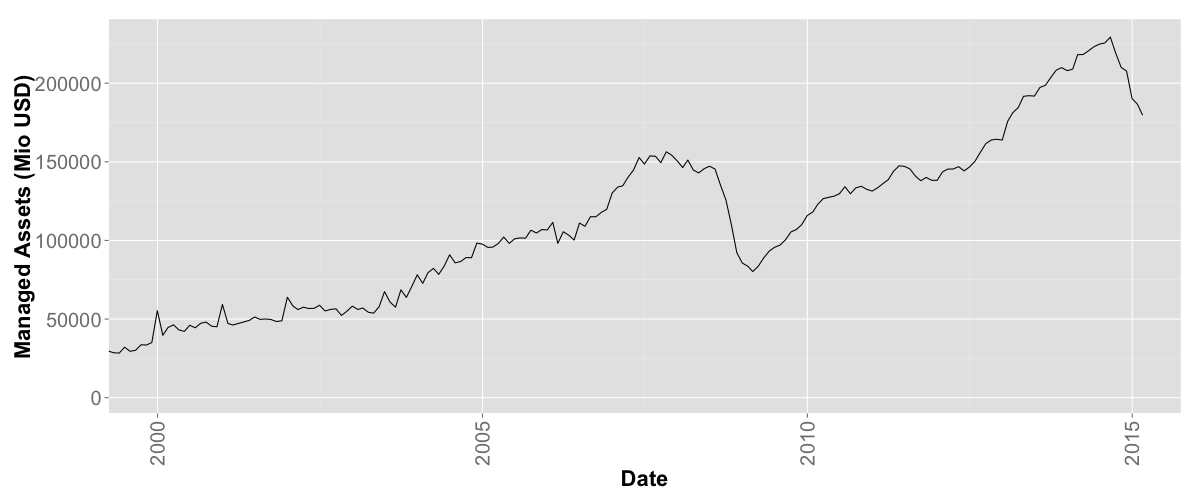}}
\label{fig:funds}
\end{figure}

The top diagram in Figure \ref{fig:funds} shows the number of live funds in the sample over time. A peak is reached in 2005 followed by a significant drop due to consolidation in the industry and the effect of the financial crisis of 2007, when a large number of funds went out of business. The bottom diagram of Figure \ref{fig:funds} shows the total amount of AuM in million USD managed by the funds included in the sample. Again, the financial crisis is clearly identifiable, marked with a substantial drop in AuM occurring in 2007 due to the double effects of capital outflows and losses on investments. Since then, there has been a continuous rise until the second half of 2014. Increased financial market volatility, which began in late Q3 2014 and was further heightened in Q4, resulted in redemptions by hedge fund investors.\footnote{HFR Global Hedge Fund Industry Report: Mid-Fourth Quarter 2014 from https://www.hedgefundresearch.com} This capital outflow is noticeable by the drop in AuM and number of live funds at the end of 2014, as shown in Figure \ref{fig:funds}.

In our sample, 753 funds have been liquidated at some point during the sample period, while 1,150 funds remain going concerns. The most popular subcategory, 'Long-Short Equity' accounts for 801 funds (including funds of hedge funds) in the sample. The summary statistics of the monthly returns are reported in panel A of Table \ref{t_basicstats3}. The average monthly return for all funds is 0.455\% and the 'Long-Short Equity' subcategory has a slightly higher mean return of 0.578\%.  The live funds have a higher average monthly return than that of the dead funds, but they also have a much higher standard deviation and a much longer left tail.

\begin{table}[!h]
\centering
{\scriptsize
\begin{tabular}{ l c  c  c  c  c  c  c  c }
\hline
 & N & Mean & Std & 10\% & 25\% & 50\% & 75\% & 90\%    \\
 \hline
\textit{Panel A: Fund Returns} & & & & & & & & \\
All Funds & 1903 & 0.455 & 8.762 & -4.33 & -1.18 & 0.5 & 2.1 & 4.68    \\
Long-Short Equity & 801 & 0.578 & 5.256 & -4.18 & -1.26 & 0.59 & 2.46 & 5.15    \\
Live & 1150 & 0.512 & 10.045 & -4.53 & -1.28 & 0.53 & 2.22 & 4.9    \\
Dead & 753 & 0.336 & 5.16 & -3.88 & -1.01 & 0.47 & 1.87 & 4.18    \\
  \hline
  \textit{Panel B: Volatility Levels} & & & & & & & & \\
VIX &  & 20.943 & 9.012 & 12.31 & 14.51 & 18.93 & 24.53 & 31.576    \\
$\sqrt{\text{RV}}$ &  & 14.888 & 9.974 & 6.754 & 8.623 & 12.353 & 17.674 & 25.584    \\
VIX-$\sqrt{\text{RV}}$ &  & 5.385 & 6.033 & 1.21 & 3.816 & 6.233 & 8.737 & 11.496    \\
  \hline
  \textit{Panel C: Entropy (N=60 days)}	 & & & & & & & & \\
  $\text{ApEn}(m=1, r=0.2 \hat \sigma)(\text{VIX}) $ &  & 0.988 & 0.238 & 0.66 & 0.829 & 1.009 & 1.163 & 1.286    \\
  $\text{ApEn}(m=1, r=0.2 \hat \sigma)(\sqrt{\text{RV}})$ &  & 1.154 & 0.284 & 0.749 & 0.99 & 1.215 & 1.376 & 1.453    \\
  $\text{ApEn}(m=1, r=0.2 \hat \sigma)(\text{VIX}-\sqrt{\text{RV}}) $ &  & 1.422 & 0.118 & 1.288 & 1.367 & 1.441 & 1.495 & 1.542    \\
    \hline
\end{tabular}}

\caption{Summary statistics}
\begin{quote}
This table presents summary statistics for the 1,903 hedge funds in our sample. Panel A summarizes returns for 'All Funds', 'Long-Short Equity' subcategory, and the 'Live' and the 'Dead' funds subsets. We report the number of funds N, mean, standard deviation and the quantiles from 10\% to 90\%. Panel B lists the statistics on the volatility measures used in percentage points and annualized. RV is the realised variance of S\&P 500, the square root of which is the realised volatility. VIX is the CBOE volatility index for S\&P 500.  Panel C shows the summary statistics for the estimated approximate entropy of the respective volatility time series calculated using a rolling window of 60 days.
\end{quote}

\label{t_basicstats3}
\end{table}

In this chapter, our focus is on the volatility of the US equity markets, for which we select the S\&P 500 index as a  representative. As a measure of actual volatility, we use the daily realized variance downloaded from the Oxford-Man Institute's Realized Library.\footnote{See http://realized.oxford-man.ox.ac.uk/. We use the daily realized varicance $RV_t$ calculated from five-minute price returns, $x_{j,t}$, as follows:
\begin{equation}
RV_t = \sum_{0 \leq t_{j-1,t} < t_{j,t} \leq 1} x_{j,t}^2 , \quad x_{j,t} = X_{t+t_{j,t}} - X_{t+t_{j-1,t}}
\end{equation}
where $X$ is the logarithmic of stock price, and $t_{j,t}$ is the normalised time of trade on $t$-th day. See \citeN{Shephard:2010ku} for details.}  Furthermore, the daily VIX index is downloaded from the CBOE (Chicago Board of Options Exchange) web site. VIX is a composite measure of market implied volatility, backed out from a series of options written on the S\&P 500. All volatility time series are annualized.\footnote{The realised volatility is annualized by multipliying $\sqrt{252}$, the square root of average trading days in a year and the result is expressed in percentage points. The VIX values reported are already annaulised.} Panel B of Table \ref{t_basicstats3} provides the summary statistics for the various volatility measures. It is clear that the VIX level is higher than the realised measure on average. Both have similar standard deviations and are right skewed, as evidenced by their means being greater than their respective medians.

 \FloatBarrier 
\section{Entropy of Realized and Implied Volatility}\label{sec_apenvol}
We estimate the ApEn of various volatility measures (i.e. VIX, $\sqrt{\text{RV}}$ and their difference) for the S\&P 500. As discussed in the previous section, ApEn suffers from a larger bias with smaller samples. To balance the need for more data points in order to produce an accurate entropy measure with the need to track the dynamics of volatility as it changes, we decided, after some trial and error, to use a rolling window of 60 days. In all cases, we set the threshold $r=0.2\hat{\sigma}$ and $m=1$ day. Panel C of Table \ref{t_basicstats3}  provides the summary statistics for the three sets of ApEn estimates. Of the three volatility measures, VIX has the lowest mean ApEn and the spread (i.e. the difference between VIX and $\sqrt{\text{RV}}$) has the highest average ApEn. In particular, the mean ApEn of the spread is high relative to its standard deviation, which may suggest a lesser degree of regularity for the spread.

\subsection{Realised Volatility}
\begin{figure}[ht]
\begin{minipage}[b]{0.5\linewidth}
\centering
\includegraphics[width=2.95in]{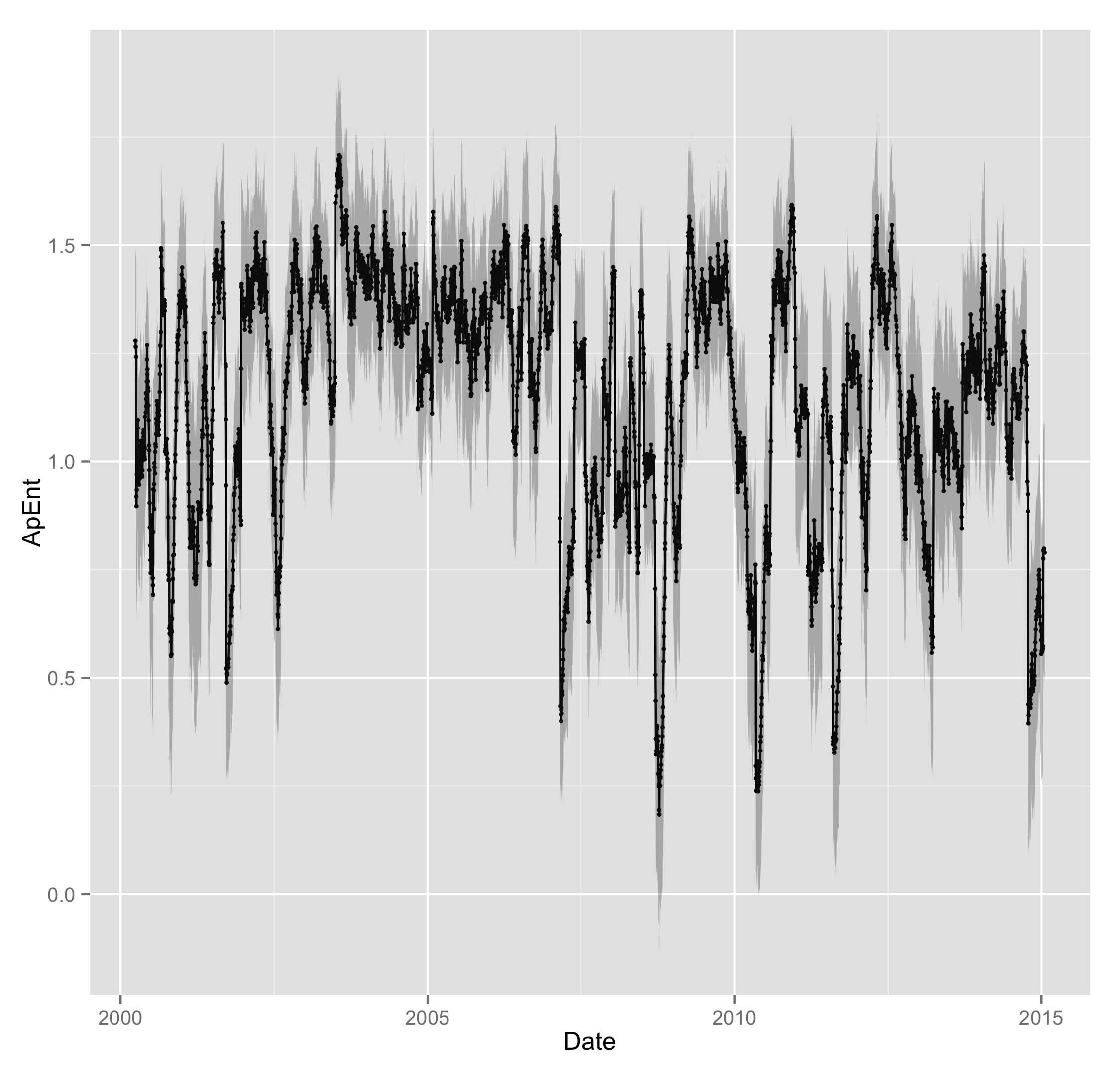}
\end{minipage}
\hspace{0.0cm}
\begin{minipage}[b]{0.5\linewidth}
\centering
   \includegraphics[width=2.95in]{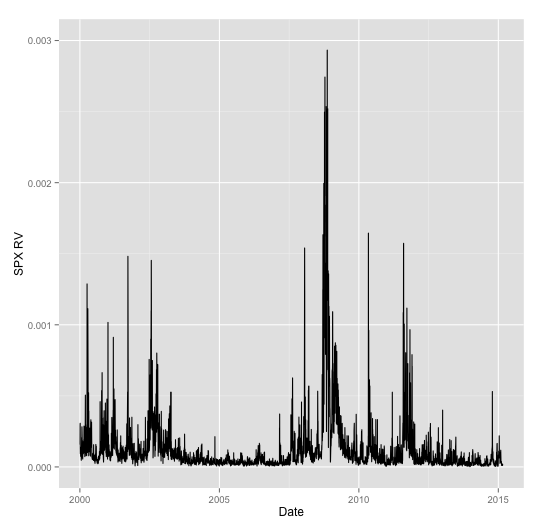}
\end{minipage}
\caption{Realized Variance}
\begin{quote}
The left diagram shows the entropy of realized variance $\text{ApEn}(m=1, r=0.2 \hat \sigma, N=60)(\text{RV})$, measured daily over a rolling window of 60 days. The right diagram shows the realized variance time series, which is truncated at $3 [10^{-3}]$, excluding the largest spike on October 10, 2010 where $\text{RV}=7.7 [10^{-3}]$.
\end{quote}
\label{fig:apentspxrv60d}
\end{figure}

Figure \ref{fig:apentspxrv60d} plots the estimated $\text{ApEn}(\text{RV})$ alongside the realized volatility time series itself over the January 2000 to February 2015 period. The entropy is calculated on the realized volatility level, i.e. it measures the conditional logarithmic likelihood that realized volatility levels that are close will remain close on the following day. A low ApEn signals higher regularity or a change in the volatility regime. From the left diagram of Figure \ref{fig:apentspxrv60d}, we can see that the $\text{ApEn}(\text{RV})$ reached the minimum of  0.18 bit on October 9, 2008, and a maximum of 1.7 bit on July 25, 2003. 

\begin{figure}[ht]
\begin{minipage}[b]{0.5\linewidth}
\centering
\includegraphics[width=2.95in]{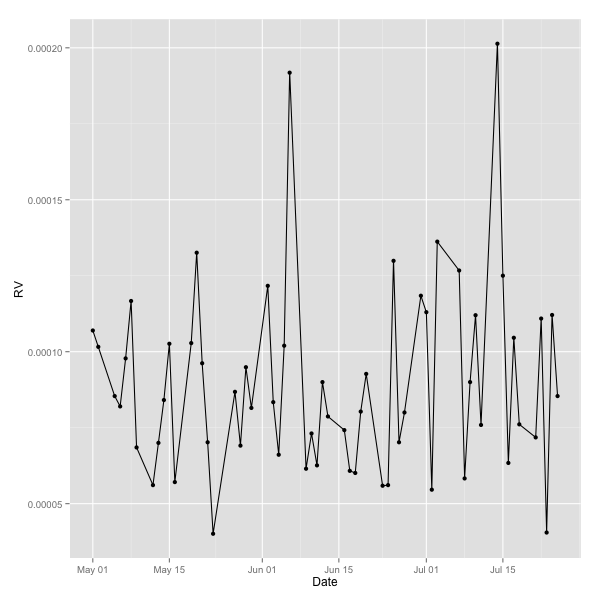}
\end{minipage}
\hspace{0.0cm}
\begin{minipage}[b]{0.5\linewidth}
\centering
   \includegraphics[width=2.95in]{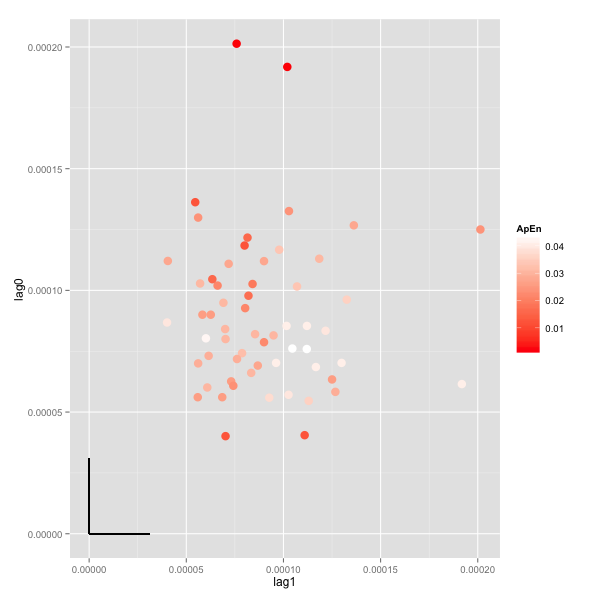}
\end{minipage}
\caption{RV: Maximum entropy episode}
\begin{quote}
Left figure shows the realized variance over the 90 days from 1 May to 25 July 2003. The entropy $\text{ApEn}(m=1, r=0.2 \hat \sigma)(\text{RV}_{\text{SP500}}) = 1.708231$ measured for this period has been the highest in the sample. The right picture shows the sequence of pairs of consecutive realized variances $(x=RV_{t-1},y=RV_t)$ as dots, coloured according to their contribution to the overall entropy of the sample.
\end{quote}
\label{fig:apentspxrv60dmax}
\end{figure}

To further investigate the above finding, Figure \ref{fig:apentspxrv60dmax} plots the RV over the 90 days period from May 1 to July 25, 2003. On the left diagram of Figure \ref{fig:apentspxrv60dmax}, the RV resembles a random walk, which is less predictable, and more irregular. The right diagram of Figure \ref{fig:apentspxrv60dmax} shows the lack of a relationship between RV on consecutive days.  These patterns are associated with a high entropy regime that fluctuates between 1 and 1.7 bit for the July 2003 to February 2007 period. Volatility during this period is consistently lower, with a few short-lived periods of slightly higher volatility. 

\begin{figure}[ht]
\begin{minipage}[b]{0.5\linewidth}
\centering
\includegraphics[width=2.95in]{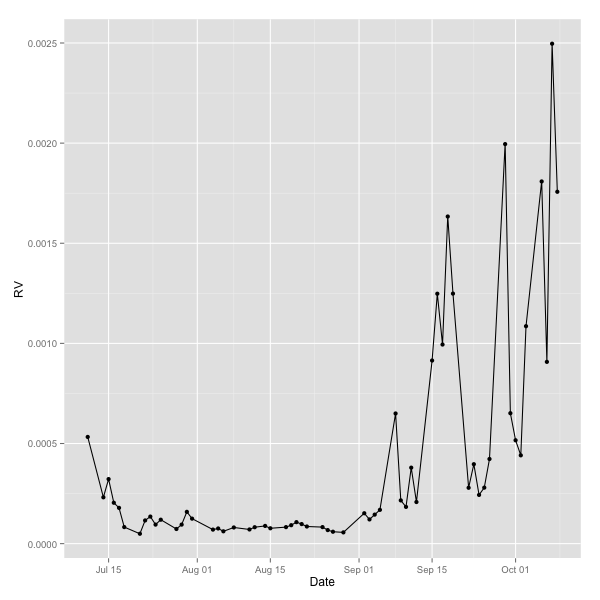}
\end{minipage}
\hspace{0.0cm}
\begin{minipage}[b]{0.5\linewidth}
\centering
   \includegraphics[width=2.95in]{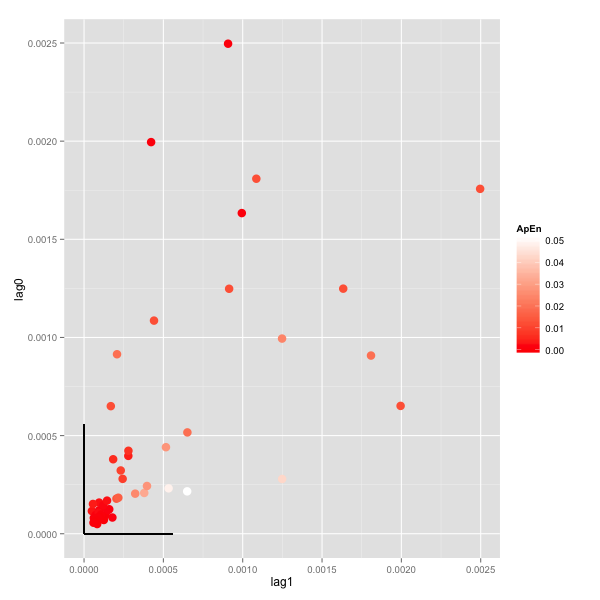}
\end{minipage}
\caption{RV: Minimum entropy episode}
\begin{quote}
The left diagram shows ApEn $(m=1, r=0.2 \hat \sigma, N=60)(\text{RV})$  from July 11 to October 9, 2008. The lowest ApEn is 0.18. The right diagram shows the sequence of pairs of consecutive realized variances $(x=RV_{t-1},y=RV_t)$ as dots, coloured according to their contribution to the overall entropy of the sample.
\end{quote}
\label{fig:apentspxrv60dmin}
\end{figure}

Figure \ref{fig:apentspxrv60dmin} plots the RV over the 60 days period from July 11 to October 9, 2008, on the left, which is marked by a low volatility initial phase as it entered into a high volatility state at the beginning of September 2008. The dramatic increase in volatility, followed by the volatility staying persistently high for a period of time is similar to that simulated in section \ref{subsec_volmodelsapen}.  The drop in entropy to 0.86 bit on the February 27, 2007 marks the beginning of a change in the volatility pattern, with significant shifts in volatility regimes. On this day, US equity markets dropped steeply in response to the concerns regarding domestic and international economic growth.  The right-hand diagram of Figure \ref{fig:apentspxrv60dmin} shows there is a great deal of dependence between consecutive days RV, which drove the ApEn values down.

\subsection{VIX}
\begin{figure}[ht]
\begin{minipage}[b]{0.5\linewidth}
\centering
\includegraphics[width=2.95in]{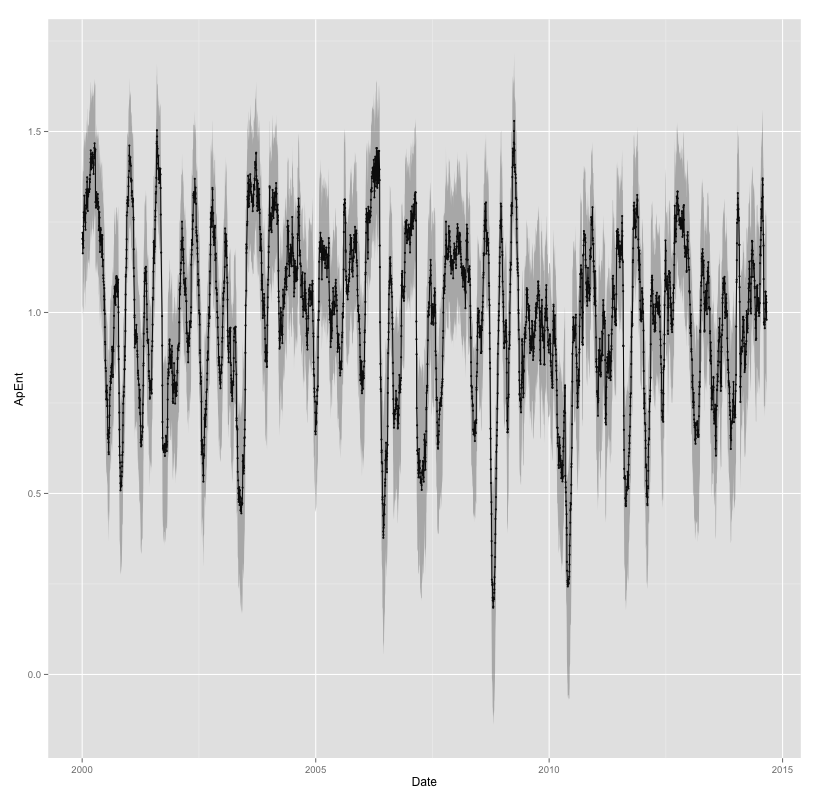}
\end{minipage}
\hspace{0.0cm}
\begin{minipage}[b]{0.5\linewidth}
\centering
   \includegraphics[width=2.95in]{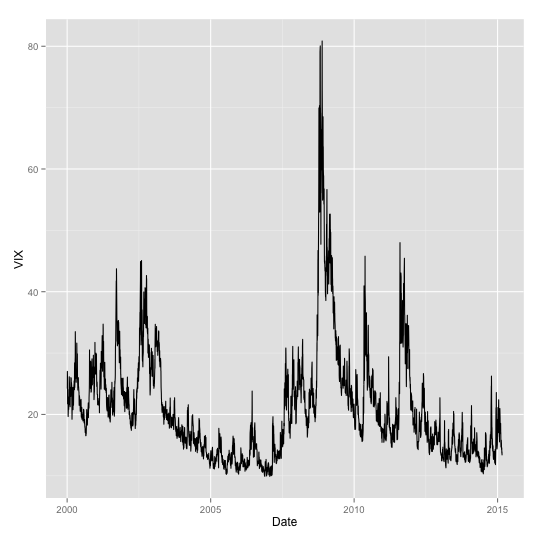}
\end{minipage}
\caption{VIX}
\begin{quote}
The left diagram shows $\text{ApEn}(m=1, r=0.2 \hat \sigma, N=60)(\text{VIX})$, measured daily over a rolling window of 60 days. The right diagram shows the VIX time series with daily close values.
\end{quote}
\label{fig:apentvix60d}
\end{figure}

The VIX index is a composite measure of the implied volatility of out-of-the-money put and call options on the S\&P 500. It is a forward-looking volatility measure, reflecting the market's expectation of volatility up to option maturity. Figure \ref{fig:apentvix60d} plots the ApEn of VIX over the 2000 to 2015 period alongside the time series of VIX level of the same period. Entropy $\text{ApEn}(\text{VIX}) $ ranges between a maximum of 1.52 bit on April 2, 2009 to a minimum of 0.18 bit on October 20, 2008. Compared with the ApEn of RV in figure  \ref{fig:apentspxrv60d}, the ApEn of VIX is consistently lower with little change. This suggests that VIX is more persistent.

\subsection{Spread}
\begin{figure}[ht]
\begin{minipage}[b]{0.5\linewidth}
\centering
\includegraphics[width=2.95in]{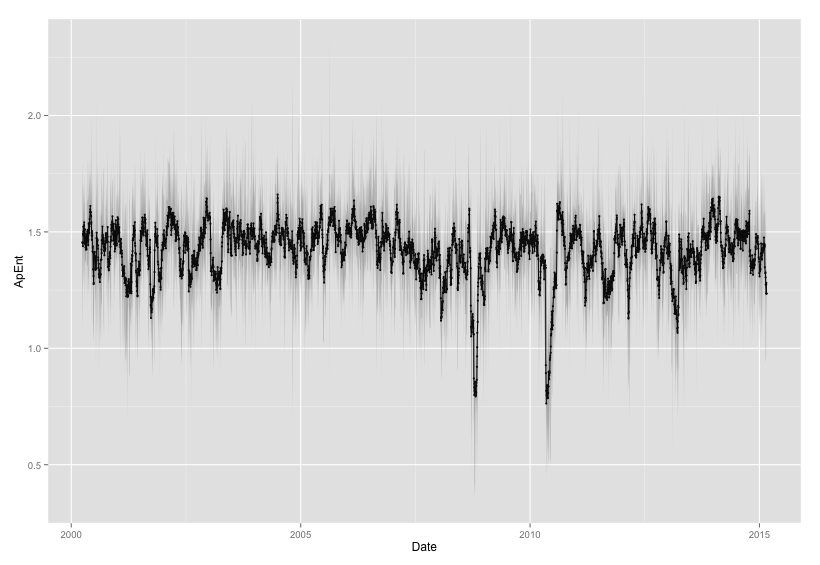}
\end{minipage}
\hspace{0.0cm}
\begin{minipage}[b]{0.5\linewidth}
\centering
   \includegraphics[width=2.95in]{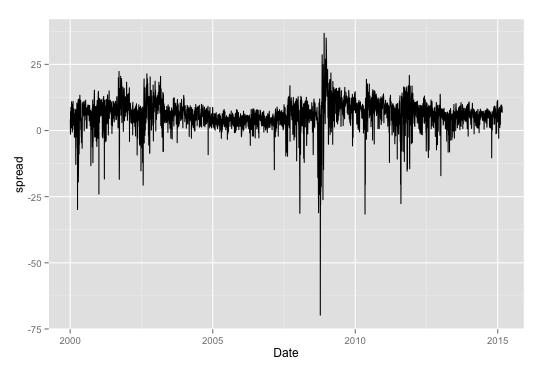}
\end{minipage}
\caption{Volatility Spread ($\sqrt{\text{RV}}- \text{VIX})$}
\begin{quote}
The left diagram shows $\text{ApEn}(m=1, r=0.2 \hat \sigma, N=60)(\sqrt{\text{RV}}- \text{VIX})$, measured daily over a rolling window of 60 days. The right diagram shows the time series of the spread.
\end{quote}
\label{fig:apentspxrvvixspread60d}
\end{figure}

We also analyse the spread between implied volatility and realized volatility. To make the two figures comparable, we annualise the realized volatility. The entropy is, as before, estimated on a rolling window of 60 days. In Figure \ref{fig:apentspxrvvixspread60d}, the left diagram shows the estimated approximate entropy and the right picture, the corresponding time series of the spread. The entropy hovers around the median of 1.4393 bit most of the time. Two periods stand out, however. In 2008, the entropy falls to 0.79 bit on October 24, 2008. Realized volatility in the markets spike around that time to 139 percentage points annualized. The VIX index spiked a few days later, leading to wild swings in the spread, and in turn this leads to lower entropy levels.  Later, the entropy falls to a low of 0.78 bit on May 25, 2010. Again, this started off as a low volatility regime for both RV and VIX. Then realized volatility first spiked, followed by the VIX index.
As a whole, there is a greater degree of irregularity in the spread series and this is confirmed by the high ApEn values. 

\FloatBarrier
\section{Empirical Results}
\subsection{Timing Model}
Linear factor models such as the Arbitrage Pricing Theory (see \citeN{Cochrane:2001tr}), or the three-factor Fama-French model (\citeN{Fama:1996ve}) are at the heart of the asset pricing literature. Given hedge fund returns, $r_t$, the standard functional form of the factor regression, used, for example, by \citeN{Fung:2004ty}, is of the form set out below:
\begin{equation}
r_t = \alpha + \beta^T \cdot r_{M,t} + \epsilon_t \label{eq_classicalfactormodel}
\end{equation}
\begin{equation}
r_t = \alpha + \sum_{i =1,\ldots N} \beta_i \cdot F_{i,t}   + \epsilon_t \label{eq_factor}
\end{equation}
where  $r_{M,t}$ is the factor returns and $\beta$ is the factor loading.

This factor model has been extended in various directions to capture the different characteristics of the investment strategies that hedge funds employ. One strand focusses on a manager's ability to forecast market conditions and on the basis of his forecasts, make adjustments to the market exposure of the fund. Timing models, such as those of \citeN{Treynor:1966ui} and \citeN{Ferson:1996je} among others, have been used to test a fund managers' ability to time markets. 

Following \citeN{Ferson:1996je}, we use a model with conditional variables that takes into account varying market conditions. The fund managers seeks exposure to the market by adjusting the market beta of his portfolio, to generate portfolio returns as follows:
\begin{equation}
r_{it} = \alpha_i + \beta_{it} \cdot r_{M,t} + \epsilon_{it} \label{eq_fersonschadt}
\end{equation}
where $r_{it}$ is the individual monthly excess return of fund $i$ at time $t$, $r_{M,t}$ is the excess return of the market over the same period, and $\epsilon_{it}$ is the residual error term. The returns are calculated to be in excess of the risk free rate, which we proxy with the returns on the one month Treasury bill. At time $t-1$, the fund manager uses the information on the market conditions available, $Z_{t-1}$, to forecast the market in the following period at $t$. Depending on the forecast the fund manager will adjust the the fund's market exposure. The time evolution of the market exposure $\beta_{it}$ is shown below: 
\begin{equation}
\beta_{it} =  \beta_{i} + \gamma_i \cdot Z_{t-1} \label{eq_fersonschadt_factor} 
\end{equation}
It is a Taylor expansion to the first linear approximation of $\beta_{it}$ in an additional variable $Z_{t-1}$. Equations (\ref{eq_fersonschadt}) and (\ref{eq_fersonschadt_factor}) combined to give the following factor model: 
\begin{equation}
r_{it} = \alpha_i + \beta_{it} \cdot r_{M,t} + \gamma_i \cdot r_{M,t} \cdot Z_{t-1}  + \epsilon_{it} \label{eq_fersonschadtcomb}
\end{equation}
The return on fund $i$ is determined by some factor $r_{M,t}$ with the factor loading depending on information $Z_{t-1}$. Using the previous period instead of the current period enables us to interpret the change in factor loading as an adjustment of the fund exposure in reaction to a change in the market conditions prediction.   

Several studies examined the fund manager's ability to forecast market returns or volatility. In this chapter, the forecasted market condition is the market volatility regularity. In particular, we focus on the regime change in volatility as measured by the approximate entropy of volatility of S\&P 500. In sections \ref{sec_apent} and \ref{sec_apent_ch3} we introduced ApEn as a model independent measure of the serial structure of a time series. Applied to volatility (see section \ref{sec_apenvol}), ApEn is sensitive to changes in the level of volatility and its persistence. We use the ApEn of volatility as the state variable of market condition, $Z$, in model  (\ref{eq_fersonschadtcomb}). 

\subsection{Panel Regression}
\citeN{Ferson:1996je} use publicly available information as conditioning factors. The selected factors, representing public information, have been shown to be useful in predicting security returns and risks over time. These include the lagged one-month Treasury bill yield, the dividend yield of the value-weighted CRSP equity index combining the New York Stock Exchange (NYSE) and the American Stock Exchange (AMEX), a measure of the credit spread of corporate bonds. 

In this chapter, we used entropy measures of volatility as conditioning variable, $Z$, to test if funds adjust their exposure in response to a change in the regularity of volatility or a volatility regime change. In particular we estimated the following models:
\begin{align}
r_{it} &= \alpha_i + \beta_{it} \cdot r_{M,t}  + \epsilon_{it} \label{eq_ch3_1}  \\
r_{it}	&= \alpha_i + \beta_{it} \cdot r_{M,t} + \gamma_i \cdot r_{M,t} \cdot Z_{\text{RV}_{t-1}}  + \epsilon_{it} \label{eq_ch3_2} \\
r_{it}	&= \alpha_i + \beta_{it} \cdot r_{M,t} + \delta_i \cdot r_{M,t} \cdot Z_{\text{ApEn}(\text{RV})_{t-1}}  + \epsilon_{it} \label{eq_ch3_3}  \\	
r_{it}	&= \alpha_i + \beta_{it} \cdot r_{M,t} + \nu_i \cdot r_{M,t} \cdot Z_{\text{ApEn}(\text{RV})_{t-1}} \cdot Z_{\text{RV}_{t-1}}   + \epsilon_{it} \label{eq_ch3_4} 
\end{align}
The conditioning variables $Z_{t}$ are derived from realized volatility $\text{RV}_t$ at $t$ and the entropy thereof. The ApEn calculated at time $t$ uses daily RV from the preceeding 60 days period. Other volatilities, such as the implied volatility, will be discussed later in this section. We follow the custom in the timing literature (\citeN{Ferson:1996je}) to demean and standardize the conditioning series, ApEn and RV, to make the inference independent of the scale of the variables.

Equation (\ref{eq_ch3_3}) is similar in structure to the Ferson Schadt model in (\ref{eq_ch3_2}), with the only difference being the choice of the conditioning variable. In equation (\ref{eq_ch3_2}), the conditioning variable is the volatility level, whereas in equation (\ref{eq_ch3_3}), it is the regularity of volatility with a low ApEn value associated with a higher regularity, indicating a possible change in volatility regime.

\begin{sidewaystable}
 \centering 
 \caption{Panel Regression Conditional CAPM}
 \begin{quote}
 Pooled OLS model estimation over the entire fund sample of various volatility timing models as defined in equations (\ref{eq_ch3_1}) to (\ref{eq_ch3_4}). The estimates for the various predictors along with the standard error are shown. The statistical significance of the various estimates is shown according to the associated p-value of a two tailed hypothesis test that the value is zero. In the lower part, statistics such as the $R^2$, adjusted $R^2$ and the F-statistics on the estimated model are shown.
\end{quote}
  \label{T:OLS_RV_1f}
  {\scriptsize
\begin{tabular}{@{\extracolsep{5pt}}lccccccc} 
\\[-1.8ex]\hline 
\hline \\[-1.8ex] 
 & \multicolumn{7}{c}{\textit{Dependent variable:}} \\ 
\cline{2-8} 
\\[-1.8ex] & \multicolumn{7}{c}{$r_{it}$} \\ 
\\[-1.8ex] & (\ref{eq_ch3_1}) & (\ref{eq_ch3_2}) & (\ref{eq_ch3_3}) & (\ref{eq_ch3_4}) & (\ref{eq_ch3_2}) \& (\ref{eq_ch3_3}) & (\ref{eq_ch3_2}) \&  (\ref{eq_ch3_4}) & (\ref{eq_ch3_3}) \& (\ref{eq_ch3_4})  \\ 
\hline \\[-1.8ex] 
  $ \beta_{it} \cdot r_{M,t}$& 0.187$^{***}$ & 0.202$^{***}$ & 0.200$^{***}$ & 0.222$^{***}$ & 0.214$^{***}$ & 0.247$^{***}$ & 0.208$^{***}$ \\ 
  & (0.006) & (0.006) & (0.006) & (0.008) & (0.006) & (0.008) & (0.011) \\ 
  $ \gamma_i \cdot r_{M,t} \cdot Z_{\text{RV}_{t-1}}$  &  & $-$0.031$^{***}$ &  &  & $-$0.028$^{***}$ &  & $-$0.023$^{*}$ \\ 
  &  & (0.004) &  &  & (0.004) &  & (0.012) \\ 
  $\delta_i \cdot r_{M,t} \cdot Z_{\text{ApEn}(\text{RV})_{t-1}}  $ &  &  & 0.035$^{***}$ &  & 0.032$^{***}$ & 0.042$^{***}$ &  \\ 
  &  &  & (0.005) &  & (0.005) & (0.005) &  \\ 
  $\nu_i \cdot r_{M,t} \cdot Z_{\text{ApEn}(\text{RV})_{t-1}} \cdot Z_{\text{RV}_{t-1}}   $  &  &  &  & $-$0.001$^{***}$ &  & $-$0.002$^{***}$ & $-$0.0004 \\ 
  &  &  &  & (0.0002) &  & (0.0002) & (0.001) \\ 
  $\alpha_i $  & 0.178$^{***}$ & 0.162$^{***}$ & 0.142$^{***}$ & 0.178$^{***}$ & 0.126$^{***}$ & 0.129$^{***}$ & 0.166$^{***}$ \\ 
  & (0.026) & (0.026) & (0.026) & (0.026) & (0.026) & (0.026) & (0.027) \\ 
 \hline \\[-1.8ex] 
Observations & 117,272 & 116,713 & 116,713 & 116,713 & 116,713 & 116,713 & 116,713 \\ 
R$^{2}$ & 0.009 & 0.009 & 0.009 & 0.009 & 0.010 & 0.010 & 0.009 \\ 
Adjusted R$^{2}$ & 0.009 & 0.009 & 0.009 & 0.009 & 0.010 & 0.010 & 0.009 \\ 
F Statistic & 1,065.425$^{***}$ & 553.230$^{***}$  & 551.065$^{***}$& 551.635$^{***}$ & 384.235$^{***}$ & 393.833$^{***}$  & 368.975$^{***}$\\ 
&  (df = 1; 117270) &  (df = 2; 116710) &  (df = 2; 116710) &  (df = 2; 116710) &  (df = 3; 116709) &  (df = 3; 116709) &  (df = 3; 116709) \\ 

\hline 
\hline \\[-1.8ex] 
\textit{Note:}  & \multicolumn{7}{r}{$^{*}$p$<$0.1; $^{**}$p$<$0.05; $^{***}$p$<$0.01} \\ 
\end{tabular}  }
\end{sidewaystable}

The regression method used is the pooled ordinary least squared regression. To test for the appropriate model specification, we ran a series of statistical tests for each regression equation. 
The Hausman test was used to decide between fixed or random effects, with the null hypothesis being that the preferred model is random effects against the fixed effects as the alternative. We also test if a pooled OLS model would be a more appropriate choice by performing a Lagrange multiplier test of individual fixed or time effects. The test results suggest that there is no substantial inter-fund variation. Accordingly, we use the pooled OLS model in all regressions. 

Table \ref{T:OLS_RV_1f} gives the empirical results for the models in  (\ref{eq_ch3_1}) to (\ref{eq_ch3_4}). The $F$ test (at the bottom of Table \ref{T:OLS_RV_1f}) rejects the null hypothesis for all regressions with combinations of conditional variables that the additional variables do not matter. 
The results for the CAPM base model, (\ref{eq_ch3_1}), as reported in the first column of Table \ref{T:OLS_RV_1f}, show a positive cross-sectional equity market exposure over time in the sample. Factor loadings for the predictor variable $\hat{\beta_{it}}=0.187 (0.006)$ and $\hat{\alpha_i }=0.178 (0.026)$ are positive and statistically significant. Given the overall equity focus of hedge fund strategies, this is as expected. The low value of the unconditional beta is explained by the fact that only a subset of the funds in the sample pursue an investment strategy which exclusively invests in equity markets. It is also plausible that the hedge funds are active in areas other than the US equity market. Following the unconditional regression, the CAPM model conditional on realized volatility is estimated. The results are shown in the second column of Table \ref{T:OLS_RV_1f}.
The estimate for the market condition variable $ r_{M,t} \cdot Z_{\text{RV}_{t-1}}$ is negative $ \gamma_i = -0.031 (0.004)$, confirming the results in \citeN{Ferson:1996je} that fund betas respond negatively to a rise in market volatility. 

We now turn to the discussion on the regression results using conditioning information from the entropy of volatility. The conditioning variable $Z_{\text{ApEn}(\text{RV})_{t-1}}$ in equation (\ref{eq_ch3_3}) is the standardized time series of $\text{ApEn}(\text{RV})$ calculated based on the realized volatility measured over a 60-day period. The results reported in the third column of Table \ref{T:OLS_RV_1f}, show similar factor loadings as the \citeN{Ferson:1996je} model in (\ref{eq_ch3_2}). All coefficients for the predictor variables test statistically significantly differently from zero.  The unconditional variables $\hat{\beta_{it}}$ and $\hat{\alpha_i }$ have lower factor loadings, whereas the conditional variable $Z_{\text{ApEn}(\text{RV})_{t-1}} $ has a higher factor loading $\delta_i = 0.035 (0.005)$, which is bigger than $\gamma_i$,  in absolute terms. The difference in sign is that a downward move in ApEn suggests a higher regularity and a move towards a change in the volatility regime, as was shown in the numerical experiments in section \ref{sec_apent_ch3}. A positive loading means the fund's beta reacts negatively (positively) to a decrease (increase) in the entropy of market volatility in the previous period. 

However, a decrease in entropy could be due either to a market volatility regime shift or less persistence in the volatility time series. To be able to distinguish between these two effects, we introduce a combined conditional variable. The combined factors in $\nu_i \cdot r_{M,t} \cdot Z_{\text{ApEn}(\text{RV})_{t-1}} \cdot Z_{\text{RV}_{t-1}}$ measure a funds beta adjustment when both entropy and volatility level change. The column marked (\ref{eq_ch3_2})  \& (\ref{eq_ch3_3}) shows the results. The estimates for $\hat{\beta_{it}}$ and $\hat{\delta_i}$ are similar compared to the simple entropy conditioning model in (\ref{eq_ch3_3}). The estimated combined conditional variable $\hat{\nu_i} =  -0.002 (0.0002)$ is statistically significant. It measures a fund beta response to a combined change in volatility and entropy of volatility. The negative estimate $\hat{\nu_i}$ is consistent with a positive beta response to a drop in volatility entropy and a negative response to an increase in volatility level, and \textit{vice versa}.  This suggests that, on average, fund managers respond to the market volatility settling into a new regime. Depending on whether the new regime has a low volatility or a high volatility regime, the fund manager will make a positive or negative market beta adjustment accordingly. In the last column, we report the results on the model with all combinations of conditional variables. All conditioning variables are statistically significant. The results show a more pronounced result due to the incorporation of all the individual market conditions; most of the factor loadings increased in absolute term. 

So, the refined model below is consistent with the empirical findings on the volatility timing ability of fund managers, and that in a situation when volatility settles into a new regime, the funds beta adjustment is the most pronounced;
\begin{align}
r_{it}	= \alpha_i + \beta_{it} \cdot r_{M,t} &+ \delta_i \cdot r_{M,t} \cdot Z_{\text{ApEn}(\text{RV})_{t-1}} \nonumber \\
 &  + \nu_i \cdot r_{M,t} \cdot Z_{\text{ApEn}(\text{RV})_{t-1}} \cdot Z_{\text{RV}_{t-1}}  + \epsilon_{it} \label{eq_refinedmodel} 
\end{align}

The basic and refined models in equations (\ref{eq_ch3_2} ) to (\ref{eq_ch3_4}) with combinations,  add an explanatory power as measured by $R^2$ and the adjusted $R^2$, which are listed in Table \ref{T:OLS_RV_1f}. The low $R^2$ could be due to some omitted variables and might be improved in a non-pooled regression. 
Both volatility timing models improved the fit marginally compared to the unconditional CAPM in terms of $R^2$.  

\subsection{Strategy and Survivorship}\label{subsec_strategy}

We now turn to the question of whether the managers, volatility timing behaviour in model (\ref{eq_refinedmodel}) is different when pursuing a specific strategy. The 'Long-Short Equity' investment strategy provides a good test in this respect. It is a strategy where the equity-related factors should be the most appropriate, and it is the most popular investment strategy represented in the fund sample with a sufficient number of funds for valid statistical inference (see Table \ref{t_basicstats3} for the summary statistics for the sample funds). The 'Long-Short Equity' returns come from directional as well as spread bets on the equity market. Exposure to the market can be either in long or short equity positions. Table \ref{T:OLS_RV_1f_subsample} show the results for the entropy volatility timing model. The funds market exposure $\hat{\beta_{it}} = 0.554 (0.006)$ is more than double of the sample containing all funds. This market beta is consistent with findings from \citeN{Fung:2011er}, who found, for a sample of long/short equity between 1994-2006, that the beta value was between $0.3$ to $0.6$. The sensitivity to a change in market volatility entropy $\hat{\nu_i}$ is smaller than in the overall sample. With a change in entropy in the market volatility, the fund manager is adjusting less compared with the overall sample. Overall, the results suggest that the 'Long-Short Equity' fund manager responds less forcefully to a change in volatility condition. 

\begin{table}[!htbp] \centering 
\caption{Panel Regression Conditional CAPM: Live, Dead and LS Equity Funds}
  \begin{quote}
  Pooled OLS model estimation of one-factor conditional model. Subsets of the fund sample used are: 'Long Short (LS) Equity', 'live' and 'dead' funds. The estimates for the various predictors along with the standard error are shown. The statistical significance of the various estimates is shown according to the associated p-value of a two tailed hypothesis test that the value is zero. In the lower part, statistics such as the $R^2$, adjusted $R^2$ and the F-statistics on the estimated model are shown.
  \end{quote} 
  \label{T:OLS_RV_1f_subsample} {\scriptsize
\begin{tabular}{@{\extracolsep{5pt}}lcccc} 
\\[-1.8ex]\hline 
\hline \\[-1.8ex] 
 & \multicolumn{4}{c}{\textit{Dependent variable:}} \\ 
\cline{2-5} 
\\[-1.8ex] & \multicolumn{4}{c}{$r_{it}$} \\ 
\\[-1.8ex] & All & LS Equity & Live & Dead\\ 
\hline \\[-1.8ex] 
 $ \beta_{it} \cdot r_{M,t}$ & 0.247$^{***}$ & 0.554$^{***}$ & 0.224$^{***}$ & 0.292$^{***}$ \\ 
  & (0.008) & (0.006) & (0.011) & (0.008) \\ 
  $\delta_i \cdot r_{M,t} \cdot Z_{\text{ApEn}(\text{RV})_{t-1}}  $ & 0.042$^{***}$ & 0.020$^{***}$ & 0.058$^{***}$ & 0.007 \\ 
  & (0.005) & (0.004) & (0.007) & (0.004) \\ 
  $\nu_i \cdot r_{M,t} \cdot Z_{\text{ApEn}(\text{RV})_{t-1}} \cdot Z_{\text{RV}_{t-1}}   $ & $-$0.002$^{***}$ & $-$0.002$^{***}$ & $-$0.001$^{***}$ & $-$0.002$^{***}$ \\ 
  & (0.0002) & (0.0001) & (0.0003) & (0.0002) \\ 
  $\alpha_i $ & 0.129$^{***}$ & 0.182$^{***}$ & 0.183$^{***}$ & 0.034 \\ 
  & (0.026) & (0.020) & (0.037) & (0.025) \\ 
 \hline \\[-1.8ex] 
Observations & 116,713 & 49,601 & 79,781 & 36,932 \\ 
R$^{2}$ & 0.010 & 0.217 & 0.006 & 0.054 \\ 
Adjusted R$^{2}$ & 0.010 & 0.217 & 0.006 & 0.054 \\ 
F Statistic & 393.833$^{***}$ & 4,585.394$^{***}$ & 156.242$^{***}$  & 706.728$^{***}$ \\ 
 &  (df = 3; 116709) &  (df = 3; 49597) &  (df = 3; 79777) &  (df = 3; 36928) \\ 

\hline 
\hline \\[-1.8ex] 
\textit{Note:}  & \multicolumn{4}{r}{$^{*}$p$<$0.1; $^{**}$p$<$0.05; $^{***}$p$<$0.01} \\ 
\end{tabular}  }
\end{table} 

We have re-estimated the models, also distinguishing funds which, by 2014, went out of business and those that are still active. The results are listed in Table \ref{T:OLS_RV_1f_subsample} and show significantly different behaviours among the two subsets. Managers of live funds do seem to adjust the portfolio beta in response to an expected change in market volatility more drastically ($\hat{\delta_i} =  0.058$) than dead funds ($\hat{\delta_i} =  0.007$). This finding is consistent with the findings in \citeN{Busse:1999hi} in the context of mutual funds, where the author shows that the market beta of surviving funds is sensitive to market volatility, which is not the case for non-surviving funds.

\subsection{Conditional Fama French Model}\label{subsec_famafrench}
In previous section \ref{subsec_strategy}, we investigated fund manager behaviour when the regularity of market volatility was changing with the conditional volatility model. We used a conditional CAPM model to find evidence of market beta adjustments when entropy is changing within the sample. Although the market beta of a fund is arguably the most important factor for an equity-focused fund, there are other risk factors that a fund is exposed to. 
The Fama French three-factor model is a widely used model in the literature  (\citeN{Fung:2011er}, and \citeN{Patton:2013fs}) to identify common risk factors in the risk exposure of equity focussed funds. It has a factor SMB, which denotes the excess returns of small cap stocks over large cap stocks. HML denotes the excess returns of a portfolio of high book-to-market stocks over returns from a portfolio with low book-to-market stocks. The third factor in the model is the market excess returns over a risk free asset. 

In the spirit of the conditional multi-factor models that were employed by \citeN{Ferson:1996je}, we extend in the current section the conditional CAPM model to a conditional three-factor model, incorporating the \citeN{Fama:1996ve} factors (SMB, HML) as additional factors into our model. 
The extended conditional model is used to show the robustness of the results from the previous section. Our Fama French volatility entropy timing models have the following specifications. 
\begin{equation} \label{eq_3F_condentropy_simple}
r_{it} = \alpha_i + \beta_{it} \cdot r_{M,t} + \beta^{\text{SMB}}_{it} \cdot r_{\text{SMB},t}  + \beta^{\text{HML}}_{it} \cdot r_{\text{HML},t} + \gamma_i \cdot r_{M,t} \cdot Z_{\text{ApEn}(\text{RV})_{t-1}} + \epsilon_{it} 
\end{equation}
\begin{align} \label{eq_3F_condentropy}
r_{it} =& \alpha_i + \beta_{it} \cdot r_{M,t} + \beta^{\text{SMB}}_{it} \cdot r_{\text{SMB},t}  + \beta^{\text{HML}}_{it} \cdot r_{\text{HML},t} + \gamma_i \cdot r_{M,t} \cdot Z_{\text{ApEn}(\text{RV})_{t-1}}  \nonumber \\
+&   \gamma_i^{\text{SMB}} \cdot r_{\text{SMB},t} \cdot Z_{\text{ApEn}(\text{RV})_{t-1}} +\gamma_i^{\text{HML}} \cdot r_{\text{HML},t}  \cdot Z_{\text{ApEn}(\text{RV})_{t-1}}  + \epsilon_{it} 
\end{align}
In Table \ref{T:OLS_RV_3f_subsample}, the results from the panel regression with this model are shown.  Column (1) of the table shows the results from the conditional CAPM of equation (\ref{eq_ch3_3}) and column (2) the from the Fama French model. A comparison of the adjusted and unadjusted $R^2$ shows that the Fama French model adds explanatory power, compared to the one factor conditional entropy model. Column (3) shows that a simplified version of the model in equation (\ref{eq_3F_condentropy_simple}), accounting only for an entropy change in the overall market, improves further the $R^2$. One notices furthermore, that the entropy factor loading $\gamma_i \cdot r_{M,t} \cdot Z_{\text{ApEn}(\text{RV})_{t-1}}$ is lower for the conditional three factor model (3) than that for the one-factor model (1), showing that the individual risk factors SMB and HML can account for a part of the factor loading. The resulting $\hat{\gamma}_i = 0.008 (0.004)$ is according to the t-test statistically significant, providing evidence that the results from the one-factor model are robust also to an enhanced risk factor model. 
The results from the fully specified model, as in equation (\ref{eq_3F_condentropy}), are shown in column (4) of Table \ref{T:OLS_RV_3f_subsample}. Taking into consideration also the changes to the risk factors SMB and HML in reaction to a change in entropy, the sensitivity to the overall market volatility, $\hat{\gamma}_i = 0.004 (0.004)$ is no longer statistically significant. However, $\hat{\gamma}_i^{\text{SMB}}$ and  $\hat{\gamma}_i^{\text{HML}}$ are statistically significant suggesting that 
the Long Short equity managers in the sample change the risk factor exposures to SMB and HML in response to an expected volatility regime change, with only marginal adjustments being made in the overall market beta of the portfolio.
This sheds also some light on our finding from section \ref{subsec_strategy}. There, we found that Long Short Equity funds responded less strongly to a change in market volatiliy regularity than the average fund in the sample. The factor loadings suggest that the adjustments in the portfolio in reaction to a change in market conditions are not picked up by the market beta, but instead by other risk factor exposure changes. 

\begin{table}[!htbp] \centering 
  \caption{Panel Regression Conditional 3F model: LS Equity funds.} 
 \begin{quote}
 Pooled OLS model estimation of conditional 1-factor and 3-factor models for  Long Short (LS) Equity funds as defined in equations (\ref{eq_ch3_3}) and (\ref{eq_3F_condentropy}). The estimates for the various predictors along with the standard error are shown. The statistical significance of the various estimates is shown according to the associated p-value of a two tailed hypothesis test that the value is zero. In the lower part, statistics such as the $R^2$, adjusted $R^2$ and the F-statistics on the estimated model are shown.
 \end{quote}
  \label{T:OLS_RV_3f_subsample} {\scriptsize
\begin{tabular}{@{\extracolsep{5pt}}lcccc} 
\\[-1.8ex]\hline 
\hline \\[-1.8ex] 
 & \multicolumn{4}{c}{\textit{Dependent variable:}} \\ 
\cline{2-5} 
\\[-1.8ex] & \multicolumn{4}{c}{$r_{it} $} \\ 
\\[-1.8ex] & (1) & (2) & (3) & (4)\\ 
\hline \\[-1.8ex] 
$\beta_{it} \cdot r_{M,t}$ & 0.507$^{***}$ & 0.460$^{***}$ & 0.463$^{***}$ & 0.465$^{***}$ \\ 
  & (0.004) & (0.005) & (0.005) & (0.005) \\ 
  $\gamma_i \cdot r_{M,t} \cdot Z_{\text{ApEn}(\text{RV})_{t-1}}$ & 0.013$^{***}$ &  & 0.008$^{**}$ & 0.004 \\ 
  & (0.004) &  & (0.004) & (0.004) \\ 
  $\beta^{\text{SMB}}_{it} \cdot r_{\text{SMB},t}$ &  & 0.185$^{***}$ & 0.185$^{***}$ & 0.182$^{***}$ \\ 
  &  & (0.007) & (0.007) & (0.007) \\ 
  $\gamma_i^{\text{SMB}} \cdot r_{\text{SMB},t} \cdot Z_{\text{ApEn}(\text{RV})_{t-1}}$ &  &  &  & 0.023$^{***}$ \\ 
  &  &  &  & (0.008) \\ 
 $ \beta^{\text{HML}}_{it} \cdot r_{\text{HML},t} $&  & 0.003 & 0.004 & 0.014$^{**}$ \\ 
  &  & (0.006) & (0.006) & (0.007) \\ 
 $\gamma_i^{\text{HML}} \cdot r_{\text{HML},t}  \cdot Z_{\text{ApEn}(\text{RV})_{t-1}}$ &  &  &  & 0.034$^{***}$ \\ 
  &  &  &  & (0.006) \\ 
  $\alpha_i$ & 0.191$^{***}$ & 0.137$^{***}$ & 0.127$^{***}$ & 0.097$^{***}$ \\ 
  & (0.020) & (0.020) & (0.020) & (0.021) \\ 
 \hline \\[-1.8ex] 
Observations & 49,601 & 49,893 & 49,601 & 49,601 \\ 
R$^{2}$ & 0.215 & 0.223 & 0.225 & 0.226 \\ 
Adjusted R$^{2}$ & 0.215 & 0.223 & 0.225 & 0.226 \\ 
F Statistic & 6,793.545$^{***}$ & 4,776.498$^{***}$ & 3,599.416$^{***}$& 2,407.782$^{***}$ \\ 
&  (df = 2; 49598) &  (df = 3; 49889) & (df = 4; 49596) & (df = 6; 49594) \\ 

\hline 
\hline \\[-1.8ex] 
\textit{Note:}  & \multicolumn{4}{r}{$^{*}$p$<$0.1; $^{**}$p$<$0.05; $^{***}$p$<$0.01} \\ 
\end{tabular}}
\end{table} 

 \FloatBarrier
\subsection{VIX and Volatility Spread}

So far we have discussed the results of estimating a conditional CAPM model using ApEn estimated from realized volatility. A different perspective is offered when turning to measures of implied volatility and the spread between realized and implied volatility. As we have seen, the entropy of implied volatility is substantially lower (see Table \ref{t_basicstats3}) than that of the realized volatility. This reflects the fact that the expectation on future volatility, which implied volatility is expressing, is moving slower than the actual market volatility, showing higher persistence, resulting in a lower entropy value. For the spread between implied and realized volatility, the entropy results in Table \ref{t_basicstats3} suggests substantially higher levels of irregularity and less persistence.

The conditional CAPM model in (\ref{eq_refinedmodel}) was re-estimated over the entire sample of funds with these other volatility measures, along with the Ferson-Schadt model of equation (\ref{eq_fersonschadtcomb}).
In Table \ref{T:OLS_AllVol_1f} the results are shown. The models display results for the unconditional beta $\beta_{it}$ in the range of 0.202-0.218. The next three lines in the table display the conditional market beta $\hat{\gamma}_i$ associated with the respective volatility measures as in the Ferson-Schadt model. The beta sensitivity is strongest to a change in implied volatility, and least strong to a change in the volatility spread. 

The reaction to a change in the volatility spread $\gamma_i \cdot r_{M,t} \cdot Z_{\text{(RV-VIX)}_{t-1}}$ is lower $\hat{\gamma_i}=-0.027 (0.005)$. The result is consistent with the fact that systematic volatility strategies that are betting on changes in the spread are a minority in the sample ('CTA', 'Managed Futures' and 'Option Arbritrage' pursuing funds make less than 11\% of overall funds). Although the average hedge fund will exploit market conditions of 'cheap' volatility relative to higher expected future volatility in buying protection for example, the result is not likely to be substantial for the overall market beta of a hedge fund. 

A slightly different aspect of the story is told by looking at the results for $\delta_i \cdot r_{M,t} \cdot Z_{\text{ApEn}(x)_{t-1}} $, where the beta adjustment to a change in volatility is measured. 
The strongest reaction from funds is seen when the entropy is calculated based on the volatility spread: $\hat{\delta_i} = 0.101 (0.007)$ for $Z_{\text{ApEn}((\text{RV}-\text{VIX})_{t-1})}$. 
If we recall, a downward (upward) movement of entropy signals that the volatility (spread) is entering a new regime (settling into a new regime).  As \citeN{Bollerslev:2011fa} have shown, compensation for rare events account for a large part of the average volatility spread. Any shock to the variance risk premium (volatility spread), for example due to heightened levels of implied volatility without changes in realised volatility (or delayed response), conveys uncertainty for the further development of the market (rare events), implied volatility or realized volatility, and possibly signals a change in regime. The funds in the sample respond to this signal by reducing market exposure in a manner stronger than to the entropy of RV or VIX. Furthermore, although the funds are most sensitive to a change in implied volatility, as measured by  $ \gamma_i \cdot r_{M,t} \cdot Z_{\text{VIX}_{t-1}}$ ($\hat{\gamma}_i  = -0.037$), they are most sensitive to a possible regime change in variance risk premium, as measured by its entropy, $Z_{\text{ApEn}((\text{RV}-\text{VIX})_{t-1})}$. 

\begin{sidewaystable}
 \centering 
 \caption{Panel Regression: RV, VIX, $\sqrt{\text{RV}} - \text{VIX}$.}
 \begin{quote}
Pooled OLS model estimation over the entire fund sample of the volatility timing models as defined in the equations (\ref{eq_ch3_1}). The conditional market variables depend on realized volatility $Z_{\text{RV}_{t-1}}$, implied volatility $Z_{\text{VIX}_{t-1}}$,  and the spread between implied and realized volatility $Z_{(\sqrt{\text{RV}}_{t-1} - \text{VIX}_{t-1})}$. The estimates for the various predictors along with the standard error are shown. The statistical significance of the various estimates is shown according to the associated p-value of a two tailed hypothesis test that the value is zero. In the second lower part, statistics such as the $R^2$, adjusted $R^2$ and the F-statistics on the estimated model are shown.
\end{quote}
  \label{T:OLS_AllVol_1f}
  {\scriptsize
\begin{tabular}{@{\extracolsep{5pt}}lcccccc} 
\\[-1.8ex]\hline 
\hline \\[-1.8ex] 
 & \multicolumn{6}{c}{\textit{Dependent variable:}} \\ 
\cline{2-7} 
\\[-1.8ex] & \multicolumn{6}{c}{rtn} \\ 
\\[-1.8ex] & (1) & (2) & (3) & (4) & (5) & (6)\\ 
\hline \\[-1.8ex] 
 $ \beta_{it} \cdot r_{M,t}$ & 0.202$^{***}$ & 0.218$^{***}$ & 0.208$^{***}$ & 0.247$^{***}$ & 0.189$^{***}$ & 0.195$^{***}$ \\ 
  & (0.006) & (0.007) & (0.007) & (0.008) & (0.006) & (0.006) \\ 
  $ \gamma_i \cdot r_{M,t} \cdot Z_{\text{RV}_{t-1}}$ & $-$0.031$^{***}$ &  &  &  &  &  \\ 
  & (0.004) &  &  &  &  &  \\ 
   $ \gamma_i \cdot r_{M,t} \cdot Z_{\text{VIX}_{t-1}}$ &  & $-$0.037$^{***}$ &  &  &  &  \\ 
  &  & (0.005) &  &  &  &  \\ 
  $ \gamma_i \cdot r_{M,t} \cdot Z_{(\sqrt{\text{RV}}-\text{VIX})_{t-1}}$ &  &  & $-$0.027$^{***}$ &  &  &  \\ 
  &  &  & (0.005) &  &  &  \\ 
  $\delta_i \cdot r_{M,t} \cdot Z_{\text{ApEn}(x)_{t-1}}  $ &  &  &  & 0.042$^{***}$ & 0.060$^{***}$ & 0.101$^{***}$ \\ 
 x:  $ \text{RV}_{t-1},\text{VIX}_{t-1},(\sqrt{\text{RV}}_{t-1}-\text{VIX}_{t-1})$ &  &  &  & (0.005) & (0.006) & (0.007) \\ 
  $\nu_i \cdot r_{M,t} \cdot Z_{\text{ApEn}(\text{RV})_{t-1}}\cdot Z_{\text{RV}_{t-1}}  $ &  &  &  & $-$0.002$^{***}$ &  &  \\ 
  &  &  &  & (0.0002) &  &  \\ 
  $\nu_i \cdot r_{M,t} \cdot Z_{\text{ApEn}(\text{VIX})_{t-1}}\cdot Z_{\text{VIX}_{t-1} } $ &  &  &  &  & $-$0.018$^{***}$ &  \\ 
  &  &  &  &  & (0.003) &  \\ 
  $\nu_i \cdot r_{M,t} \cdot Z_{\text{ApEn}(\sqrt{\text{RV}}-\text{VIX})_{t-1}}\cdot Z_{(\sqrt{\text{RV}}_{t-1} -\text{VIX}_{t-1} )} $ &  &  &  &  &  & $-$0.023$^{***}$ \\ 
  &  &  &  &  &  & (0.002) \\ 
  $\alpha_i$& 0.162$^{***}$ & 0.161$^{***}$ & 0.157$^{***}$ & 0.129$^{***}$ & 0.200$^{***}$ & 0.155$^{***}$ \\ 
  & (0.026) & (0.026) & (0.026) & (0.026) & (0.026) & (0.026) \\ 
 \hline \\[-1.8ex] 
Observations & 116,713 & 117,272 & 117,272 & 116,713 & 117,272 & 117,272 \\ 
R$^{2}$ & 0.009 & 0.010 & 0.009 & 0.010 & 0.010 & 0.011 \\ 
Adjusted R$^{2}$ & 0.009 & 0.010 & 0.009 & 0.010 & 0.010 & 0.011 \\ 
F Statistic & 553.230$^{***}$ & 565.588$^{***}$ & 549.599$^{***}$ & 393.833$^{***}$  & 386.244$^{***}$ & 427.753$^{***}$ \\ 
 &  (df = 2; 116710) &  (df = 2; 117269) &  (df = 2; 117269) &  (df = 3; 116709) &  (df = 3; 117268) &  (df = 3; 117268) \\ 

\hline 
\hline \\[-1.8ex] 
\textit{Note:}  & \multicolumn{6}{r}{$^{*}$p$<$0.1; $^{**}$p$<$0.05; $^{***}$p$<$0.01} \\ 
\end{tabular}  }
\end{sidewaystable}

\FloatBarrier
 \section{Conclusion}
In this chapter, we use for the first time approximate entropy (ApEn) in a conditional CAPM model to identify volatility timing behaviour in hedge funds. We show the use of ApEn in measuring changes in the serial structure of volatility. Various numerical simulations showcase the pattern of ApEn in economic scenarios such as a change of volatility regime. We derive conditions for low levels of entropy analytically in popular volatility models and processes, demonstrating the usefulness of ApEn in distinguishing different volatility patterns. 
We find that entropy is more sensitive to volatility shocks than to serial correlation. This way ApEn is a useful and model free measure of a change in volatiltity regime. 

On a sample of hedge funds, we found that manager adjust the funds' market beta when volatility is settling into a new regime. Dead funds show no such market beta adjustment in response to a change in volatility regularity, differing from live funds where such an adjustment is much stronger. The results are consistent with the findings in \citeN{Busse:1999hi} on mutual funds. For 'Long-Short Equity' funds, managers adjust their risk exposure to SMB and HML, but not the market beta, when there is a change in volatility regime. The results we obtained with the conditional CAPM are robust when extending the model to a conditional Fama-French model. The results are robust when the entropy is calculated using VIX and spread (defined as the difference between VIX and realised volatility), though the qualitative interpretation of the results is slightly different. Our research also showcase the usefulness of entropy as a model independent measure of volatility regime change.

\chapter{Conclusion}\label{chap_conclusion}
In this thesis, three studies have been conducted, which have the application of information theoretic tools as a common thread. The first study formulates a model to analyse, for the first time, \textit{draws} in a bivariate setting. Encoding the draw information into a time series of discrete values, we used entropy on the derived series to analyse the correlation and cross-correlations of drawdowns and drawups. We investigated the daily and hourly exchange rates of EUR/USD and GBP/USD from 2001 to 2012. For the hourly and the daily data, we found evidence of dependence among the largest draws. We were also able to identify information flows between drawup/drawdown of the two exchange rates, both for the hourly and daily data. Both daily and hourly series show clear evidence of information flowing from EUR/USD to GBP/USD and, in a slightly more pronounced way in the reverse direction. Robustness tests show that the information measured is not due to noise, from which we conclude that there is a measurable information transfer between the large draws of the two exchange rates, which is potentially useful for forecasting and risk management.

In the second study, we defined a two-state hidden Markov model to identify volatility regimes in a set of currencies quoted against the US dollar. The estimated state process is a process taking discrete values, which lends itself to analysis with information theoretic tools. Following the procedures in the previous study, we used the entropy framework with mutual information and transfer entropy to examine interdependence and volatility spillover among the currencies in the sample in a novel way. We found evidence of various volatility regime relationships between the currency pairs in the sample. Among the European currencies (EUR/USD, GBP/USD, CHF/USD), volatility regime co-movements were identified. We also found evidence of a co-movement relationship betwen CAD/USD and AUD/USD volatility states. We were able to find evidence of information flows between the volatility states of currency pairs, most notably from EUR/USD to GBP/USD, which indicates a causal volatility spillover relationship, confirming findings in the literature (\citeN{Inagaki:2006kj}), but using a different model and set-up. The work showcased the usefulness of the concepts of mutual information and transfer entropy when applying them to volatility spillover studies.

For the last study, we extended the market timing model of \citeN{Ferson:1996je}, using the approximate entropy (ApEn) of market volatility series as a state variable. We demonstrated the sensitivity of ApEn to volatility regime changes and serial correlation patterns, and used it in the conditional factor model. The analysis adds to the market volatility timing research for hedge funds using ApEn as a model independent measure for volatility regime changes. On a sample of hedge funds, we found that the managers adjust a fund's market beta when volatility is settling into a new regime. Dead funds show no such market beta adjustment skill, differing from live funds, where such an adjustment can be identified.

\clearpage
\onehalfspacing
\bibliography{DBA20150501}

\end{document}